\documentclass[apj,numberedappendix]{emulateapj}

\usepackage{morefloats}
\usepackage{graphicx}
\usepackage{multirow}
\usepackage{epsfig}
\usepackage{color}
\usepackage{journals}
\usepackage{amsmath,amssymb,latexsym}
\usepackage{listings}

\newcommand{\Msun}{M_{\odot}}
\newcommand{\Zsun}{Z_{\odot}}

\allowdisplaybreaks

\shorttitle{Chemical evolution library}
\shortauthors{Saitoh}

\begin{document}

\title{Chemical evolution library for galaxy formation simulation}

\author{Takayuki \textsc{R.Saitoh}\altaffilmark{1}
}
\altaffiltext{1}{Earth-Life Science Institute, Tokyo Institute of
Technology, 2--12--1, Ookayama, Meguro, Tokyo, 152-8551, Japan}
\email{saitoh@elsi.jp}

\begin{abstract}
We have developed a software library for chemical evolution simulations of
galaxy formation under the simple stellar population (SSP) approximation. In
this library, all of the necessary components concerning chemical evolution,
such as initial mass functions, stellar lifetimes, yields from type II and Ia
supernovae, asymptotic giant branch stars, and neutron star mergers, are
compiled from the literature. Various models are pre-implemented in this library
so that users can choose their favorite combination of models.  Subroutines of
this library return released energy and masses of individual elements depending
on a given event type.  Since the redistribution manner of these quantities
depends on the implementation of users' simulation codes, this library leaves it
up to the simulation code.  As demonstrations, we carry out both one-zone,
closed box simulations and three-dimensional simulations of a collapsing gas and
dark matter system using this library. In these simulations, we can easily
compare the impact of individual models on the chemical evolution of galaxies,
just by changing the control flags and parameters of the library.  Since this
library only deals with the part of chemical evolution under the SSP
approximation, any simulation codes that use the SSP approximation -- namely
particle-base and mesh codes, as well as semi-analytical models -- can use it.
This library is named ``CELib'' after the term ``Chemical Evolution Library''
and is made available to the community.
\end{abstract}


%@arxiver{fig1.eps,fig25.eps,fig33.eps}

\section{Introduction} \label{sec:intro}

% What is chemical evolution?
Our Universe starts with hydrogen (H), helium (He) and a small amount of
Lithium (Li). These elements are synthesized during the Big Bang and all the
other elements equal to or heavier than carbon (C) are synthesized in stars and
during supernovae.  In astrophysics, the history of heavy element generation via
the nucleosynthesis in stars and succeeding pollution of the interstellar medium
(ISM) are expressed as ``chemical evolution''.

During stellar evolution, nucleosynthesis progresses from H to iron (Fe) where
the final products depend on the mass of a star.  The massive stars heavier than
$8~\Msun$ are thought to explode as type II supernovae (hereafter SNe II). In
this phase, heavy elements ($>$ Fe) are synthesized.  If a star is not so
massive, say $\le 8~\Msun$, it experiences an asymptotic giant branch (AGB)
phase after the main sequence phase.  During the AGB phase, the neutron capture
process takes place with a slow speed (slow means that the timescale of the
$\beta$ decay, $\tau_\beta$, is comparable to that of the neutron capture,
$\tau_n$).  This slow process ($s$-process) generates elements heavier than Fe
efficiently.  In addition, a part of the intermediate mass stars' binary systems 
is thought to explode as a type Ia SNe (hereafter SNe Ia).  SNe Ia are
triggered by the Roche-lobe overflows or the final coalescences of binary
systems.  In the former situation, one of the progenitor star in a binary is a
degenerated star and, in the latter, both are degenerated stars.  Thus, these
two scenarios are called single-degenerate or double-degenerate scenarios
\citep{WhelanIben1973, Nomoto1982, IbenTutukov1984, Webbink1984}.  Again,
heavier elements are synthesized, such as Fe and nickel (Ni). Numerical studies
tell us that neutron star-neutron star mergers (NSMs) could be a possible site
of $r$-process ($\tau_\beta \gg \tau_n$) elements, in particular, the heavier
r-process elements \citep[e.g.,][]{Freiburghaus+1999, Goriely+2011, Wanajo+2014}.

% Chemical evolution in galaxies; time log of their evolution.
Each process has its own characteristic time-scale and abundance pattern. Thus,
they work as the chronometer for the formation history of the galactic stellar
system. For instance, in an early phase ($<10^8~{\rm yr}$), the chemical
evolution progresses mainly through the pollution by SNe II. With this process,
the $\alpha$ elements of which the masses are multiples of He, such as
oxygen (O), magnesium (Mg), silicon (Si), and calcium (Ca), are ejected to the
interstellar medium (ISM). On the other hand, in SNe Ia,
Fe, and Ni are efficiently released to the ISM. The combination of these events
makes the observed features of the plateau (flat part) of the [$\alpha$/Fe] as a function
of [Fe/H]{\footnote{$[{\rm A/B}] \equiv \log_{10} (n_{\rm A}/n_{\rm B}) -
\log_{10} (n_{\rm A}/n_{\rm B})_{\odot}$, where $n_{\rm A}$ and $n_{\rm B}$ are
the number abundance of element A and B, respectively.}} in a low metallicity
part, and causes the distribution of [$\alpha$/Fe] as a function
of [Fe/H] \citep[e.g.,][]{Tinsley1980}.  The flat part represents the abundance
pattern of SNe II. The decreasing part is composed of the mixture of SNe II and SNe
Ia where the contribution of SNe II is gradually decreasing, reflecting
its time-scale, and that of SNe Ia is gradually increasing.  The breaking point
of the flat part reflects the time scale of the enrichment of the chemical
composition in the ISM. The breaking point is [Fe/H] $= -1 \sim -0.5$ in the
Milky Way galaxy \citep[e.g.,][]{Hayden+2015}, whereas it is much lower in the
local dwarf galaxies \citep[e.g.,][]{Tolstoy+2009ARAA}.  AGBs do not mainly
contribute to the distributions of [$\alpha$/Fe]-[Fe/H] relations while they can
contribute rather low mass elements such as C and nitrogen (N).

There are many galactic surveys which resolve the chemical compositions and
kinematic information of a lot of stars, e.g., RAVE \citep{Steinmetz+2006},
APOGEE \citep{Majewski+2016}, Gaia-ESO \citep{Gilmore+2012}, HERMES (GRASH)
\citep{DeSilva+2015}.  The data obtained by these surveys can make a strong
constraint on the formation and evolution of the Milky Way galaxy and thus, it
becomes a clue to understanding the formation and evolution of galaxies in
general.

It has been almost a quarter of a century since the first 3-dimensional
simulations of galaxy formation, including dark matter and baryon.  The first
galaxy formation simulation including chemical evolution was carried out by
\cite{SteinmetzMueller1994, SteinmetzMuller1995}. In this simulation, they only
solved the evolution of the metallicity $Z$, the mass fraction of the heavy
elements, synthesized by SNe II. Then, \cite{Raiteri+1996} introduced both SNe
II and Ia.  The models have improved further and the current concordance
simulations include about 10 elements \citep [e.g.,][]{Mosconi+2001,
KawataGibson2003, Scannapieco+2005, Okamoto+2005, Okamoto+2008, Wiersma+2009,
KobayashiNakasato2011, Rahimi+2011, Few+2012, Vogelsberger+2013, Brook+2014,
Few+2014, Snaith+2016}, as well as the mixing of them in the ISM
\citep{Greif+2009, Shen+2010}.  Semi-analytical models are also considered, in
which various elements are released from different feedback models
\citep[e.g.,][]{Nagashima+2005, Cora2006, Arrigoni+2010, Yates+2013,
Gargiulo+2015}.  Unlike solving the evolution of individual stars, simulations
of galaxy formation use a simple stellar populations (SSP) approximation, where
star particles consist of a cluster of stars sharing the same age and
metallicity, and whose mass function follows a certain initial mass function
(hereafter IMF).

In order to solve chemical evolution, the amount of newly synthesized elements
in stars, ``yields'', and the return mass fraction of SSP particles are
necessary. These data are obtained from the studies of stellar evolution and
explosions. There are flexibilities to choose yields tables as well as the
functional forms of IMF and its mass range. The most popular yields for SNe II
are \cite{WoosleyWeaver1995} and its improved version supplied by
\cite{Portinari+1998}.  For SNe Ia, the yields of \cite{Nomoto+1997} and its
updated version given by \cite{Iwamoto+1999} are well-used. For AGBs, the yields
tables of \cite{vandenHoekGroenewegen1997}, \cite{Portinari+1998}, and
\cite{Marigo2001} are widely used. The combinations of yields tables can lead to
different chemical abundance patterns and thus it is intensively studied
\citep[e.g.,][]{Francois+2004, Wiersma+2009, Romano+2010, Few+2012}.

It is pointed out by the galactic chemical evolution model of \cite{Timmes+1995}
that the Fe yields of \cite{WoosleyWeaver1995} are slightly large in order to
obtain good agreement with observations.  {\footnote {\cite{Nomoto+2013}
summarized the problems of \citet{Portinari+1998}'s yields tables as (1) the Fe
yield is overestimated, (2) the mass loss affects the final C+O core structure,
but they ignored the evolution effects, (3) the Mg yield with a mass of
$40~\Msun$ is too small.}} Usually the Fe yield is reduced by a factor of 2 from
the original values \citep[e.g.,][]{Timmes+1995, Gibson1997, Gibson+1997}.  The
yields tables of \cite{Portinari+1998}, which is an extended version of
\cite{WoosleyWeaver1995}, also have the same problem.  Sometimes ad hoc
modifications are used to halve the yield of Fe and double those of Mg and C in
order to fit the observations \cite[see][]{Wiersma+2009}. Since these yield
tables are widely used in simulations of galaxy formation, these modifications
are commonly used.

An  essential solution to solve this problem is to update the yields tables following
the progress of stellar evolution models and reaction rates.  However, it might
be difficult to update yields tables frequently since, in simulation code, the
chemical evolution part is deeply connected to them. Moreover, the size and
range of mass and metallicity grids of the yields tables are generally different
from each other. This inhibits the smooth replacement of yields tables, although
it is a theoretically straightforward task.

If the chemical evolution part is implemented separately in a software library
which can be linked from simulation codes directly and it has clearly defined
application interfaces (APIs), the chemical evolution part is developed
independently and it becomes easy to take the latest models.  Insofar as we
know, such software library does not exist.

% SB99 & NuGrid
There are several softwares which can generate stellar yields tables.  {\tt
Starburst99}{\footnote{{\tt http://www.stsci.edu/science/starburst99/}}} is a
widely-used software to model spectrophotometric properties of star-forming
galaxies \citep{Leitherer+1999, VazquezLeitherer2005, Leitherer+2010,
Leitherer+2014}.  This software contains a huge database of stellar evolutions
and thus, it can output yields.  Since this software is not designed to link
from other codes, it is necessary to pre-generate a yields table if one uses it
in one's simulation code. Thus, when parameters change, e.g., the functional
form and the mass range of IMF, the reconstruction of the yields table is
necessary.
The NuGrid collaboration{\footnote{{\tt http://nugridstars.org/}}} has developed
a Python code, {\tt SYGMA}, which can deal with the yields of the NuGrid project
and can be used in chemical evolution of galaxies \citep{Cote+2016OMEGA}.  Since
this is implemented as an iPython notebook, users need to pre-generate yields in
order to use the outcomes from {\tt SYGMA}.

% Aim of this paper
The aim of this paper is to describe a software library 
which deals with chemical evolution via APIs and to demonstrate the capability
of this library.  This library is named ``CELib'' after the term ``Chemical
Evolution Library''.  Since the primary purpose of this library is to use it
with simulation codes of galaxy formation, this library works under the SSP
approximation.  Note that its main function is to return yields (and energy) of
various feedback types from SSP particles.  Redistribution of these quantities
is not within its scope because there are a number of redistribution manners and
the implementation of them is deeply coupled with simulation codes. Models and
parameters of this library are selectable at the runtime and the IMF weighted
yields are automatically generated in the initialization process.

CElib is an open-source software library and is released under the MIT license.
The whole source codes including some examples are distributed through the
following website: {{\tt https://bitbucket.org/tsaitoh/celib}}.  A documentation
is also available in this site.

% Structure of this paper
The structure of this paper is as follows.  First, we briefly describe the
design concept of this library in \S \ref{sec:design}. Then we describe the
notations and definitions in \S \ref{sec:Notations} and we introduce components
which this library consists of in \S \ref{sec:Components}.  The reference
feedback models of SNe II/Ia, AGBs and NSMs are described in \S
\ref{sec:modeling}.  In \S \ref{sec:Implementation}, we explain our
implementation of this library.  In \S \ref{sec:Applications}, we show the
results of the simple one-zone model and galaxy simulation using this library.
Finally, we provide the summary in \S \ref{sec:Summary}.

\section{Design Concept} \label{sec:design}

The key concepts of the design of CELib are as follows.
\begin{itemize}
\item CELib returns yields and energy from SSP particles, based on the adopted
IMF and yields tables.  All returned quantities are IMF weighted ones.  Useful
data, such as the functional form of IMF and stellar lifetimes, are also
implemented.
\item CELib supplies reference feedback models of SNe II/Ia, AGBs and NSMs.
Reference feedback models can provide the event time and the amount of released
mass, metal and energy. A redistribution manner of these quantities is outside
its scope. 
\item The functional form of the IMF, its mass ranges, yields tables, etc. are
changed by using control flags and parameters. All changes of flags
and parameters are reflected when the initializer of CELib is called.  
\item All communication between a user simulation code and CELib is carried out
through CELib APIs.
\end{itemize}

\begin{figure}
\centering
\epsscale{1.0}
\plotone{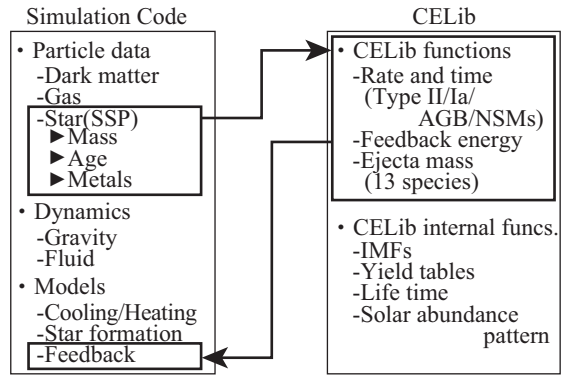}
\caption{Relation between a user's simulation code and this chemical evolution
library, CELib.  The left rectangular box with the thin line represents the
simulation code.  It has particles data, functions which evaluate dynamical
interactions, and models for the baryon physics. The right rectangular box with
the thin line is CELib. CELib consists of functions for chemical evolution and
they can use via APIs.  For instance, when an SSP particle is born in the
simulation code, it asks CELib to get the time of SNe II. When the SSP particle
reaches the event epoch, the simulation code sends information regarding the SSP
particles and gets the amount of released energy and ejected metals of 13
elements.  When these event times come, the code again sends information of the
SSP particle in order to obtain feedback results.  The procedures of feedback
from SNe Ia, AGB and NSMs are identical.
}
\label{fig:Concept}
\end{figure}

Figure \ref{fig:Concept} shows the schematic picture of the relation between a
user's simulation code and CELib.  The simulation code deals with the time
integration of a system solving gravitational interactions among DM, gas, and
star (SSP) particles and hydrodynamics, involving baryon physics such as
radiative cooling/heating, star formation, and feedbacks.  CELib is responsible
for the chemical evolution part.  Event time, released energy and ejected metals
are evaluated when the simulation code sends data of an SSP particle to CELib
via APIs. What the simulation code needs to do is to add the released energy and
the ejected metals to the surrounding ISM and to reduce the mass of the SSP
particle so that the mass is conserved.

\section{Notations and Definitions} \label{sec:Notations}

% Yields 
Before explaining details of CELib, we need to confirm the notations and what
quantities are used and obtained by using the library. We describe them in this
section.

The primary purpose of this library is to calculate the stellar yields from
various evolution phases of stars under the SSP approximation.  The stellar
yield, the ejected mass of the $i$-th element from a star whose mass is $m$, is
expressed as  
\begin{equation}
Y_i (m) = y_i(m) + Z_i^0 m_{\rm ej} (m), \label{eq:StellarYields}
\end{equation}
where $y_i(m)$ is the {\it net yield} which represents that the newly
synthesized mass of $i$-th element, $Z_i^0$ is the metallicity of the $i$-th
element originally presented in the progenitor star and $m_{\rm ej} (m)$ is the
total ejecta mass of the star. This quantity, $Y_i (m)$, is always larger than
or equal to zero.  The total ejecta mass is 
\begin{equation}
m_{\rm ej} (m) = m-m_{\rm rem}(m),
\end{equation}
where $m_{\rm rem}(m)$ is the remnant mass of a star whose mass is $m$.

It is essential to use the net yields for the study of the galactic
chemical evolution. Rearranging Eq. \eqref{eq:StellarYields}, we can obtain
\begin{equation}
y_i (m) = Y_i(m) - Z_i^0 m_{\rm ej} (m). \label{eq:NetYields}
\end{equation}
We gather this quantity from the literature. However, some studies provide not
$y_i (m)$ but $Y_i (m)$ or $Y_i(m)-m_{{\rm sw},i} (m)$ where $m_{{\rm sw},i}
(m)$ is the ejected mass of the $i$-th element via stellar winds. In such cases,
we reconstruct $y$ carefully with several reasonable assumptions.  This quantity
can be positive and negative, unlike the stellar yield $Y$.  We use $y$ for SNe
II and AGBs, whereas we adopt $Y$ for SNe Ia and NSMs.

Since this library is used under the SSP approximation, the IMF weighted yield
is useful. Here, we use
\begin{equation}
p_i (m_{\rm L} < m < m_{\rm U}) = \int_{m_{\rm L}}^{m_{\rm U}} y_i(m) \frac{\xi (\log_{10} m)}{m} dm,
\label{eq:IMFWeightedYileds}
\end{equation}
where $\xi (\log_{10} m)$ is the IMF function (see below for the concrete functional forms).
With this quantity and the return mass fraction,
\begin{equation}
E_{\rm ret} (m_{\rm L} < m < m_{\rm U}) = \int_{m_{\rm L}}^{m_{\rm U}} m_{\rm ej} (m)
\frac{\xi (\log_{10} m)}{m} dm,
\label{eq:ReturnMassFraction}
\end{equation}
we can evaluate the yield of an SSP particle as
\begin{align}
E_i (m_{\rm L} < m < m_{\rm U}) &= p_i (m_{\rm L} < m < m_{\rm U}) \nonumber\\
&   + Z_i^0 E_{\rm ret} (m_{\rm L} < m < m_{\rm U}). \label{eq:SSPYields}
\end{align}
By multiplying the mass of the SSP particle to Eq. \eqref{eq:SSPYields}, we can
obtain the amount of the released mass of $i$-th element from a feedback
event.

The IMF weighted released amount of energy from massive stars can be evaluated
by the following equation:
\begin{equation}
E_{\rm en} (m_{\rm L} < m < m_{\rm U}) = \int_{m_{\rm L}}^{m_{\rm U}} e_{\rm en} (m)
\frac{\xi (\log_{10} m)}{m} dm,
\label{eq:ReturnEnergy}
\end{equation}
where, $e_{\rm en}(m)$ is the released energy from a star whose mass is $m$.

Strictly speaking, the released mass of $i$-th element changes depending on the
amount of the other elements.  The $Q_{ij}$ formulation \citep{TalbotArnett1973,
Ferrini+1992} is known as a model which considers the effect of the change in
amounts of other elements. We do not implement the $Q_{ij}$ formulation
here in CELib, for simplicity.

\section{Components} \label{sec:Components}

In this section, we describe key components of CELib.  IMFs which are
pre-implemented are shown in \S \ref{sec:IMF}. The metallicity dependent stellar
lifetimes are explained in \S \ref{sec:LifeTime}.  Then, thirteen elements in
CELib are explained in \S \ref{sec:Elements}. Yields for SNe II/Ia, AGBs and
NSMs are described in \S \ref{sec:SNII}, \S \ref{sec:SNIa}, \S \ref{sec:AGB},
and \S \ref{sec:NSM}, respectively.  Released energy and return mass from an SSP
particle with various IMFs are summarized in \S \ref{sec:Misc}.  In \S
\ref{sec:abundance}, the implemented solar abundance patterns are shown.

\subsection{Selectable IMFs in this library} \label{sec:IMF}

CELib pre-implements seven popular IMFs for ordinary population stars
(population I/II; hereafter Pop I/II) and users can choose which one to use.
Moreover, CELib supports an IMF for population III (hereafter Pop III) stars,
which is utilized only for extremely low metal SSPs.  CELib does not allow the
functional form of the IMF to be changed depending on the situation in a single
simulation, except for the case using a Pop III IMF.  When the contribution of
Pop III stars is adopted, the functional form of the IMF and mass range are
changed depending on metallicity.

The definition of the IMF is
\begin{equation}
\xi(\log_{10} m) = \frac{dN}{d\log_{10} m}, \label{eq:IMF:IMF}
\end{equation}
where $N$ is the number of stars in a given mass bin ($d\log_{10} m$). 
The explicit functional forms of the IMF users can choose will be explained below.

The normalization of the IMF is done by the following equation:
\begin{equation}
\int_{m_{\rm IMF,L}}^{m_{\rm IMF,U}} \xi(\log_{10} m) dm = 1.
\label{eq:IMF:Normalization}
\end{equation}
Here, $m_{\rm IMF,L}$ and $m_{\rm IMF,U}$ are the lower and upper mass
boundaries of the IMF.  Substituting Eq. \eqref{eq:IMF:IMF} into Eq.
\eqref{eq:IMF:Normalization}, we have
\begin{align}
\int_{m_{\rm IMF,L}}^{m_{\rm IMF,U}} \xi(\log_{10} m) dm &=
\int_{m_{\rm IMF,L}}^{m_{\rm IMF,U}}
\frac{dN}{d\log_{10} m} dm, \\
&=1.
\end{align}
Thus, the normalization is ``the mass weighted number of stars''.  Because of
this definition, the number of stars per $1 \Msun$ of a given mass range
($m_{\rm L} < m <m_{\rm U}$) can be calculated as
\begin{equation}
N = \int_{m_{\rm L}}^{m_{\rm U}} \frac{\xi(\log_{10} m)}{m} dm  = 
\int_{m_{\rm L}}^{m_{\rm U}} dN. \label{eq:IMF:Number}
\end{equation}

The current version of CELib supports seven well used IMFs for Pop I/II
and one IMF for Pop III stars.  They are the Salpeter, Diet Salpeter,
Miller-Scalo, Kennicutt, Kroupa(2001 \& 1993), Chabrier, and Susa IMFs.  We
describe the functional form of each IMF below.

The Salpeter IMF \citep{Salpeter1955} is the most fundamental one
which has a power law form with a single segment:
\begin{equation}
\xi(\log_{10} m) \propto m^{-1.35}  \\ \hspace{0.5cm}0.1~\Msun < m < 120~\Msun.
\end{equation}

The second IMF is the diet Salpeter IMF.  The lower part of the IMF is reduced
to be a flat profile, whereas the higher part is the same as the original
Salpeter IMF.  This IMF is originally introduced in order to explain the
observed mass-to-light ratio of disk galaxies \citep{BelldeJong2001}.  The form
of the diet Salpeter IMF is 
\begin{equation}
\xi(\log_{10} m) \propto
\begin{cases}
m^{0} & 0.1~\Msun < m < 0.6~\Msun, \\ 
m^{-1.35} & 0.6~\Msun  < m < 120~\Msun.
\end{cases}
\end{equation}

The Miller-Scalo IMF \citep{MillerScalo1979} has a power law form with three
segments:
\begin{equation}
\xi(\log_{10} m) \propto
\begin{cases}
m^{-0.4} & 0.1~\Msun < m < 1~\Msun, \\ 
m^{-1.5} & 1~\Msun < m < 10~\Msun, \\ 
m^{-2.3} & 10~\Msun  < m < 120~\Msun.
\end{cases}
\end{equation}
The high (low) mass end has a deeper (shallower) power index when we compare it
with that of the Salpeter IMF.

The functional form of the Kroupa IMF \citep{Kroupa2001} CELib adopts is as
follows:
\begin{equation}
\xi(\log_{10} m) \propto
\begin{cases}
m^{-0.3} & 0.1~\Msun < m < 0.5~\Msun, \\ 
m^{-1.3} & 0.5~\Msun < m < 120~\Msun. \\ 
\end{cases}
\end{equation}
Note that the minimum mass of a star which is followed by the original form of
the Kroupa's IMF is $0.01~\Msun$, but we do not take them %for simplicity 
and adopt an IMF greater than $0.1~\Msun$.

In addition to the Kroupa IMF \citep{Kroupa2001}, CELib also has the IMF of
\cite{Kroupa+1993} (hereafter Kroupa1993 IMF). The functional form of this IMF
is 
\begin{equation}
\xi(\log_{10} m) \propto
\begin{cases}
m^{-0.3} & 0.1~\Msun < m < 0.5~\Msun, \\ 
m^{-1.2} & 0.5~\Msun < m < 1.0~\Msun, \\ 
m^{-1.7} & 1.0~\Msun < m < 120~\Msun. \\ 
\end{cases}
\end{equation}
As is obvious, the slope of this IMF in the mass regime at $>1~\Msun$ is steeper
than that of the \cite{Kroupa2001}'s IMF.

\citet{Kennicutt1983} has provided the IMF with the following form:
\begin{equation}
\xi(\log_{10} m) \propto 
\begin{cases}
m^{-0.4}, & 0.1~\Msun < m < 1~\Msun \\ 
m^{-1.5}. & 1~\Msun < m < 120~\Msun 
\end{cases}
\end{equation}

The final IMF for ordinary populations of stars is the Chabrier IMF
\citep{Chabrier2003}. This IMF has a rather complicated form since the lower
part of it is expressed by a log-normal function.  The values of the
normalization parameters are given by \cite{Chabrier2003}.  The functional form
of the IMF is
\begin{equation}
\xi(\log_{10} m) \propto
\begin{cases}
m^{0} \exp \left \{ -\frac{ \left [\log_{10} \left (\frac{m}{m_{\rm crit}} \right )\right ]^2}{2 \sigma^2} \right \}  
\\ \hspace{1cm} 0.1~\Msun < m < 1~\Msun, \\ 
m^{-1.3} \\ \hspace{1cm} 1~\Msun < m < 100~\Msun,
\end{cases}
\end{equation}
where $m_{\rm crit} = 0.079$ and $\sigma = 0.69$.

For the IMF of Pop III stars, we use the following function:
\begin{align}
\xi(\log_{10} m) &\propto 
\exp \left \{ -\frac{ \left [\log_{10} \left (\frac{m}{m_{\rm popIII}} \right )\right ]^2}
{2 \sigma_{\rm popIII}^2} \right \}  \\
&\hspace{1cm} 0.7~\Msun < m < 300~\Msun, \nonumber
\end{align}
where $m_{\rm popIII} = 22.0$ and $\sigma_{\rm popIII} = 0.5$.  This functional
form is evaluated by using the eye-ball fitting of the mass distribution of
\cite{Susa+2014} (their figure 9). This IMF is used only when
the Pop III mode turns on.  It should be noted that there is no consensus on the
shape of the Pop III IMF and other researchers obtain different functional forms
of Pop III IMFs \cite[e.g.,][]{Komiya+2007, Greif+2011, Hirano+2014,
Hartwig+2015, Hirano+2015, Fraser+2015, Stacy+2016}.  It is formally not
difficult to extend this library to use other functional forms for Pop III
stars.
 
The normalization coefficient of each IMF for the given mass range is evaluated
at the initialization phase of the library following Eq.
\eqref{eq:IMF:Normalization}.

% Minimum and maximum mass of stars.
The contributions of the lower and upper mass boundaries of Pop I/II IMFs are
not crucial for the whole evolution of a system.  For the lower mass boundary, we
adopted $0.1~\Msun$ for all these IMFs for simplicity.  There are some
variations, but the contribution of the lower mass limit is almost negligible
because of their long lifetimes (see \S \ref{sec:LifeTime}). For the upper mass
boundary, we assumed $100~\Msun$ for the Chabrier IMF and $120~ \Msun$ for the
others.  This difference is, again, negligible, because the contribution of
massive stars is insignificant due to steeply decreasing profiles.  Note that
the lower and upper mass ends of the IMF are changeable from the fiducial ones
in this library.

For Pop III stars, we adopt an entirely different mass range,
$0.7$--$300~\Msun$, reflecting the potential difference in formation of Pop III
stars from ordinary stars due to their inherent cooling processes
\citep[e.g.,][]{OmukaiNishi1998, Abel+2002, Bromm+2002, Yoshida+2008,
Omukai+2010, Hosokawa+2011, Bromm2013, Glover2013, Susa+2014, Hirano+2015}.

\begin{figure}
\centering
\epsscale{1.0}
\plotone{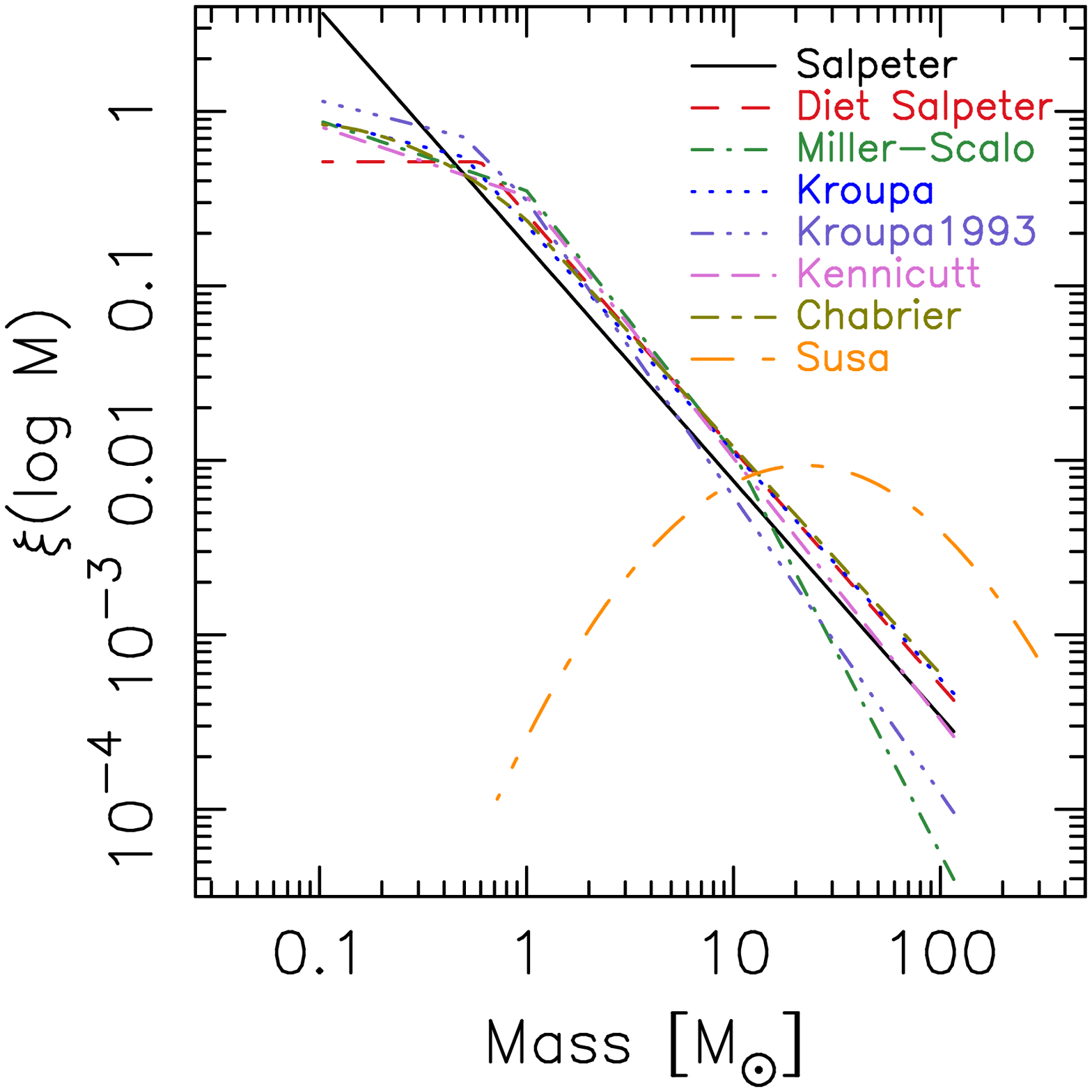}
\plotone{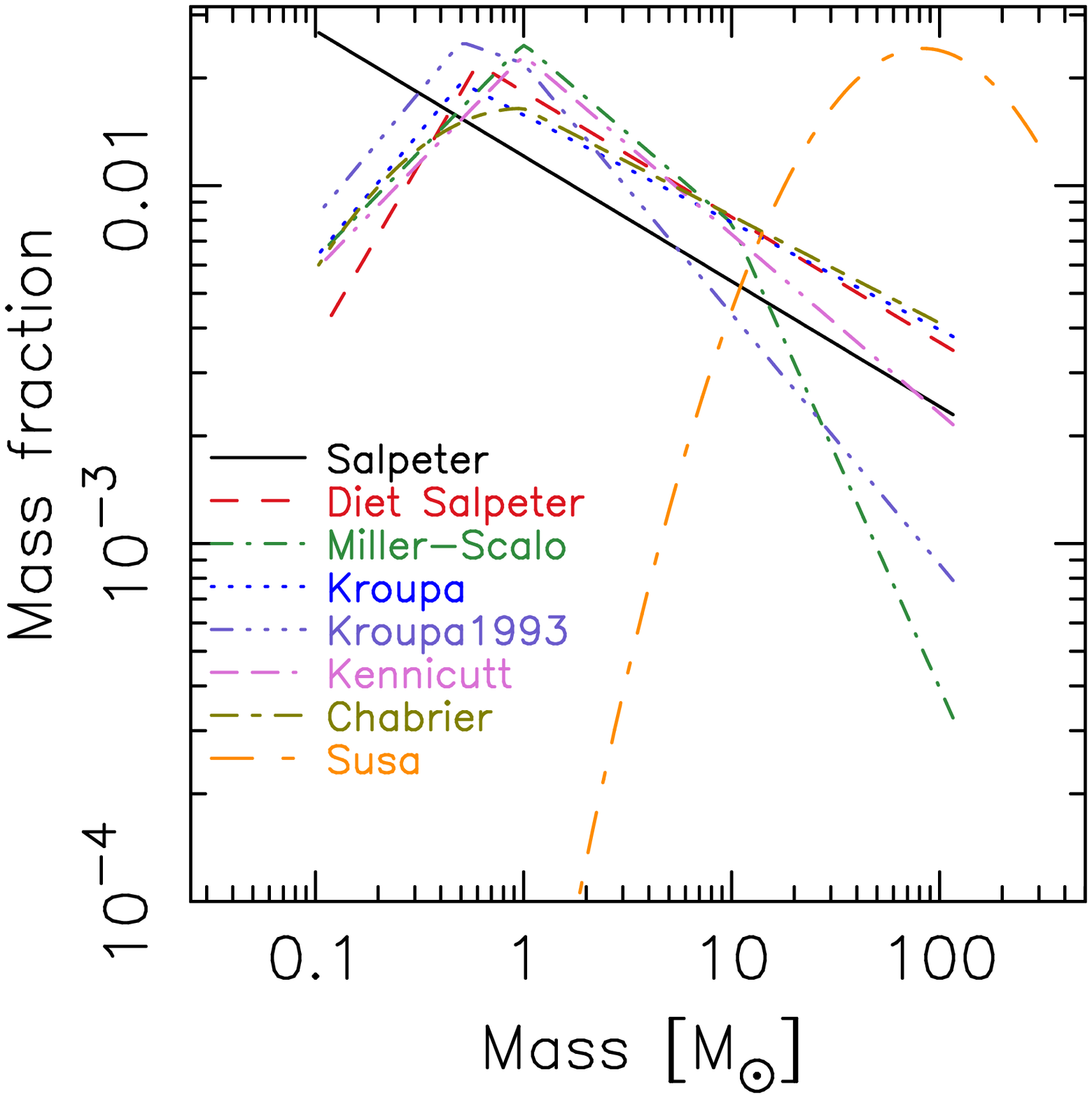}
\caption{$\xi(\log_{10} m)$ (top panel) and mass fraction (bottom panel) of seven
IMFs. The normalization of each IMF is carried out by using Eq
\eqref{eq:IMF:Normalization}.
\label{fig:IMFs}
}
\end{figure}

In figure \ref{fig:IMFs}, we compare the shapes of these seven IMFs.  For
ordinary populations, we can see that stars with masses of $\sim 1~{\Msun}$
dominate in the six IMFs except for the Salpeter IMF; it has many more low-mass
stars.  We also find a substantial variation at the high mass end; high mass end
Pop I/II stars make the largest contribution in the Chabrier IMF, and the
smallest in the Miller-Scalo and Kroupa1993 IMFs.  These differences reflect the
amount of energy released by SNe II (see \S \ref{sec:modeling:SNII} and table
\ref{tab:SNIIEnergy}).

The Pop III IMF has a peak at $\sim 20~\Msun$ which surely follows the mass
spectrum of the Pop III in \cite{Susa+2014}.  It is noted that the very massive
stars ($>140~\Msun$) with $Z=0$ results in pair-creation instability SNe
\citep{Barkat+1967} and so far there is no clear relic which predicts the
existence of the pair-creation instability SNe \citep{UmedaNomoto2002,
UmedaNomoto2005, Kobayashi+2011PISNe}. With our Pop III IMF, the contribution of
the pair-creation instability SNe is negligible and thus, there is no
discrepancy. 

There is a possibility that the IMF is not universal.  For instance, the top-heavy
IMF is favored to obtain a sufficient number of faint sub-mm galaxies in
semi-analytical models \citep{Baugh+2005, Nagashima+2005} \citep[see also][which
does not require a top-heavy IMF to reproduce the number count of sub-mm
galaxies]{Hayward+2013}.  Also, it is used in galaxy formation simulations
\citep{Okamoto+2005}. As is described in the above studies, the top-heavy IMF is
utilized for a case with violent events such as mergers and when strong shocks
take place.  Recent observations indicate that IMF slopes alter depending on
host galaxy mass \citep{ConroyvanDokkum2012, Kalirai+2013, Spiniello+2014}.
\cite{Suda+2013} reported that there is a tension between binary population
synthesis results and observed AGB stars in the galactic halo if the present-day
IMF is applied.  In the current version of CELib, the IMF does not allow
situation-dependent changes to its slopes and mass range, except for in the case
of extremely metal poor stars.

\subsection{Stellar Lifetime} \label{sec:LifeTime}

In this library, we adopt the metallicity dependent lifetime table provided by
\cite{Portinari+1998} (see their table 14).  The definition of the lifetime in
\cite{Portinari+1998} is the sum of the H-burning and He-burning time scales of
the stellar tracks of the Padova library.  The lifetimes depend slightly on the
stellar metallicity.  Their table covers stars from $0.6~\Msun$ to $120~\Msun$
and from $Z=0.0004$ to $Z=0.05$.  Note that we use the linear interpolation in
$Z$ and the lifetime of $Z=0.0004$ ($Z=0.05$) if the metallicity is below
(above) the value.  It is also worth noting that there are other stellar
lifetime data: both analytical forms \citep[e.g.,][]{Tinsley1980,
GreggioRenzini1983, MaederMeynet1989, Hurley+2000} and grid data
\citep{Schaller+1992}.  The differences between them are not so large.

For Pop III stars, we need the lifetime of stars $>120~\Msun$. Unfortunately,
the lifetime table of \cite{Portinari+1998} did not provide the lifetime of this
mass range. Here, we use the lifetime data obtained by \cite{Schaerer2002}.  He
calculated the lifetime of the Pop III stars within a mass range of $80~\Msun
\le m \le 1000~\Msun$. To make the lifetime data of $Z=0$ stars, we
combine the \cite{Portinari+1998}'s lifetime data of $Z=0.0004$ and the
\cite{Schaerer2002}'s lifetime data of $Z=0$.  We employ the
\cite{Portinari+1998}'s lifetime data to less than $120~\Msun$ stars whereas we
adopt the \cite{Schaerer2002}'s lifetime data of $150$--$300~\Msun$ stars.  With
these two sets of data, we can cover the full range of Pop III stars
($0.7$--$300~\Msun$).

\begin{figure}
\centering
\epsscale{1.0}
\plotone{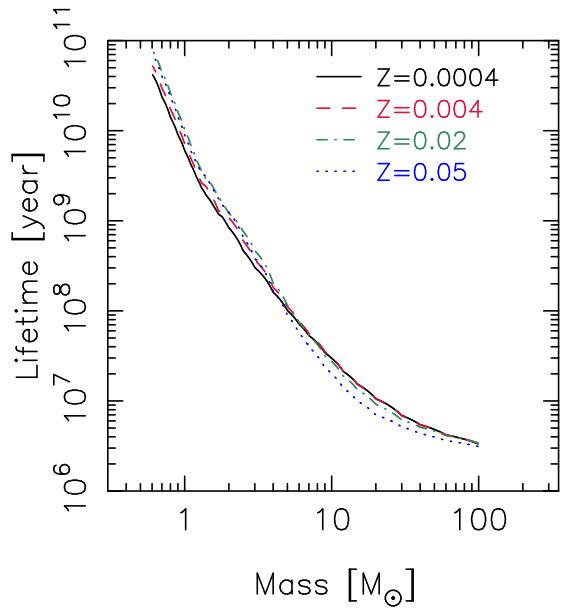}
\caption{Stellar lifetime as a function of its mass. Curves represent the cases
for five different metallicities, i.e., $Z=0.0004, 0.004, 0.02$, and $0.05$.}
\label{fig:LifeTime}
\end{figure}

Figure \ref{fig:LifeTime} shows the stellar lifetime as a function of mass.
Here, we interpolate both mass and metallicity and the linear interpolation is
used for this process. Apparently, the curve of the lifetime is not smooth.
These ``bumps'' come from the original data, and they will propagate other
quantities. Thus, instead of using this data, we make polynomial functions using
the least square method and use them as the fiducial ones.  The procedure is
described below.

We assume that the lifetime distribution function can be described using the
following polynomial function:
\begin{equation}
\log_{10} (t_*) = \sum_{k=0}^{k\le6} c_{{\rm LF},k} \left [ \log_{10} (m_*) \right] ^k,
\label{eq:LifeTime:Polynomial}
\end{equation}
where $m_*$ is the mass of stars in units of $\Msun$, $t_*$ is the lifetime
of a star with $m_*$ in the unit of year, and $c_{{\rm LF},k}$, is the
coefficient of $k$-th power of $\log_{10} (m_*)$. Coefficients are summarized in
table \ref{tab:LifeTime:Polynomial}. The first three coefficients for $Z=0.02$
(9.99, -3.55, 2.60) are not far from those found in the other fitting formula
for the stellar lifetime of Eq. (5) in \cite{GreggioRenzini1983} (10, -4.319, 1.543)
which was based on \cite{Rood1972} and \cite{Becker1979}.

\begin{figure}
\centering
\epsscale{1.0}
\plotone{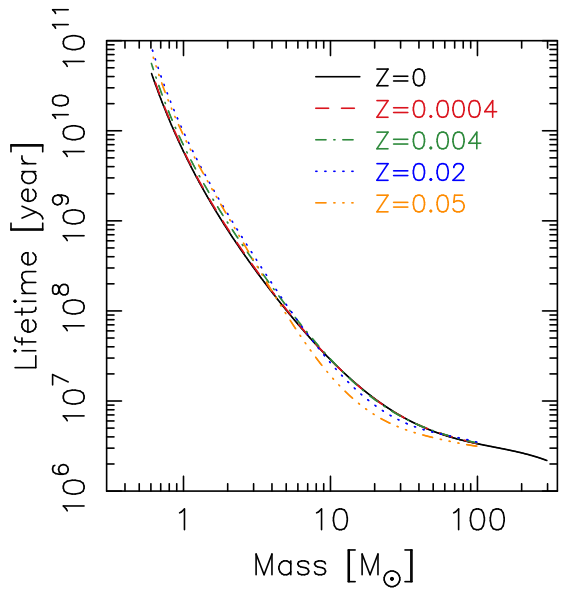}
\caption{Same as figure \ref{fig:LifeTime}, but reconstructed by the polynomial functions.
The lifetime of zero metal stars, which is the combination of the
\cite{Portinari+1998}'s lifetime data ($\le 120~\Msun$) and the
\cite{Schaerer2002}'s lifetime data ($\ge150~\Msun$), are also shown.
}
\label{fig:LifeTime:Polynomial}
\end{figure}

Using the five fitted functions and the results of their interpolations in the
direction of the metallicity (we use the linear interpolation), we obtain 
smoothed lifetime distribution functions.  We show the lifetime distributions of
five representative metallicities in figure \ref{fig:LifeTime:Polynomial}.  The
typical difference between the original and the smoothed lifetimes at a given
mass is less than $10\%$.  CELib adopts these smoothed lifetime distribution
functions as the fiducial ones.

\begin{table}[htb]
\begin{center}
\caption{Coefficients of stellar lifetime distribution functions used in Eq. \eqref{eq:LifeTime:Polynomial}.
}\label{tab:LifeTime:Polynomial}
\begin{tabular}{ccccccc}
\hline
\hline
& \multicolumn{6}{c}{Metallicity} \\
 & $0.0$& $0.0004$  & $0.004$ & $0.008$ & $0.02$ & $0.05$ \\
\hline
$c_{{\rm LF},0}$ &  9.77  &  9.77  &  9.84  &  9.90  &  9.99   &  9.95  \\
$c_{{\rm LF},1}$ & -3.40  & -3.40  & -3.40  & -3.47  & -3.55   & -3.41  \\
$c_{{\rm LF},2}$ &  2.18  &  2.17  &  2.62  &  2.75  &  2.60   &  2.60  \\
$c_{{\rm LF},3}$ & -2.05  & -1.95  & -3.51  & -3.84  & -3.60   & -5.39  \\
$c_{{\rm LF},4}$ &  1.29  &  1.11  &  2.76  &  3.06  &  2.81   &  5.53  \\
$c_{{\rm LF},5}$ & -0.362 & -0.250 & -0.973 & -1.08  & -0.911  & -2.35  \\
$c_{{\rm LF},6}$ & 0.0353 & 0.0112 &  0.126 & 0.138  &  0.0988 &  0.358 \\
\hline
\end{tabular}\\
\end{center}
\end{table}

\begin{figure}
\centering
\epsscale{1.0}
\plotone{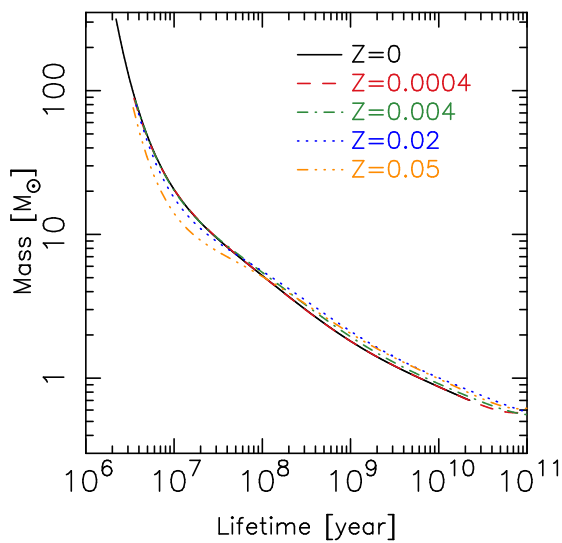}
\caption{Mass of stars who just finished their lifetime as a function of time.
}
\label{fig:DyingStellarMass:Polynomial}
\end{figure}

It is convenient if we have an inverse function that can tell us the mass of a
star which just died when we enter its lifetime.  Again, we use the polynomial
fitting so that we can obtain such functions,
\begin{equation}
\log_{10} (m_*) = \sum_{k=0}^{k\le6} c_{{\rm MD},k} \left [ \log_{10} (t_*) \right] ^k,
\label{eq:DyingStellarMass:Polynomial}
\end{equation}
where $c_{{\rm MD},k}$ is the coefficients and they are listed in table
\ref{tab:Mass:Polynomial}.  Figure \ref{fig:DyingStellarMass:Polynomial} shows
the functions of Eq.  \eqref{eq:DyingStellarMass:Polynomial}. In the metallicity
direction, we again use the linear interpolation. 

\begin{table}[htb]
\begin{center}
\caption{Coefficients of mass who died at a given time used in Eq.
\ref{eq:DyingStellarMass:Polynomial}.
}\label{tab:Mass:Polynomial}
\begin{tabular}{ccccccc}
\hline
\hline
& \multicolumn{6}{c}{Metallicity} \\
 & $0.0$ & $0.0004$  & $0.004$ & $0.008$ & $0.02$ & $0.05$ \\
\hline
$c_{{\rm MD},0}$ &   2590   &   2560   &   2320  &   2470  &   2560  &   3640 \\
$c_{{\rm MD},1}$ &  -1750   &  -1740   &  -1550  &  -1650  &  -1690  &  -2460 \\
$c_{{\rm MD},2}$ &    493   &    492   &    430  &    454  &    466  &    688 \\
$c_{{\rm MD},3}$ &  -73.7   &  -73.8   &  -63.4  &  -66.6  &  -68.0  &   -102 \\
$c_{{\rm MD},4}$ &   6.17   &   6.21   &   5.23  &   5.47  &   5.55  &   8.53 \\
$c_{{\rm MD},5}$ &  -0.274  &  -0.277  & -0.229  & -0.238  & -0.240  & -0.377 \\
$c_{{\rm MD},6}$ &  0.00506 &  0.00514 & 0.00417 & 0.00430 & 0.00432 & 0.00692 \\
\hline
\end{tabular}\\
\end{center}
\end{table}

\subsection{Elements considered in this library}\label{sec:Elements}  

CELib can follow the evolution of the thirteen representative elements.
These are hydrogen (H), helium (He), carbon (C), nitrogen (N), oxygen (O), neon
(Ne), magnesium (Mg), silicon (Si), sulfur (S), calcium (Ca), iron (Fe), nickel
(Ni) and europium (Eu). They cover the major products synthesized by SNe II (C,
N, O, Mg, S, Ca, etc.), by SNe Ia (Fe, Ni, etc.) and by AGBs (C, N, O, etc.).
The Eu is the tracer of the $r$-process in NSMs since most of the Eu in the
solar neighborhood comes from the $r$-process \citep{Burris+2000}.  These
elements, except Eu, dominate in the mass of the local ISM.  Hence, these
elements play a major role in the ISM as coolants \citep{Wiersma+2009Cooling}.
For each element, all of the isotopes listed on yields tables are combined into
one.  For instance, carbon represents three isotopes, $^{12}$C, $^{13}$C and
$^{14}$C, and in this library, their masses are combined and expressed simply as
``C''.  Table \ref{tab:yieldsfull} summarizes the yields tables and elements
adopted in the current version of CELib. 

\begin{table*}[htb]
\begin{center}
\caption{Yields and elements adopted in the current version of CELib.
}\label{tab:yieldsfull}
\begin{tabular}{c|c|ccccccccccccc}
\hline
\hline
Reference & Type & H & He  & C & N & O & Ne & Mg & Si & S & Ca & Fe & Ni & Eu \\
\hline
\cite{Portinari+1998}& SNe II & \checkmark &\checkmark &\checkmark &\checkmark &\checkmark 
&\checkmark &\checkmark &\checkmark &\checkmark &\checkmark &\checkmark &$\times$  &$\times$  \\
\cite{Nomoto+2013}$^a$& SNe II \& HNe & \checkmark &\checkmark &\checkmark &\checkmark &\checkmark &\checkmark 
&\checkmark &\checkmark &\checkmark &\checkmark &\checkmark &$\checkmark$  &$\times$  \\
\hline
\cite{Iwamoto+1999}& SNe Ia & $\times$ &$\times$ &\checkmark &\checkmark &\checkmark 
&\checkmark &\checkmark &\checkmark &\checkmark &\checkmark &\checkmark &$\checkmark$  &$\times$  \\
\cite{Travaglio+2004}& SNe Ia & $\times$ &$\times$ &\checkmark &\checkmark &\checkmark 
&\checkmark &\checkmark &\checkmark &\checkmark &\checkmark &\checkmark &$\checkmark$  &$\times$  \\
\cite{Maeda+2010}& SNe Ia & $\times$ &$\times$ &\checkmark &\checkmark &\checkmark 
&\checkmark &\checkmark &\checkmark &\checkmark &\checkmark &\checkmark &$\checkmark$  &$\times$  \\
\cite{Seitenzahl+2013}& SNe Ia & $\times$ &$\times$ &\checkmark &\checkmark &\checkmark 
&\checkmark &\checkmark &\checkmark &\checkmark &\checkmark &\checkmark &$\checkmark$  &$\times$  \\
\hline
\cite{Karakas2010}& AGBs & \checkmark &\checkmark &\checkmark &\checkmark &\checkmark &\checkmark 
&\checkmark &\checkmark &\checkmark &$\times$ &\checkmark &$\checkmark$  &$\times$  \\
\cite{Doherty+2014}& AGBs & \checkmark &\checkmark &\checkmark &\checkmark &\checkmark &\checkmark 
&\checkmark &\checkmark &\checkmark &$\times$ &\checkmark &$\checkmark$  &$\times$  \\
\cite{CampbellLattanzio2008}& AGBs & \checkmark &\checkmark &\checkmark &\checkmark &\checkmark &\checkmark 
&\checkmark &\checkmark &\checkmark &$\times$ &$\times$ &$\times$  &$\times$  \\
\cite{Gil-Pons+2013}& AGBs & \checkmark &\checkmark &\checkmark &\checkmark &\checkmark &\checkmark 
&$\times$ &$\times$ &$\times$ &$\times$ &$\times$ &$\times$  &$\times$  \\
\hline
\cite{Wanajo+2014}$^b$
& NSMs & $\times$ &$\times$ &$\times$ &$\times$ &$\times$ &$\times$ &$\times$ &$\times$ &$\times$ 
&$\times$ &$\times$ &$\times$  &$\checkmark$  \\
\hline
\end{tabular}\\
\end{center}
$^a$ The yields table of \cite{Nomoto+2013} gives not only SNe II, but also AGBs
and hypernovae. We do not adopt their AGB yields in this library.\\
$^b$ In order to obtain the absolute yield value of the $r$-process elements, we
follow the argument of \cite{Ishimaru+2015, Hirai+2015}. See text for detail.
\end{table*}

\subsection{Yields for SNe II}\label{sec:SNII}  

There are a number of yields tables of SNe II \citep[e.g.,][]{Maeder1992,
WoosleyWeaver1995, Portinari+1998, Rauscher+2002, Hirschi+2005, Nomoto+2006,
Kobayashi+2006, Nomoto+2013, Pignatari+2016}.  In the current version of CELib,
two yields tables are implemented; those provided by \cite{Portinari+1998} and
\cite{Nomoto+2013}.  These two yields tables are selectable, and users can
choose one of them at the beginning of the simulation.

The yields table provided by \cite{Portinari+1998} has been used very widely in
galaxy formation simulations.  It is based on the calculations of
\cite{WoosleyWeaver1995} and it took into account the effect of the pre-SN mass
loss due to stellar winds.  The original yields table covers the mass range of
$1$--$1000~\Msun$ and the metallicity range of $0.0004<Z<0.05$.  Here, we only
adopt the data of the mass range above $9$--$120~\Msun$ as SNe II yields.

It is known that this yields table does not match the observations when one uses
it for a galactic chemical evolution model, as is first described in the
original paper \citep{Portinari+1998}. To make it fit with observations, one
generally applies ad hoc modifications to the yields, i.e., multiplying factors
of 0.5, 2, and 0.5 for C, Mg, and Fe yields, respectively  \citep[see \S A 3.2
in][]{Wiersma+2009}. We also follow this modification.  Although we implemented
these yields, we do not use them as the default yields for SNe II. We use them
mainly for comparison.

The fiducial yields table for SNe II in this library is that of
\cite{Nomoto+2013}.  The yields table of \cite{Nomoto+2013} for the high mass
regime is based on \cite{Kobayashi+2006} and \cite{UmedaNomoto2002}.  The
yields table covers the mass range of $13$--$40~\Msun$ for $Z=0.001, 0.004,
0.02$ and 0.05. {\footnote{This yields table has the data at $Z=0.008$.
Since the data of this metallicity are obtained by the linear interpolation
of their data of $Z = 0.004$ and $Z = 0.02$, we do not adopt it. CELib can
recompute the yields at $Z=0.008$ internally.}} This yields table also has the
date for Pop III stars. The mass range is $11$--$300~\Msun$.  These data are
used by combining the Pop III IMF. We use the Pop III yields when $Z_{\rm pop
III} = 10^{-5} \Zsun$.

Since they also listed the yields of hypernovae (HNe) on their yields table, we
adopted it and users can use the HNe mode.  The HNe are characterized by their large
energy release (typically ten times more energy is released) and a significant
amount of iron production \citep{Nomoto+2006}.  The HNe yields table covers
$20$--$140~\Msun$ for $Z=0$ and $20$--$40~\Msun$ for $Z>0$. We introduce a
parameter: the HN blending fraction $f_{\rm HN}$. The definition of $f_{\rm HN}$
is that the number of HNe divided the number of SNe in the mass ranges of HNe.
Thus, $f_{\rm HN}$ alters from zero to unity. If it has a value larger than
zero, we blend the HNe yields to the normal SNe yields in the given mass range
depending on metallicity.  Since the mass range of HNe is limited, we only
consider the contribution of the normal SNe when the mass range exceeds that of
HNe.  The remnant mass, ejecta mass, and released energy are also evaluated by
taking into account the contribution of HNe. 

In this paper, we show the results with $f_{\rm HN} = 0.05$ and $f_{\rm HN} =
0.5$, as well as $f_{\rm HN} = 0$.  The value of $f_{\rm HN} = 0.5$ is based on
\cite{Kobayashi+2006}.  In \cite{Kobayashi+2006}, they used this value in order
to reproduce the zinc abundance.  We note that this value is one or two orders
of magnitude larger than those evaluated by observations
\citep{Podsiadlowski+2004, GuettaDellaValle2007}.  Moreover, there is another
potential site of zinc synthesis \citep{Wanajo+2011} and thus, a high HN
fraction ($f_{\rm HN} = 0.5$) might not be necessarily required.  We therefore
adopt $f_{\rm HN} = 0.05$ and CELib employs it as a fiducial value.

We pick up twelve elements (H, He, C, N, O, Ne, Mg, Si, S, Ca, Fe, and Ni) from
these yields tables. The other elements listed in the original yields tables are
ignored because of their minor contribution to the total amount of released
metals. Note that \cite{Portinari+1998} did not provide the yield of Ni.
Therefore, we assume that the contribution of Ni is zero when we adopt the
yields table of \cite{Portinari+1998}.

The yields table supplied by \cite{Portinari+1998}, table 10 in their paper,
lists the stellar yields, $Y$s. Therefore, we need to convert $Y$s into $y$s in
order to use galactic chemical evolution. Since the yields table also
includes the remnant mass, we can evaluate $y$s if we know the chemical
abundance pattern of the pre-existing materials using Eq. \eqref{eq:NetYields}.
For this purpose, we assume the solar abundance pattern of
\cite{AndersGrevesse1989} and simply change the total mass of heavy elements
while keeping the original abundance ratios. The change in the total metal mass
corresponds to the change in masses of H and He, while the mass ratio of H and
He remains the same.  We do not apply any correction to the H/He ratio to
reproduce the cosmic primordial He fraction \citep{Plank2014XVI} at $Z = 0$ in
this process.

For the yields table of \cite{Nomoto+2013}, the data conversion procedure we
used is rather complicated. The high mass regime of this yields table consists
of the stellar yields from SNe II, HNe and remnant masses of each progenitor's
mass and metallicity. In this table, the amount of the lost mass due to pre-SNe
stellar winds is not explicitly included.  In other words, the sum of the total
mass of the stellar yields by SNe II/HNe and the remnant mass is not equal to
the progenitor's mass.  We, thus, need to complement the amount of pre-SNe mass loss.
Based on \cite{Kobayashi+2006}, we assumed no enrichment in stellar winds and
complemented the lost mass with heavy elements that have the solar abundance
ratio of \cite{AndersGrevesse1989}.
{\footnote {
There is another way to fill the stellar wind yields by using the data of
pre-SNe yields of massive stars including metallicity dependent mass loss
and rotation effects \cite[see][]{Romano+2010}. However, we do not adopt this 
mainly because they (models which give stellar yields and pre-supernova yields)
are not generated by the same model and thus they might be inconsistent when we
connect them without some tuning.}}
From these reconstructed $Y$s and remnant masses, we derived net yields $y$s.
For Pop III stars, we assumed the primordial composition of \cite{Plank2014XVI}
for the progenitors' chemical composition.

\begin{figure}
\centering
\epsscale{1.0}
\plotone{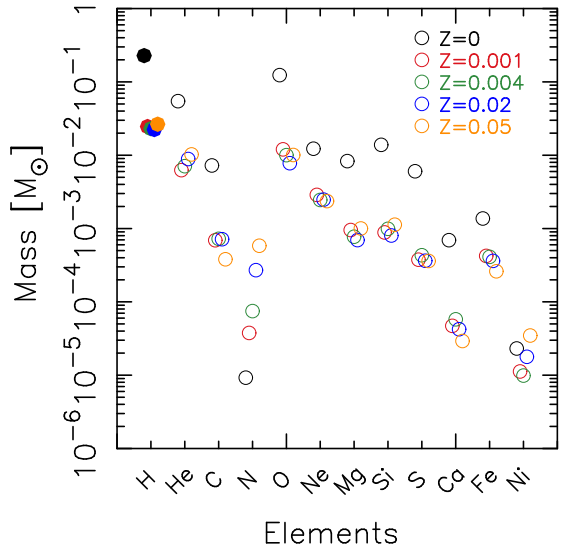}
\caption{Yield of each element per $1~{\Msun}$ SSP particle.  The yields
table of \cite{Nomoto+2013} is used.  Open circles are the positive yields
whereas filled circles are the negative yields but their absolute values.  The
Chabrier IMF with the mass range of $0.1$--$100~\Msun$ is assumed. Note that the
yields table of \cite{Nomoto+2013} provides the SNe II yields of $13$--$40~\Msun$
stars for Pop I/II.  For the case of $Z = 0$, the Susa IMF with a mass range of
$0.7$--$300~\Msun$ is adopted. The yields table of \cite{Nomoto+2013} supplies the
yields of $11$--$300~\Msun$ stars for $Z = 0$. $f_{\rm HN} = 0$ is assumed.
}
\label{fig:SNII:Yields:N13}
\end{figure}

% IMF weighted released metals 
The IMF weighted yields per $1~\Msun$ SSP particle with the yields table
of \cite{Nomoto+2013} are shown in figure \ref{fig:SNII:Yields:N13}. No
contributions from HNe are assumed in this figure. Here, we assumed the Chabrier
IMF of which the mass range is $0.1$--$100~\Msun$.  The difference induced by
the metallicity of the progenitor SSP particle is also shown in this figure.
For the $Z=0$ case, the Susa IMF is adopted.  We used Eq.
\ref{eq:IMFWeightedYileds} to depict this figure.  We note that the ejecta due
to stellar winds are included in the SN ejecta, for simplicity.  We can see that
the oxygen release amount is the most prominent and other elements, such as He,
Ne, Mg, and Si, follow it. When the SSP particle's metallicity increases, the
yields also increase. The IMF weighted yields of Pop III stars are typically a
factor of five larger than those of Pop I/II stars.  However, their contribution
is rather limited (see \S \ref{sec:Applications}).

\begin{figure}
\centering
\epsscale{1.0}
\plotone{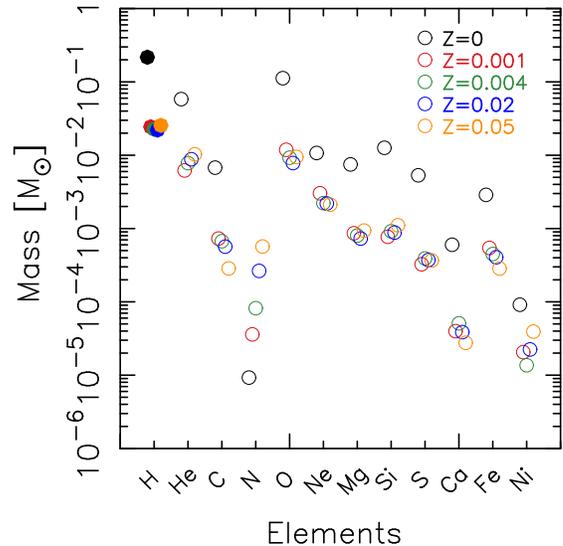}
\caption{
Same as figure \ref{fig:SNII:Yields:N13}, but $f_{\rm HN} = 0.5$.
}
\label{fig:SNII:Yields:N13HN}
\end{figure}

In figure \ref{fig:SNII:Yields:N13HN}, we show the case with $f_{\rm HN} = 0.5$.
The most prominent difference is found in the yields of Fe and Ni. They increase
several times when we compare those found in the case with $f_{\rm HN} = 0$.
Since the difference between the yields with $f_{\rm HN} = 0$ and 
$f_{\rm HN} = 0.05$ are almost identical, we do not show here.

\begin{figure}
\centering
\epsscale{1.0}
\plotone{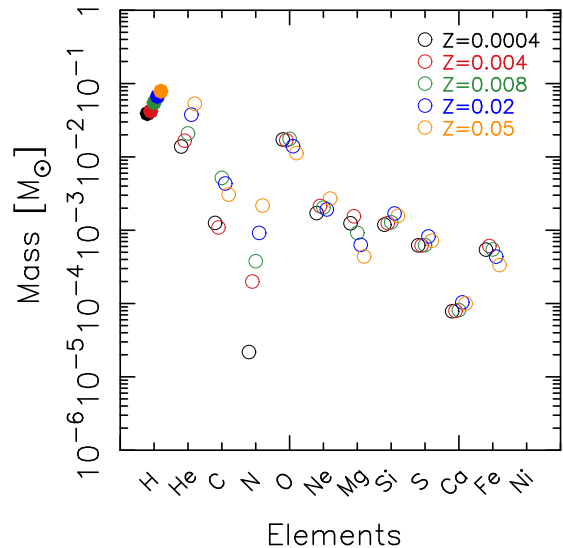}
\caption{Same as figure \ref{fig:SNII:Yields:N13}, but with the yields of
\cite{Portinari+1998}. The yields data of $8$--$100~\Msun$ is used since the
Chabrier IMF is used. The modifications of \cite{Wiersma+2009} are applied.
}
\label{fig:SNII:Yields:P98mod}
\end{figure}

Figure \ref{fig:SNII:Yields:P98mod} shows the IMF weighted yields of
\cite{Portinari+1998}. The modifications of \cite{Wiersma+2009} were applied.
From these figures (figures \ref{fig:SNII:Yields:N13},
\ref{fig:SNII:Yields:N13HN} and \ref{fig:SNII:Yields:P98mod}), we see that the
differences between them are insignificant since we applied the modifications.
We will investigate the difference between these two yields when applying
simulations of chemical evolution in \S \ref{sec:Applications}.

\subsection{Yields for SNe Ia}\label{sec:SNIa}  

For SNe Ia, CELib implemented four different yields tables based on one-
\citep{Iwamoto+1999}, two- \citep{Travaglio+2004, Maeda+2010}, and three-
\citep{Travaglio+2004, Seitenzahl+2013} dimensional simulations.  In total,
CELib holds six one-dimensional models, four two-dimensional models, and 17
three-dimensional models and it is possible to use any of them (users should
select their preferred model before calling the initialization routine of
CELib). An extra model which consists of metallicity dependent yields based on
the three-dimensional models of \cite{Seitenzahl+2013} can be selected.  All
models are of a thermonuclear explosion of a carbon-oxygen white dwarf that
reaches the Chandrasekhar-mass.  Thus, the yields tables consist of elements
which are equal to or heavier than carbon.

The first one is the yields table given by \cite{Iwamoto+1999} which was
computed by one-dimensional simulations.  In their table, seven different models
of SN Ia were listed. In their models, the W7 and the WDD2
(deflagration-detonation transition model with the transition density of
$2.2\times10^7~{\rm g}$) models agree with the observational constraint on the
$^{56}$Ni production.  In the literature, the W7 model is the most widely used.
CELib takes all of the seven models listed in the table so that users can choose
their preferred model.  The original table includes 66 elements and isotopes
(see table 3 in their paper). We adopted ten elements, i.e., C, N, O, Ne, Mg,
Si, S, Ca, Fe and Ni.  We summed up all isotopes of each element.

The second SN Ia yields table is that supplied by \cite{Travaglio+2004}, which
was derived from their two- and three-dimensional simulations of thermonuclear
burning. Unlike one-dimensional simulations where the flame speed is a free
parameter, there is no need to adjust this parameter in multi-dimensional
simulations.  Depending on the spatial resolution, model dimensions, and
ignition configurations (centered v.s. multipoint ignition), there are five
models (one two-dimensional model and four three-dimensional models).  While all
models satisfy the observed $^{56}$Ni production rate \citep{Stritzinger+2006},
they noted that the three-dimensional model with the highest spatial resolution,
b30\_3d\_768, is the best one of which the synthesized amount of $^{56}$Ni is
the largest. Ten elements the same as those in the one-dimensional case are
adopted.

The third is the yields of \cite{Maeda+2010}. They gave yields of four models
such as the pure-deflagration explosion model (C-DEF) and the delayed detonation
models with the center and off-center deflagration (C-DDT and O-DDT), as well as
the standard W7 model.  All four models are implemented in CELib. Note that
their W7 model is essentially the one-dimensional model and is not important in
this context.  There are $\sim 70$ elements and isotopes, but we adopt ten
elements which are the same as those adopted in \cite{Iwamoto+1999}.  According
to their comparisons between their yields and solar abundance ratios of Ni over
Fe and Si over Fe, they remarked that O-DDT is the most favorable model.

The fourth and last one is the yields table evaluated by three-dimensional
simulations of SN Ia provided by \cite{Seitenzahl+2013}.  In their
study, they mainly changed the number of ignition points and solved the
nucleosynthesis using test particles. The number of ignition points tested is
$1$--$1600$. Moreover, they changed the central density and the initial
metallicity of the progenitor star.  Although the ejecta mass of $^{56}$Ni in
all these models is a factor of two more/less than the amount estimated from
observations \citep{Stritzinger+2006}, the model with 100 ignition points is
almost the same as the observational estimate and is regarded as the
fiducial model.  As with previous yields tables, CELib adopts the ten primary
elements.

In \cite{Seitenzahl+2013}, they also assessed the dependence of the progenitor
metallicity. They employed the N100 model and changed the progenitor metallicity
from $0.5$ to $0.01$ times of the canonical $^{22}$Ne mass fraction of 0.025.
These models are labeled as N100\_Z0.5, N100\_Z0.1, and N100\_Z0.01.  In CELib,
the model, which returns the metallicity dependent yields of SN Ia, is
implemented using these yields.  By reducing the initial amount of $^{22}$Ne,
the production rate of $^{56}$Ni decreases. This tendency is preferable because
overproduction is observed in the canonical models \citep[See details in
section 3.2 in][]{Seitenzahl+2013}.

In all yields tables, the ejecta mass of each element, $Y_i$, is listed.  We use
it when SN Ia takes place in an SSP particle.

\begin{figure}
\centering
\epsscale{1.0}
\plotone{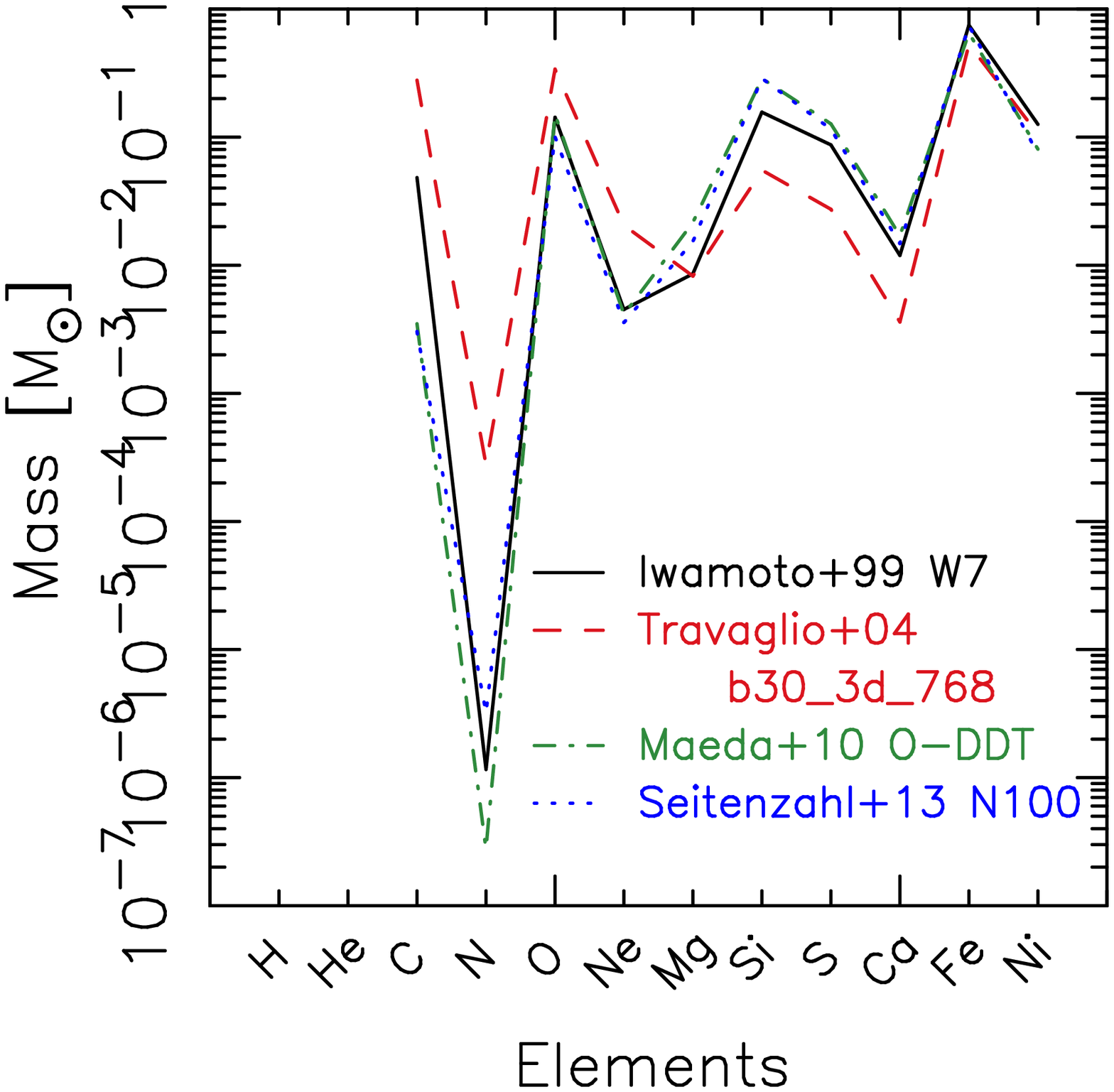}
\caption{Stellar yields of SN Ia. Four representative models, the W7 model from
\cite{Iwamoto+1999}, the b30\_3d\_768 model from \cite{Travaglio+2004}, the O-DDT
model from \cite{Maeda+2010}, and the N100 model from \cite{Seitenzahl+2013},
are shown. }
\label{fig:SNIa:Yields4Models}
\end{figure}

Figure \ref{fig:SNIa:Yields4Models} displays the distributions of metals in
the ejecta of a single SN Ia. Four representative models are selected from the
implemented yields tables. In these four models, the Fe and Ni production rates
are almost identical. These models reproduce or are close to the observed amount
of $^{56}$Ni production.  Three models \citep[models from][]{Iwamoto+1999,
Maeda+2010, Seitenzahl+2013} are almost identical even in the light elements.
For light elements (C,N,O, Ne), the model of \cite{Travaglio+2004} has more
leftovers. It is mentioned in \cite{Travaglio+2004} that the amount of the
leftovers increases when increasing the resolution and changing the dimensions
from two to three.  We need to keep in mind that there are fluctuations from
model to model.

Multidimensionality can remove a free parameter (the flame velocity). Thus, such
models are more realistic.  As fiducial models, we chose the N100 model of
\cite{Seitenzahl+2013}.  The metallicity dependent yields we made from models of
\cite{Seitenzahl+2013} are also one of the fiducial models.  Differences due to
the adopted SN Ia yields are studied in \S \ref{sec:Applications}.

\begin{figure}
\centering
\epsscale{1.0}
\plotone{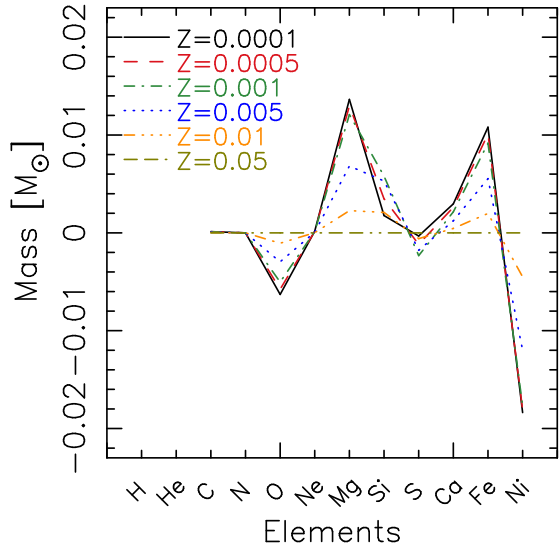}
\caption{Mass difference of each element from the SN Ia yields of the solar
metallicity. Metallicity dependent yields are based on \cite{Seitenzahl+2013}. 
}
\label{fig:SNIa:YieldsS13Diff}
\end{figure}

The metallicity dependencies of the yields in \cite{Seitenzahl+2013} are shown in
figure \ref{fig:SNIa:YieldsS13Diff}. Here we show the quantities of
\begin{equation}
Y_{{\rm diff},i}(Z) = Y_i (Z) - Y_i (Z_\odot),
\end{equation}
where $Z_\odot$ is the solar metallicity and we adopt the value of
\cite{Asplund+2009} ($Z=0.0134$).  There is a tendency for the amounts of O and
Ni to decrease with decreasing metallicity, whereas those of Mg, Si, Ca and Fe
increase with increasing metallicity. It is noted by \cite{Seitenzahl+2013} that
the metallicity dependent yields are favorable because it reduces the
overproduction of $^{58}$Ni and $^{54}$Fe.

\subsection{Yields for AGBs}\label{sec:AGB}  

For the AGB yields table, we combine four yields tables of AGBs given by the
Monash group \citep{CampbellLattanzio2008, Karakas2010, Gil-Pons+2013,
Doherty+2014} so that we can cover wide mass and metallicity ranges.

We refer to \cite{Karakas2010} for the yields from low- and intermediate-mass
AGB stars. These yields are the successor to those obtained by their previous
study \citep{KarakasLattanzio2007} with an improved reaction rate network. The
covering range of this yields table in both mass and metallicity is the broadest
\citep[see table 2 in][]{KarakasLattanzio2014}.  It covers the metallicity range
of $Z=0.0001, 0.004, 0.008$ and $0.02$ and the mass range of $1$--$6~\Msun$.

The yields table provided by \cite{Karakas2010} does not include
the yields from super-AGB stars ($>6~\Msun$). Therefore, there is a gap between
SNe II and AGBs \citep[`super-AGB star gap'][]{Romano+2010, Kobayashi+2011}.  In
order to diminish this gap, we combine the Karakas's yields table with the
yields table of super-AGB stars given by \cite{Doherty+2014}. In
\cite{Doherty+2014}, they tested different mass-loss rate models and gave two of
them in the electric form. Here we adopted their VM93 model where the mass loss
rate follows the model of \cite{VassiliadisWood1993} and which was regarded as
the standard model in their study.  The yields table of \cite{Doherty+2014}
covers $Z=0.004, 0.008$ and $0.02$ and $6.5$--$9~\Msun$. The maximum mass listed
on the table changes depending on the metallicity from $8~\Msun$ at $Z=0.004$ to
$9~\Msun$ at $Z=0.02$.  We adopt the data of $6.5$--$8~\Msun$ with the three
metallicities. In order to fit the \cite{Karakas2010}'s table, we use the yields
at $Z=0.004$ to the yields at $Z=0.0001$ because the \cite{Doherty+2014}'s
yields table did not provide data at $Z=0.0001$.  Since both the yields tables
of \cite{Karakas2010} and \cite{Doherty+2014} were generated by the Monash
University stellar evolution program ({\tt MONSTER}) \citep{FrostLattanzio1996},
they would have a high affinity.

These two yields tables did not cover the extremely low-metallicity AGBs.
In CELib, the yields at $Z=0$ in \cite{CampbellLattanzio2008} and
those at $Z=10^{-5}$ in \cite{Gil-Pons+2013} are used for this metallicity
range.  The yields table of \cite{CampbellLattanzio2008} covers a mass range
$\le 3~\Msun$ and a metallicity range $Z=0$--$10^{-3}$.  On the other hand,
the yields table of \cite{Gil-Pons+2013} gives yields of low-metallicity AGB
stars ($Z=10^{-5}$) in the mass range of $4$--$9~\Msun$. We connect these two
yields tables and then generate a single AGB yields for extremely
low-metallicity AGB stars of $1$--$8~\Msun$.  Again, these two yields tables
were computed by {\tt MONSTER}. We regard these low-metallicity AGB yields as
the AGB yields of $Z=0$, for simplicity.  Since the adopted metallicity of
tables is slightly different, this seems to be inconsistent. However, the
difference is not so large compared to the other factors in the chemical
evolution model, such as the form of the IMF.

Overall, the covered range of the combined  AGB yields is $1$--$8~\Msun$ and
$Z=0$--$0.02$.  Practically, when the metallicity of an SSP particle is below
$Z_{\rm popIII}$, CELib returns the yields of the Pop III AGBs.  On the other
hand, when the metallicity of an SSP particle is above $0.02$, CELib returns
yields of $Z=0.02$.

%% Metals
The two yields tables \citep{Karakas2010, Doherty+2014} include yields of 40--80
elements and their isotopes.  CELib adopts 11 elements (H, He, C, O, N, Ne, Mg,
Si, S, Fe, and Ni) from them. All isotopes of each element are summed up. Since
Ca and its isotopes are not included in their models, the zero yield is assumed
for Ca.  The Pop III AGB tables did not contain elements heavier than S
\citep{CampbellLattanzio2008} and Ne \citep{Gil-Pons+2013}. We hence assumed
zero yields for the unlisted elements.

Although the yields tables of \cite{Karakas2010} and \cite{Doherty+2014} provide
the net yields $y$s, the other two yields tables do not give net yields $y$s
but stellar yields $Y$s.  Therefore, as a preprocess, we converted these
table data to the net yields using Eq. \eqref{eq:NetYields}.

\begin{figure}
\centering
\epsscale{1.0}
\plotone{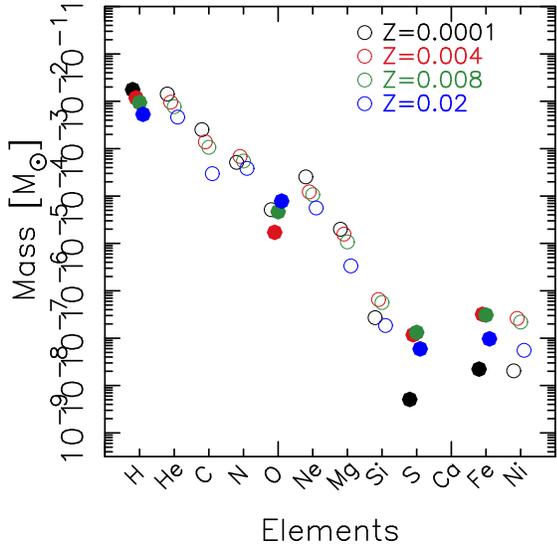}
\caption{AGB yields per $1~{\Msun}$ SSP particle. The Chabrier IMF with a mass
range of $0.1$--$100~\Msun$ is adopted and AGB stars of $1$--$6~\Msun$ are
considered.  Open circles indicate the positive yields whereas filled circles
show the negative yields and the absolute values are used.
}
\label{fig:AGB:Yields0}
\end{figure}

Figure \ref{fig:AGB:Yields0} shows the IMF weighted yields (Eq.
\ref{eq:IMFWeightedYileds}) of 11 elements considered in AGBs in this library.
Only AGBs from $1$--$6~\Msun$ are taken into account. We can see that the
elements of H, O, S and Fe are destroyed while those of the others are
synthesized.  When we compare the IMF weighted AGB yields to those obtained by
SNe II (figures \ref{fig:SNII:Yields:N13} and \ref{fig:SNII:Yields:P98mod}), we
find that the C and N yields of AGBs are greater than those of SNe II.  Hence,
lighter elements are strongly affected by AGBs.

\begin{figure}
\centering
\epsscale{1.0}
\plotone{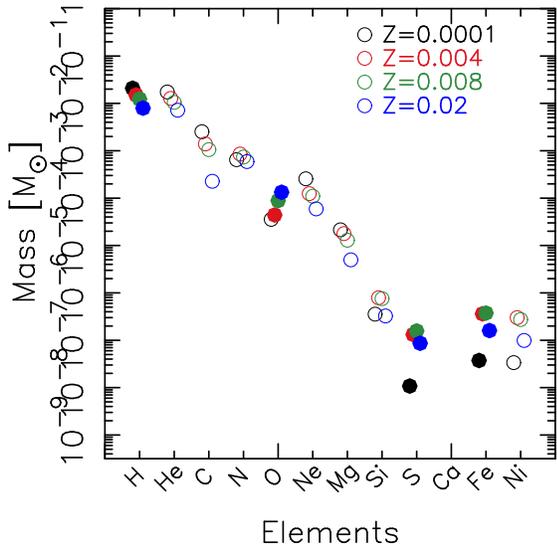}
\caption{Same as figure \ref{fig:AGB:Yields0}, but for AGB stars of
$1$--$8~\Msun$.  The contributions from the yields of super-AGB stars
\citep{Doherty+2014} are taken into account. 
}
\label{fig:AGB:Yields1}
\end{figure}

The IMF weighted yields of AGBs from $1$--$8~\Msun$ are displayed in figure
\ref{fig:AGB:Yields1}. Here the super-AGB yields provided by \cite{Doherty+2014}
are taken into account as well as the yields provided by \cite{Karakas2010}.
The differences between figures \ref{fig:AGB:Yields0} and \ref{fig:AGB:Yields1}
are non-negligible. The reason is simply that the return mass fraction of the
massive AGBs is larger than those of less massive AGBs, regardless of the
relatively small number of the super-AGBs.  Depending on the elements, the
increase in yields is about several times larger, when we adopt the effects of
super AGBs.  This effect can be seen in simulations of chemical evolution (see
\S \ref{sec:Applications}) while it is not significant.

\begin{figure}
\centering
\epsscale{1.0}
\plotone{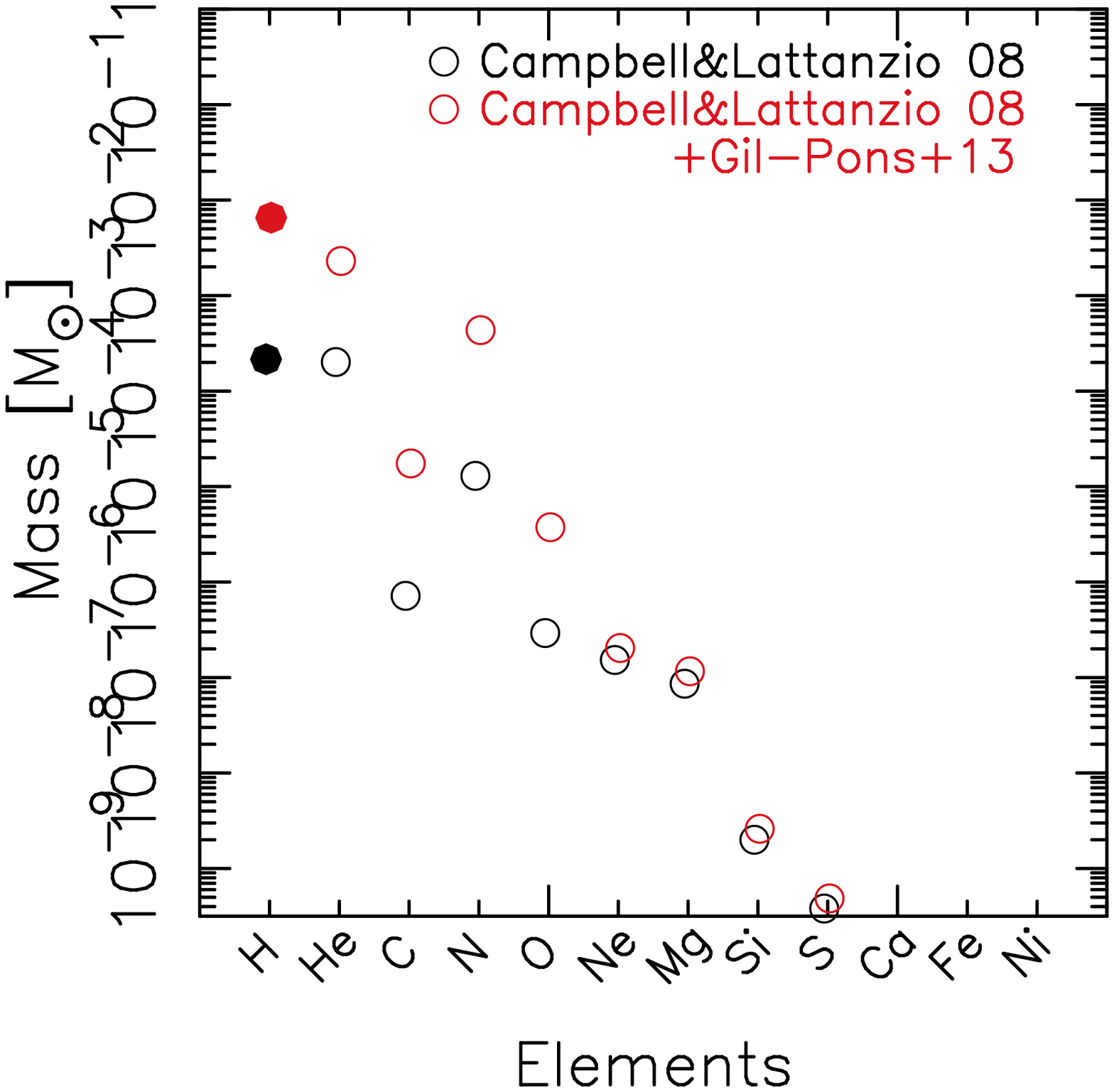}
\caption{Same as figure \ref{fig:AGB:Yields0}, but for Pop III AGB stars of
$1$--$3~\Msun$ \citep{CampbellLattanzio2008} and 
$1$--$8~\Msun$ \citep{CampbellLattanzio2008,Gil-Pons+2013}.
}
\label{fig:AGB:Yields2}
\end{figure}

The yields of Pop III AGBs are shown in figure \ref{fig:AGB:Yields2}.  They have
their own characteristic abundance pattern among C, N and O.  This is due to the
effect of the thermal pulsation in relatively massive stars
\citep{Gil-Pons+2013}.  The contributions of the Pop III AGBs are, however,
rather limited because of a very low chance of this feedback due to the limited
metallicity range.

There are other commonly used yields tables of AGBs
\citep{vandenHoekGroenewegen1997, Marigo2001, Izzard+2004}.  These tables are
compared in \cite{Wiersma+2009}.  They found that these agreed with each other
on solar metallicity. However, they also found that there are large gaps among
them in the low-Z regime.  In the current version of CELib these tables are not
implemented.

\subsection{Yields for NSMs}\label{sec:NSM}  

NSMs have gained considerable attention as the potential sites of $r$-process
elements synthesis \citep[e.g.,][]{LattimerSchramm1976, Lattimer+1977,
Eichler+1989}.
Recent numerical simulations \citep{Goriely+2011, Wanajo+2014} imply that NSMs
are a potential synthesis site of $r$-process elements. Thus, it would be desirable to
take them into account in numerical simulations in order to get a comprehensive
understanding of the stellar feedback in regards to galaxy formation. 

This library adopts Eu as a tracer of the NSMs in which the massive $r$-process
elements are synthesized.  Following the argument in \cite{Ishimaru+2015}, we
fix the Eu yield at $2\times10^{-5}~\Msun$ per NSM.  This value is calibrated by
the typical event rate of NSMs [$2\times10^{-3}$ times the event rate of SNe II
\citep{Dominik+2012}] and the nucleosynthesis results of \cite{Wanajo+2014,
Goriely+2011}.  The variation of the Eu yield is rather insensitive to the
configuration of mergers \citep{Goriely+2011}.

Although many heavy elements are synthesized in NSMs, the current version of
CELib ignores other elements.  Their contributions are negligible for the whole
chemical evolution because of a relatively low event rate of NSMs (see the next
section for the argument regarding the NSM rate).

\subsection{Released energy and return mass fraction} \label{sec:Misc}

The amount of released energy and return mass from individual events are
also described in the yields tables. Here we summarize these quantities. 

For SNe II, the released energy of each SN is set to $10^{51}~{\rm erg}$ when
the yields table of \cite{Portinari+1998} is selected.  When the yields table of
\cite{Nomoto+2013} is selected, the released energy listed in the table is used.
The normal SNe releases $10^{51}~{\rm erg}$ per SN whereas the Pop III SNe and
HNe release more than 10 times more energy than the normal SNe and the released
energy depends on the progenitor mass.

Table \ref{tab:SNIIEnergy} shows IMF weighted released energy from SNe II.
These quantities are calculated by using Eq.
\ref{eq:ReturnEnergy}.  The integrated mass ranges for the yields table of
\cite{Nomoto+2013} are $13$--$40~\Msun$ for $Z>0$ and $11$--$300~\Msun$ for
$Z=0$, whereas that of \cite{Portinari+1998} is $9$--$120~\Msun$.  When we
compare the released energies evaluated with the yields tables of
\cite{Nomoto+2013} with $f_{\rm HN} = 0$ and \cite{Portinari+1998} we find that
the case with \cite{Portinari+1998} releases about twice as much energy as that
with \cite{Nomoto+2013}, reflecting the narrow mass range for SNe II in
\cite{Nomoto+2013}.

An SSP particle with the Pop III IMF releases a significant amount of energy
because they have a lot of massive stars. The released energy from Pop III stars
is two to five times greater than those with IMFs for Pop I/II. 
In the case with $f_{\rm HN} = 0.05$, the increase of the released energy is
only $\sim~25\%$.  If we adopt $f_{\rm HN} = 0.5$, the integrated released
energy increases 3-4 times depending on the adopted IMF.  Hence, significant
impacts from these modes are expected for galaxy formation.

By dividing the released energy listed in table \ref{tab:SNIIEnergy} by
$10^{51}~{\rm erg}$, we can obtain the typical number of SNe II from a $1~\Msun$
SSP particle.  For IMFs of Pop I/II, typical numbers of SNe II are
$0.0023$--$0.0058~\Msun^{-1}$ for the yields table of \cite{Nomoto+2013} with
$f_{\rm HN} = 0.0$ and $0.0043$--$0.01~\Msun^{-1}$ for that of
\cite{Portinari+1998}.  These are comparable to those used in the previous
galaxy formation simulations.

\begin{table*}[htb]
\begin{center}
\caption{Release energy of SNe II per a unit mass (${\rm erg}~\Msun^{-1}$).}\label{tab:SNIIEnergy}
\scriptsize
\begin{tabular}{lcccccccc}
\hline
\hline
& Salpeter &Diet Salpeter& Miller-Scalo& Kroupa & Kroupa1993 &Kennicutt & Chabrier & Susa\\
\hline
\cite{Nomoto+2013}$^a$ & $3.56\times10^{48}$ & $5.38\times10^{48}$ & $3.00\times10^{48}$ & $5.37\times10^{48}$ & $2.34\times10^{48}$ & $4.40\times10^{48}$ & $5.75\times10^{48}$ & $5.64\times10^{49}$ \\
\cite{Nomoto+2013}$^b$ & $4.44\times10^{48}$ & $6.71\times10^{48}$ & $3.51\times10^{48}$ & $6.72\times10^{48}$ & $2.84\times10^{48}$ & $5.42\times10^{48}$ & $7.19\times10^{48}$ & $7.63\times10^{49}$ \\
\cite{Nomoto+2013}$^c$ & $1.23\times10^{49}$ & $1.86\times10^{49}$ & $8.08\times10^{48}$ & $1.88\times10^{49}$ & $7.37\times10^{48}$ & $1.46\times10^{49}$ & $2.02\times10^{49}$ & $2.55\times10^{50}$ \\
\cite{Portinari+1998} & $6.31\times10^{48}$ & $9.54\times10^{48}$ & $6.08\times10^{48}$ & $9.45\times10^{48}$ & $4.29\times10^{48}$ & $7.89\times10^{48}$ & $1.02\times10^{49}$ & N/A\\
\hline
\end{tabular}\\
\end{center}
$^a$ $f_{\rm HN} = 0$, $^b$ $f_{\rm HN} = 0.05$, and $^c$ $f_{\rm HN} = 0.5$.
\end{table*}

Table \ref{tab:SNII:ReturnMass:N} summarizes the return mass fraction of each
IMF by SNe II with the yields of \cite{Nomoto+2013}.  To evaluate these values,
we used Eq. \eqref{eq:ReturnMassFraction}.  Here, we assumed $f_{\rm HN} = 0.0$.
We note that the return mass fractions with $f_{\rm HN} = 0.5$ are almost
identical to those with $f_{\rm HN} = 0.0$.  The IMF to IMF variation is about a
factor of 2 and generally, the return mass fraction increases with increasing
metallicity. When we adopt the Pop III IMF, the return mass fraction reaches
$89$\% which is about 10 to 20 times larger than those with IMFs for Pop I/II.

\begin{table*}[htb]
\begin{center}
\caption{Return mass fraction of SNe II per unit mass 
with the yields table of \cite{Nomoto+2013}. $f_{\rm HN} = 0$ is assumed.}
\label{tab:SNII:ReturnMass:N}
\begin{tabular}{lcccccccc}
\hline
\hline
& Salpeter &Diet Salpeter& Miller-Scalo& Kroupa & Kroupa1993 & Kennicutt & Chabrier & Susa \\
\hline
$Z = 0.0   $  & N/A & N/A & N/A & N/A & N/A & N/A & N/A & $0.892$ \\
$Z = 0.001 $  & $0.0643$ & $0.0971$ & $0.0487$ & $0.0975$ & $0.0404$ & $0.0779$ & $0.104$ & N/A\\
$Z = 0.004 $  & $0.0641$ & $0.0970$ & $0.0486$ & $0.0972$ & $0.0403$ & $0.0777$ & $0.104$ & N/A\\
$Z = 0.02  $  & $0.0647$ & $0.0978$ & $0.0490$ & $0.0892$ & $0.0407$ & $0.0785$ & $0.105$ & N/A\\
$Z = 0.05  $  & $0.0646$ & $0.0976$ & $0.0490$ & $0.0980$ & $0.0406$ & $0.0783$ & $0.105$ & N/A\\
\hline
\end{tabular}\\
\end{center}
\end{table*}

The return mass fraction of SNe II with the yields table of
\cite{Portinari+1998} is shown in table \ref{tab:SNII:ReturnMass:P}. Typically,
the return mass fraction increases $\sim 60\%$ when we compare it to that
summarized in table \ref{tab:SNII:ReturnMass:N}.  This is because the yields
table of \cite{Nomoto+2013} covers a narrow mass range ($13~\Msun \le m \le
40~\Msun$).

\begin{table*}[htb]
\begin{center}
\caption{Return mass fraction of SNe II per unit mass 
with the yields table of \cite{Portinari+1998}.}
\label{tab:SNII:ReturnMass:P}
\begin{tabular}{lccccccc}
\hline
\hline
& Salpeter &Diet Salpeter& Miller-Scalo& Kroupa & Kroupa1993 & Kennicutt & Chabrier \\
\hline
$Z=0.0004$ & 0.104 & 0.157 & 0.0778 & 0.160 & 0.0632 & 0.124 & 0.170 \\
$Z=0.004$  & 0.105 & 0.159 & 0.0783 & 0.161 & 0.0640 & 0.125 & 0.173 \\
$Z=0.008$  & 0.114 & 0.173 & 0.0804 & 0.176 & 0.0676 & 0.134 & 0.188 \\
$Z=0.02$   & 0.120 & 0.182 & 0.0824 & 0.185 & 0.0704 & 0.141 & 0.199 \\
$Z=0.05$   & 0.124 & 0.187 & 0.0839 & 0.190 & 0.0720 & 0.144 & 0.204 \\
\hline
\end{tabular}\\
\end{center}
\end{table*}

For SNe Ia, a $10^{51}~{\rm erg}$ per SN is assumed for simplicity, although the
typical energy of a SN Ia in models is slightly larger than $10^{51}~{\rm erg}$.
The return mass of each event is typically $1~\Msun$. As is expected, the amount
of the return mass from SNe Ia is insignificant.

The return mass fraction due to AGBs is shown in table \ref{tab:AGB:ReturnMass}.
This quantity is also evaluated by using Eq.  \eqref{eq:ReturnMassFraction}.
Reflecting the shape of the IMF in the lower mass regime, the return mass
fraction changes when we change the adopted IMF.  However, the variation is less
than a factor of 2.  The return mass fraction of AGBs is $1.5$--$2.5$ times as
much as those of SNe II. On the other hand, since the net yields in AGBs are
smaller than those in SNe II (figures \ref{fig:SNII:Yields:N13},
\ref{fig:SNII:Yields:P98mod}, \ref{fig:AGB:Yields0} and \ref{fig:AGB:Yields1}),
the contribution from AGBs to chemical evolution is not necessarily greater than
that from SNe II. We confirm this in \S \ref{sec:Applications}.

\begin{table*}[htb]
\begin{center}
\caption{Return mass fraction of AGBs per unit mass.}\label{tab:AGB:ReturnMass}
\begin{tabular}{lcccccccc}
\hline
\hline
& Salpeter &Diet Salpeter& Miller-Scalo& Kroupa & Kroupa1993 & Kennicutt & Chabrier & Susa\\
\hline
$Z = 0.0$     & N/A      & N/A     & N/A     & N/A     & N/A     & N/A     & N/A    & $0.0186$\\
$Z = 0.0001$  & $0.165 $ & $0.250$ & $0.292$ & $0.227$ & $0.212$ & $0.271$ & $0.243$ & N/A     \\
$Z = 0.004 $  & $0.171 $ & $0.258$ & $0.303$ & $0.235$ & $0.221$ & $0.281$ & $0.251$ & N/A     \\
$Z = 0.008 $  & $0.174 $ & $0.262$ & $0.307$ & $0.238$ & $0.225$ & $0.286$ & $0.255$ & N/A     \\
$Z = 0.02  $  & $0.176 $ & $0.266$ & $0.312$ & $0.241$ & $0.228$ & $0.290$ & $0.258$ & N/A     \\
\hline
\end{tabular}\\
\end{center}
\end{table*}

For AGBs and NSMs, we assume that they do not release energy for simplicity,
although the energy and momentum release via stellar winds and jets might affect
the evolution of galaxies.

\begin{figure}
\centering
\epsscale{1.0}
\plotone{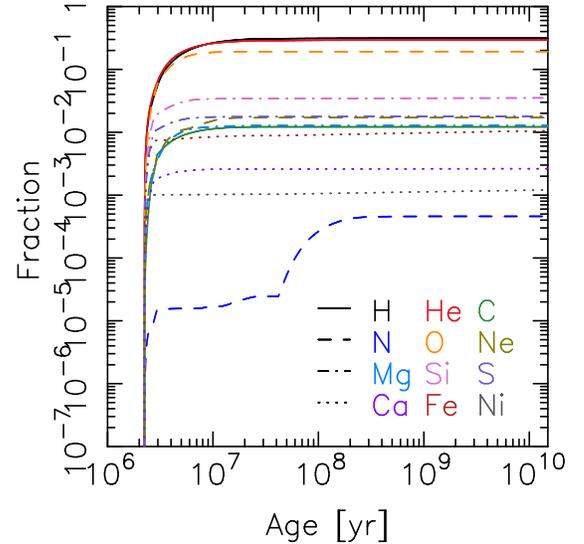}
\caption{Cumulative return mass fractions of 12 primary elements as a function
of SSP particle age. Solid, dashed, dot-dashed and dotted curves correspond to
groups of (H, He, C), (N, O, Ne), (Mg, Si, S) and (Ca, Fe, Ni), respectively.
$Z=0$ is assumed, and thus the Susa IMF ($0.7$--$300~\Msun$) is used.  SNe II,
SNe Ia, and AGBs are taken into account.  For SNe II, we use the yields table of
\cite{Nomoto+2013}, while for SNe Ia, we use that of \cite{Seitenzahl+2013}
(N100).  The AGB yields are the combination of \cite{CampbellLattanzio2008} and
\cite{Gil-Pons+2013}.  We assume that the amount of mass loss in each massive
star is released at the same time as its SN explosion.
}
\label{fig:Library:Z0}
\end{figure}

\begin{figure}
\centering
\epsscale{1.0}
\plotone{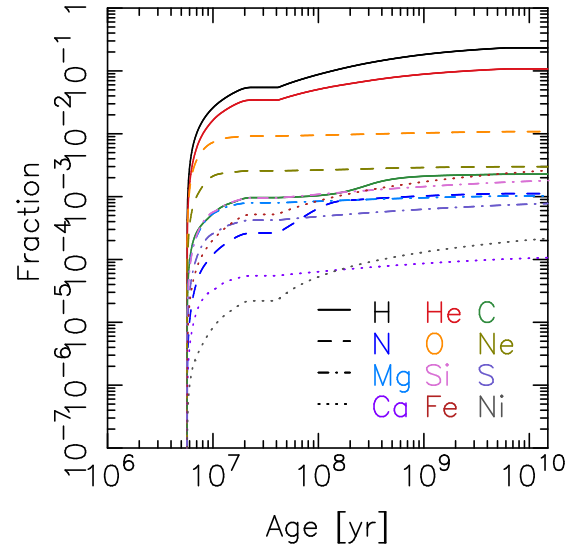}
\caption{Same as figure \ref{fig:Library:Z0}, but with $Z=0.0134$.
The Chabrier IMF with the mass range of $0.1$--$100~\Msun$ is assumed.  Yields
tables for SNe II and SNe Ia are the same as those used in figure
\ref{fig:Library:Z0}.  For AGBs, we use the combined yields table of
\cite{Karakas2010} and \cite{Doherty+2014}.
}
\label{fig:Library:Zsun}
\end{figure}

Figures \ref{fig:Library:Z0} and \ref{fig:Library:Zsun} show the evolutions of
cumulative return mass fractions of 12 primary elements as a function of age.
In these figures, the amounts of return masses from SNe II, SNe Ia, and AGBs are
taken into account.  For SNe II and AGBs, the mass release occurs at the end of
progenitors' lifetimes whereas, for SNe Ia, the mass release follows a power-law
type event rate (see \S \ref{sec:modeling:SNIa}).  The evolution of the return
mass fractions is characterized by two phases: the early phase ($\sim10~{\rm
Myr}$) is due to SNe II and the late phase ($t>4\times10^7~{\rm yr}$) is due to
both SNe Ia and AGBs.

In the zero-metallicity case (figure \ref{fig:Library:Z0}), we can see the
evolution return masses, except for N, are characterized only by SNe II because
of their specific IMF and high return mass fraction.  On the other hand, in the
$Z=\Zsun$ case (figure \ref{fig:Library:Zsun}), we can clearly read the
contributions of three different feedbacks.  For instance, the evolutions of the
return O, Mg and Si are almost unchanged after SNe II ($>3\times 10^7~{\rm
yr}$), since these $\alpha$ elements are mainly released from SNe II and others
contributions are almost negligible.  Those of Ca, Fe and Ni have two phases and
the late phase is due to the SNe Ia. The late stage evolutions of H, He, C and N
are owing to AGBs.

\subsection{Solar abundance patterns}\label{sec:abundance}  

CELib equips three popular solar abundance patterns. These are provided by
\cite{AndersGrevesse1989}, \cite{GrevesseSauval1998}, and \cite{Asplund+2009}.
In the three solar abundance patterns above, that of \cite{Asplund+2009} is
regarded as the fiducial one.  The fiducial value of $\Zsun$ is $0.0134$ in
CELib.  Users can choose the other two if necessary by changing a control flag.

\section{Reference Models of Stellar Feedback} \label{sec:modeling}

\begin{figure*}
\centering
\epsscale{1.0}
\plotone{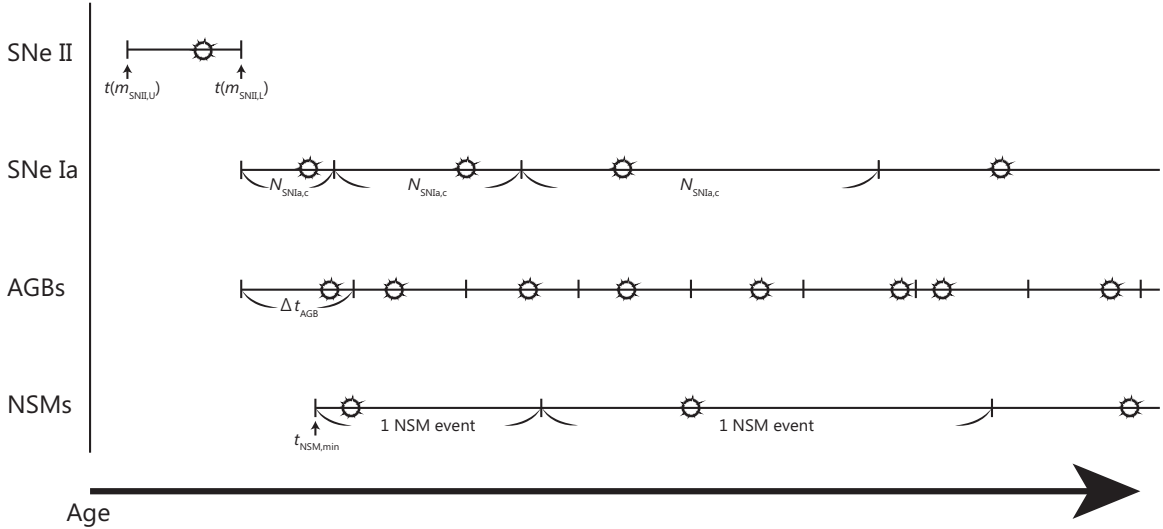}
\caption{The schematic picture of our reference feedback models.
From top to bottom, we describe feedbacks of SNe II, SNe Ia, AGBs, and NSMs for
a single SSP particle. Time passes from left to right. Explosion marks indicate
time of feedback. These times are evaluated by using random numbers. The
functional form of the adopted IMF is used for the weight to determine the
time.  Further details are described in the text.
}
\label{fig:feedbacks}
\end{figure*}

In this library, feedback models are also implemented which utilize CELib yields
tables. We describe these ``reference'' feedback models in this section. These
are optional functions of this library, and thus, it does not necessarily use
them. However, these are good references for using this library in one's own
code.  The models we described here give the time of events and return mass of
each element, and the released energy, depending on feedback types.  How to
redistribute these quantities is left to the user code.  The summary of our
reference feedback models is shown in figure \ref{fig:feedbacks}.

\subsection{Feedback Model of Type II SNe} \label{sec:modeling:SNII}

The reference model of the feedback from SNe II follows the SNe II model of
\cite{Okamoto+2008}.  In this model, all of the SNe in an SSP particle are
assumed to have exploded at the same time. Therefore, the number of explosion
events of SNe II is only one (see figure \ref{fig:feedbacks}).  We implement
this model with a slight modification to determine the explosion time.

In order to obtain the explosion time of an SSP particle, we use the
following equation:
\begin{align}
{\mathcal P}_{\rm SNII}(m_*) &= \frac{1}{N_{\rm SNII}(m_{\rm SNII,L};m_{\rm SNII,U})} \nonumber \\
& \times \int_{m_*}^{m_{\rm SNII,U}} \frac{\xi(\log_{10} m)}{m} dm, \label{eq:SNII:Rate:P}
\end{align}
and 
\begin{equation}
N_{\rm SNII}(m_{\rm SNII,L};m_{\rm SNII,U})
%\epsilon_{\rm SNII} 
= {\int^{m_{\rm SNII,U}}_{m_{\rm SNII,L}} \frac{\xi(\log_{10} m)}{m} dm},
\end{equation}
where $m_{\rm SNII,L}$ and $m_{\rm SNII,U}$ are the lower and upper mass for the
SNe II, respectively, and these quantities are defined by the adopted yields
table.  The value of ${\mathcal P}_{\rm SNII}(m_*)$ ranges from zero to unity, since $m_{\rm
SNII,L} \le m \le m_{\rm SNII,U}$.  When an SSP particle is born, we generate a
random real number with a domain of $[0,1)$, $A_{\mathcal R}$. We then find the
mass $m_*$ which satisfies $A_{\mathcal R} = {\mathcal P}_{\rm SNII}(m_*)$, and then we
compute the lifetime of the star whose mass is $m_*$ using Eq.
\eqref{eq:LifeTime:Polynomial}. We adopt this lifetime as the explosion time of
the SSP particle.  The explosion time depends on both the IMF and the
metallicity, and it is automatically taken care of in this library.
Practically, CELib generates lookup tables at the initialization phase to reduce
the evaluation cost of Eq. \eqref{eq:SNII:Rate:P}.

This model can obtain the explosion time in one trial.  Thus, this
implementation can reduce the number of evaluations of explosion time greatly,
compared to the original implementation in \cite{Okamoto+2008}.  As a trade-off,
we need to prepare an argument in a simulation code to store the explosion time.

\begin{figure}
\centering
\epsscale{1.0}
\plotone{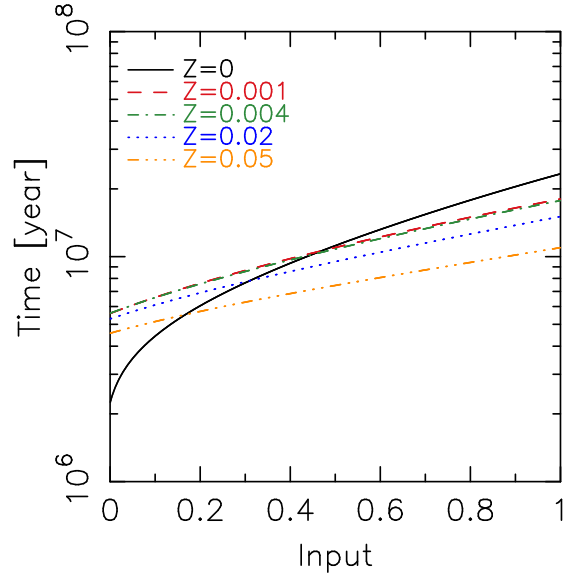}
\caption{Explosion time of SNe II in our SNe II feedback model as a
function of input parameter, $A_{\mathcal R}$. The yields table of \cite{Nomoto+2013}
is used. The Chabrier IMF is adopted for SSPs with $Z \ge 0.001$, whereas the
Susa IMF is employed for SSPs with $Z = 0$.
}
\label{fig:SNII:Rate:N13}
\end{figure}

\begin{figure}
\centering
\epsscale{1.0}
\plotone{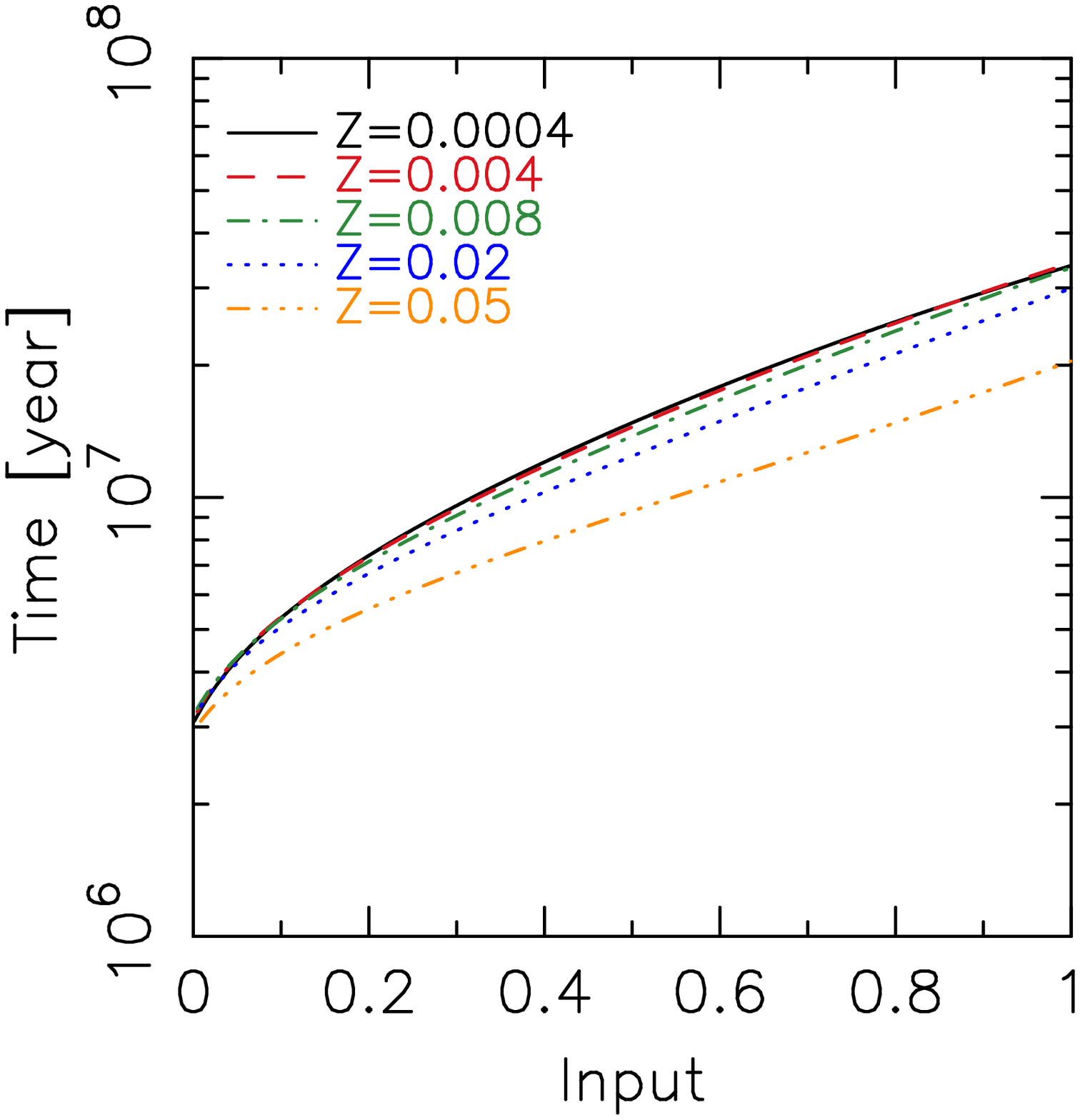}
\caption{Same as figure \ref{fig:SNII:Rate:N13}, but with the yields table of
\cite{Portinari+1998}.
}
\label{fig:SNII:Rate:P98}
\end{figure}

Figures \ref{fig:SNII:Rate:N13} and \ref{fig:SNII:Rate:P98} show the time
distributions of our SNe II feedback model.  When we set a random real number
$A_{\mathcal R}$ of $[0,1)$ and the metallicity of an SSP particle $Z$, we can
obtain the corresponding explosion time.  Since less massive stars dominate in
the Pop I/II IMFs, the explosion time is weighted by the lifetimes of the less
massive stars.  The distribution of the explosion time of Pop III stars has a
different shape if we compare it with those of Pop I/II stars. This is because
the functional forms of IMFs are different.

\subsection{Feedback Model of Type Ia SNe}
\label{sec:modeling:SNIa}

The SN Ia is usually considered to be an explosion that occurs in low-mass
binary systems. The candidates of the progenitors are (1) white
dwarf-giant binaries \citep[the SD scenario;][]{WhelanIben1973, Nomoto1982} and
(2) white dwarf-white dwarf binaries \citep[the DD
scenario;][]{IbenTutukov1984, Webbink1984}.  In both cases, the explosion occurs
when the white dwarf's mass exceeds the Chandrasekhar limit.  They each
have their own advantages/disadvantages and there is no consensus \citep[For
those who are interested in the details regarding SNe Ia, see recent
reviews:][]{Parrent+2014, Maoz+2014ARAA}.  Thus, an empirical modeling of the SN
rate would currently be the best solution.

Here, we implemented two types of feedback models for SNe Ia.  One is an
analytical model based on the SD scenario. The implementation is identical to
that of \cite{GreggioRenzini1983} with updated parameters by
\cite{Portinari+1998}.

The rate function of \cite{GreggioRenzini1983} is based on the evolution of
binary systems.  Assume that $m_{\rm B}$ is the total mass of a binary system
and it is $m_{\rm B} = m_1 +m_2$ where $m_1$ and $m_2$ are masses of the primary
and secondary stars, respectively. The functional form of the SNe Ia rate (the
cumulative number of SNe Ia explosions) up to time $t$ is 
\begin{equation}
{\mathcal R}_{\rm SNIa}(\le t) = b \int_{m_{\rm B,L}}^{m_{\rm B,U}}
\frac{\xi (\log_{10} m_{\rm B})}{m_{\rm B}}
\left [ \int_{\mu_{\rm min}(t)}^{0.5} f(\mu) d\mu \right ] dm_{\rm B},
\label{eq:SNIa:Rate:GR83}
\end{equation}
where $m_{\rm B,L}$ and $m_{\rm B,U}$ are the minimum and maximum mass of a
binary system, respectively, $\mu_{\rm min}(t)$ is the minimum mass fraction
which contributes to SNe Ia:
\begin{equation}
\mu_{\rm min}(t) = \max \left [ \frac{m_2(t)}{m_{\rm B}}, 
\frac{m_{\rm B}-0.5 m_{\rm B,U} }{m_{\rm B}} \right],
\end{equation}
and $f(\mu) = 24 \mu^2$ is the distribution function of the functionary mass of
the secondary star.  According to \cite{Portinari+1998}, we assume that $m_{\rm
B,L} = 3~\Msun$, $m_{\rm B,U} = 12~\Msun$, and $b = 0.05$. Note that the value
of $b$ is determined in order to make the number of SNe Ia $\sim0.2$ times that
of SNe II \citep{vandenBerghTammann1991, Cappellaro+1997}.  Apparently, it is
necessary to recalibrate this value when one changes the assumed IMF shape, mass
range, and/or the values of $m_{\rm B,L}$ and $m_{\rm B,U}$.  However, we only
use the fixed value for simplicity.

There are other analytical formulations of the SNe Ia rate for different
scenarios. For instance, \cite{Hachisu+1996, Hachisu+1999} gave the extended
model for the SD scenario where both white dwarf-main sequence and white
dwarf-red giant binary systems are taken into account, and this is used in
chemical evolution studies \cite[e.g.,][]{Kobayashi+1998, Kobayashi+2000}.  For
the DD scenario, \cite{Greggio2005} showed the analytical formulations.  Note
that the DD scenario based rate of \cite{Greggio2005} is similar to that of the
SD scenario \citep{GreggioRenzini1983}, as it is shown in \cite{Matteucci+2009}.
Here, we do not implement further analytical models. We move to the empirical
model.

The other model we implemented is the empirical model based on recent
observations of SNe Ia.  Observations imply that the delay time distribution
(DTD) function of SNe Ia follows the power law of $t^{\sim -1}$
\citep[e.g.,][]{Totani+2008, MaozMannucci2012}.  The power law DTD we employed
here is expressed as 
\begin{equation}
\frac{dN_{\rm SNIa}}{dt} = \epsilon_{\rm SNIa} \left ( \frac{t}{10^9~{\rm yr}} \right )^{p_{\rm SNIa}}, 
~~~~(t > 4\times10^7~{\rm yr})
\label{eq:SNIa:DTD}
\end{equation}
where $\epsilon_{\rm SNIa} = 4\times10^{-13}$ and $p_{\rm SNIa} = -1$ according to
\cite{MaozMannucci2012}.  The time offset of $4\times10^7~{\rm yr}$ comes from
the typical lifetime of a $8~\Msun$ star (we ignore the metallicity dependence
of the stellar lifetime because of its weak dependence).  The integrated SNe Ia
number per $1~\Msun$ during the first $10~{\rm Gyr}$ is $2.2\times10^{-3}$
\citep{MaozMannucci2012}.  This even rate is $1/5\sim1/2$ of those of the SNe
II.  By multiplying the mass of an SSP particle, we can obtain the event number
of SNe Ia in the SSP particle. 

By integrating Eq. \eqref{eq:SNIa:DTD}, we can obtain the cumulative numbers of
the SNe Ia up to time $t$ for this model and it is expressed as
\begin{equation}
{\mathcal R}_{\rm SNIa}(\le t) = \int_{0}^{t} \frac{dN_{\rm SNIa}}{dt'} dt'.
\label{eq:SNIa:Rate:MM12}
\end{equation}

The event rate of the SNe Ia is very low (0.002 event per$1~\Msun$ per $10~{\rm
Gyr}$) and the duration time is very long ($\sim 10~{\rm Gyr}$).  For instance,
an SSP particle of $10^4~\Msun$ has only 20 events in the time span of $10~{\rm
Gyr}$.  Thus, in the current feedback model of SNe Ia, either an individual event 
or a cluster of events follows. The explosion time of SN Ia is evaluated by the
probabilistic manner which is described below.

First, we calculate the cumulative event rate of the SNe Ia as a function
of time (Eqs.  \ref{eq:SNIa:Rate:GR83} or \ref{eq:SNIa:Rate:MM12}).  Inversely
solving this relation, we obtain 
\begin{equation}
t_{\rm SNIa} (N_{\rm SNIa}) = \mathcal R_{\rm SNIa}^{-1} (N_{\rm SNIa}).
\end{equation}
This function returns the time where the cumulative number of SNe Ia becomes
$N_{\rm SNIa}$.  Using this equation, we obtain two epochs, $t_{\rm SNIa}(N_{\rm
SNIa})$ and $t_{\rm SNIa}(N_{\rm SNIa}+N_{\rm SNIa,c})$, where $N_{\rm SNIa}$ is
the cumulative SNe Ia event count which is finished before this step in an SSP
particle. The number, $N_{\rm SNIa,c}$, is the ``size of SNe Ia cluster''. If
the value is larger than unity, a SN Ia event in a simulation makes an
association of $N_{\rm SNIa,c}$ SNe Ia. This can reduce the number of evaluation
times of SNe Ia and can also reduce the computation time.  Then, we find the
next explosion time between these two epochs using a random real number and the
cumulative SNe Ia rate.  This is also implemented as lookup tables.

\begin{figure}
\centering
\epsscale{1.0}
\plotone{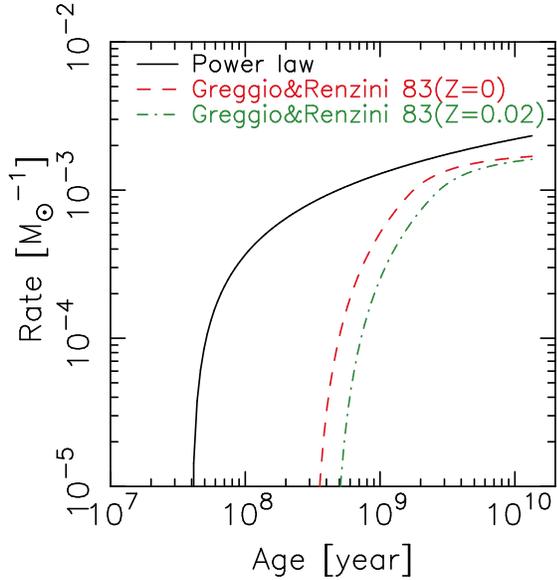}
\caption{Cumulative event number of SNe Ia per a $1~\Msun$ SSP particle as a
function of time.  Solid curve is based on the power-law type DTD (Eq.
\ref{eq:SNIa:DTD}).  Dashed and dot-dashed curves are derived from the models of
\cite{GreggioRenzini1983} with $Z=0$ and $Z=0.02$, respectively.
}
\label{fig:SNIa:Rate}
\end{figure}

We show the cumulative event numbers of SNe Ia in figure \ref{fig:SNIa:Rate}.
The difference in event starting times of the three models is due to the
differences in lifetime between $8~\Msun$ and $6~\Msun$ stars and the
metallicity dependence of stellar lifetime.  This effect can be observed in the
distribution of chemical composition (see \S \ref{sec:OneZone}).  In this case,
the final event numbers of these two models are almost comparable.  As noted
above, it is necessary to recalibrate the normalization value $b$ in Eq.
\ref{eq:SNIa:Rate:GR83} if one wants to fit the observational value more
closely.  As is expected from figure \ref{fig:SNIa:Rate}, since we assume a
constant number interval of SNe Ia, $N_{\rm SNIa,c}$, the time between two
neighboring SNe Ia feedback events becomes longer with increasing age (see
figure \ref{fig:feedbacks}).

Here we only implemented two models while there are a lot of models of DTDs.  We
refer the readers to \cite{Matteucci+2009} for further detailed comparison among
DTD functions.

\subsection{Feedback Model of AGBs}
\label{sec:modeling:AGB}

For the modeling of the AGB mass loss from an SSP particle, we assume the
following conditions: (1) the AGB mass loss takes place at the end of their
lifetime, (2) all AGB mass losses, which take place in a given time interval ($t
\rightarrow t+\Delta t_{\rm AGB}$), occur once in the time interval, and (3) the
time is selected randomly from the time interval.  The schematic picture of this
feedback model is shown in figure \ref{fig:feedbacks}.

\begin{figure}
\centering
\epsscale{1.0}
\plotone{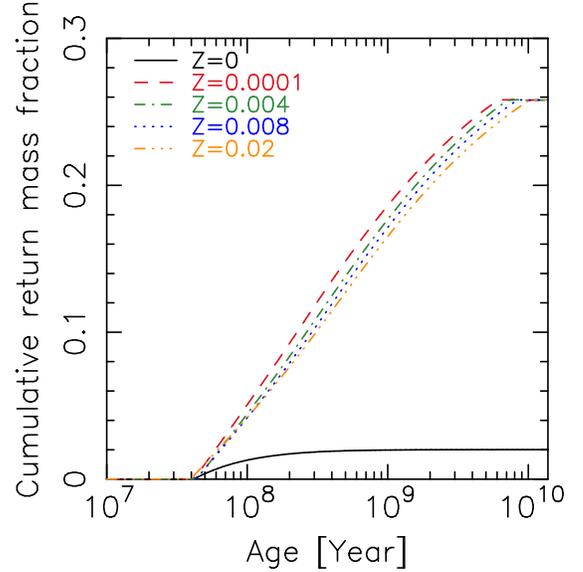}
\caption{Cumulative return mass fractions of AGBs as a function of the age of
the SSP particle are shown. The Chabrier IMF is assumed for the case with $Z>0$
while the Susa IMF is employed for the case with $Z=0$.
}
\label{fig:AGBs:CumulativeReturnMass}
\end{figure}

Using Eq. \eqref{eq:IMF:Number}, the number of AGB mass loss events per
$1~\Msun$ at a given time interval ($t \rightarrow t+\Delta t_{\rm AGB}$) is 
\begin{equation}
N_{\rm AGB}(t;t+\Delta t_{\rm AGB}) 
= \int_{m(t)}^{m(t+\Delta t_{\rm AGB})} \frac{\xi(\log_{10} m)}{m} dm,
\label{eq:AGB:Number}
\end{equation}
where $m(t)$ is the mass of star whose lifetime is $t$.  In order to evaluate
the time of the AGB mass loss from an SSP particle, we generate a random real
number $A_{\mathcal R}$ with a range of $[0,1)$ and solve the following equation
for $t_{\rm AGB}$,
\begin{equation}
A_{\mathcal R} = \mathcal P_{\rm AGB} (t_{\rm AGB};t;t+\Delta t_{\rm AGB}),
\label{eq:AGB:EventTiming}
\end{equation}
where
\begin{align}
\mathcal P_{\rm AGB}(t_{\rm AGB};t;t+\Delta t_{\rm AGB}) &= \frac{1}{N_{\rm AGB}(t;t+\Delta t_{\rm AGB})} \nonumber \\
& \times \int_{m(t)}^{m(t+t_{\rm AGB})} \frac{\xi(\log_{10} m)}{m} dm.
%\label{eq:AGB:EventTiming}
\end{align}
This $t+t_{\rm AGB}$ is the next event time and $0 \le t_{\rm AGB} \le \Delta
t_{\rm AGB}$.  Practically, we prepare the lookup table of the cumulative number
of $N_{\rm AGB}$ from $t=0$ to $t=t_H$ with a constant time interval, $\Delta
t_{\rm AGB}$.  Note that $t_H$ is the current age of the Universe
\citep{Plank2014XVI} and we adopt $\Delta t_{\rm AGB} = 3\times10^{\rm 8}~{\rm
yr}$ as a fiducial value so that it can follow the AGB mass loss time scale (see
below and figure \ref{fig:AGBs:CumulativeReturnMass}).

The evaluation of the event time using Eq. \eqref{eq:AGB:EventTiming} is carried
out just after star formation and every time after the AGB mass loss event.

Within our reference AGB feedback model, we assume that there is no energy
release, although it has been pointed out that stellar winds of AGB stars would
have a large impact on galaxy formation, in massive galaxies in particular
\citep{Conroy+2015}.

The cumulative return mass fractions with different metallicities as a function
of time are shown in figure \ref{fig:AGBs:CumulativeReturnMass}.  It is evident
that for the case with $Z>0$ the release timescale of the AGBs is very long.
It starts at $4\times 10^6~{\rm yr}$ and continues up to
$5$--$9\times10^{9}~{\rm yr}$.  The end time depends on the metallicity.
Contrary to this, the return mass of the zero metal AGBs is very small, and the
release timescale is very short. The mass release stops $\sim 10^8~{\rm yr}$,
reflecting the IMF with a peak at $\sim 20~\Msun$ for Pop III stars.

\subsection{Feedback Model of NSMs} \label{sec:modeling:NSM}

Here, we adopt a power-law type DTD \citep{Dominik+2012, Shen+2015}.  The event
rate of the NSMs, $\epsilon_{\rm NSM}$, is set to be proportional to the event
number of the core-collapse SNe in the mass range of $20$--$40~\Msun$ with the
proportional factor of $0.01$ at $10~{\rm Gyr}$ \citep{Ishimaru+2015,
Hirai+2015}. Since this value changes in accordance with the IMF type, it is
calculated at the time of initialization.  The method of evaluation of the event
time is the same as that used in SNe Ia.

The functional form of the DTD of NSMs is 
\begin{equation}
\frac{dN_{\rm NSM}}{dt} = \epsilon_{\rm NSM} \times  \left ( \frac{t}{1~{\rm yr}} \right )^{p_{\rm NSM}},
~~~~{\rm for}~t > t_{\rm NSM,min}
\label{eq:NSM:DTD}
\end{equation}
where $p_{\rm NSM}=-1$ is the fiducial value but we deal with it as a parameter.
$t_{\rm NSM,min}$ is the delay-time and is also a parameter.  The fiducial value
is set to $10^8~{\rm yr}$.  This is comparable to the averaged binary lifetime
of the NSMs obtained through observation \citep{Lorimer2008} and the binary
population synthesis model \citep{Dominik+2012}.  The parameter dependencies are
studied in \S \ref{sec:Applications}. The cumulative number of NSMs is
\begin{equation}
{\mathcal R}_{\rm NSM}(\le t) = \int_{0}^{t} \frac{dN_{\rm NSM}}{dt'} dt'.
\label{eq:NSM:Rate}
\end{equation}

\begin{figure}
\centering
\epsscale{1.0}
\plotone{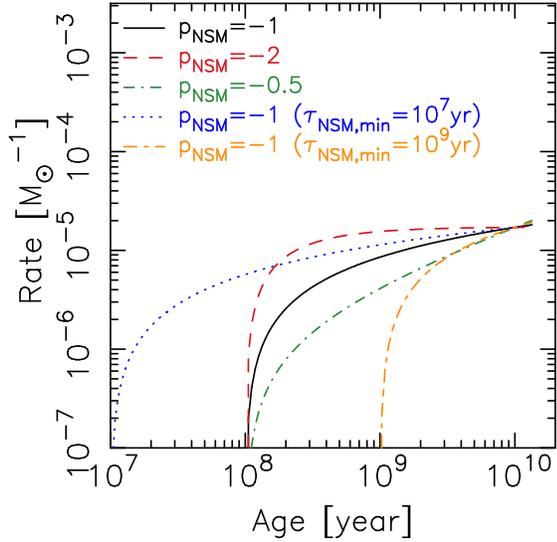}
\caption{Cumulative number of NSMs as a function of the age of an SSP particle.
The Chabrier IMF is assumed in order to evaluate the normalization. 
}
\label{fig:NSM:Rate}
\end{figure}

Figure \ref{fig:NSM:Rate} shows the cumulative numbers of NSMs as a function of
the age of an SSP particle with different power-law indexes, $p_{\rm NSM}$.
When the power-law index decreases from $-0.5$ to $-2$, the enrichment of Eu
becomes faster. The minimum event time, $t_{\rm NSM,min}$, is also expected to
affect the enrichment history. We will show results of the parameter survey of
the NSM feedback in \S \ref{sec:Applications}.

The event time is evaluated in the same way as those of SNe Ia.  We always deal
with the individual events of NSM (see figure \ref{fig:feedbacks}). In other
words, the ``cluster'' model used in the SNe Ia model is not introduced in the
NSM feedback model because of its very low event rate. As is the case with
other models, CELib generates a lookup table to evaluate the event time.

\section{Implementation} \label{sec:Implementation}

This library is written in C language. In particular, the standard of C99 is
used. To guarantee the portability, this library does not require any external
library. Thanks to ``\verb+extern C+'', C++ programs can also used this library.

This library consists of four parts: IMFs, stellar lifetimes, yields, and the
reference feedback models. When the subroutine for the initialization is called
by a simulation code, the library computes normalizations of the IMFs, lookup
tables of stellar lifetimes for different metallicities and their inverse lookup
tables, IMF weighted yields for SNe II and AGBs, and lookup tables used for the
reference feedback models. If a user changes control flags and parameters, 
the subroutine for initialization should be re-called.  After initialization,
all functions are ready to use.  See appendix \ref{sec:APIs} where we explain
the usage of the primary functions of CELib.

\section{Applications} \label{sec:Applications}

In this section, we show the results of numerical simulations using CELib.
First, we apply this library to the one-zone calculation.  It is a good
demonstration and benchmark test.  We then show the chemical evolution of a disk
galaxy which is evolving in a static Navarro-Frenk-White (NFW) halo
\citep{Navarro+1997}. The results of these tests prove that this library is a
powerful tool for studying the chemical evolution of galaxies.  

We here emphasize that the primary purpose of these tests is to demonstrate the
ability of CELib. We do not carry out the detailed comparison between the
numerical results and observations.

\subsection{One-zone model} \label{sec:OneZone}

As an initial demonstration of this library, we use it to evaluate the metal
enrichment of a one-zone model. There is a long history of the one-zone model
and there are many variants (e.g., \citeauthor{Tinsley1980}
\citeyear{Tinsley1980}; for reviews, see \citeauthor{Matteucci2003}
\citeyear{Matteucci2003} and \citeauthor{Prantzos2008} \citeyear{Prantzos2008}).
Since the aim of this test is only to check the capability of this library and
to know the typical differences which come from the adopted yields and
parameters, we carry out a closed-box simulation which does not take into
account both the inflow and outflow. Thus, the history of metal enrichment that
we show here has some inconsistency with observations (e.g., the G-dwarf
problem). We do not investigate this discrepancy thoroughly.  Since we assume a
closed system, we ignore the energy releases from feedback.

A closed system with an initial gas mass of $10^{11}~\Msun$ is considered.  The
governing equation of this system is 
\begin{equation}
M_{\rm gas}^{n+1} = M_{\rm gas}^{n} - M_{\rm *}^n + M_{\rm ej}^n
\label{eq:onezone:gov}
\end{equation}
where $n$ is the time step, $M_{\rm gas}^n$ is the gas mass, $M_{\rm *}^n$ is
the stellar mass formed in this step, and $M_{\rm ej}^n$ the ejecta mass
released from stars formed until this step. We integrate the system governed by
this equation through $13~{\rm Gyr}$. The total number of steps is 10000 and a
constant time interval of $\Delta t_{\rm oz} = 1.3~{\rm Myr}$ is adopted. This
time-interval is sufficient to resolve the time delay of SNe from the epoch of
the formation of the progenitor stars. The initial mass of the stellar component
is set to zero.

The gas component, whose mass is $M_{\rm gas}$, consists of thirteen elements
and the following relation is held for every step:
\begin{equation}
M_{\rm gas}^n = \sum_i M_{{\rm el},i}^n,
\end{equation}
where the mass of $i$-th element at the $n$-th step is expressed as $M_{{\rm
el},i}^n$.

The second term on the right-hand side of Eq. \eqref{eq:onezone:gov} is the
stellar mass formed in this step. In order to evaluate the stellar mass formed
in a given step, we assumed for simplicity that the star formation rate of this
system follows an exponential form function with a timescale of $\tau_{\rm
sf} = 8~{\rm Gyr}$. This timescale is comparable to the prediction from other
studies for the solar neighborhood \citep[e.g.,][]{Chiappini+2001}. The
normalization of star formation, $N_{\rm *}$, is evaluated by the following
equation
\begin{equation}
M_{\rm *, total} = N_{\rm *} \int_{t=0}^{t=13~{\rm Gyr}} \exp(-t/\tau_{\rm sf}) dt,
\end{equation}
where $M_{\rm *, total}$ is the stellar mass formed by star formation until the
end of the simulation and it is set to half of the initial gas mass, i.e.,
$5\times 10^{10}~\Msun$.  Thus, the stellar mass formed during $n
\rightarrow n+1$ is 
\begin{equation}
M_{\rm *}^n = N_{\rm *} \int_{t_{\rm oz} (n)}^{t_{\rm oz} (n+1)} \exp(-t/\tau_{\rm sf}) dt,
\end{equation}
where $t_{\rm oz} (n) = n \Delta t_{\rm oz}$.
We regard these stars, $M_{\rm *}^n$, as the SSP with the Chabrier IMF. 
If we turn on the Pop III mode, the Susa IMF is used for $Z<Z_{\rm popIII}$.
Note that if the IMF (for Pop I/II stars) changes it alters the evolution of the
system.  However, its typical contribution would be small \citep{Romano+2005}.
As described below, the stellar mass at $n$ decreases when time passes because
of SNe and stellar mass loss. We thus denote the initial mass of the stars
formed at step $n$, $M_{\rm *,init}^n$.  The metal distribution in newly born
stars at a given step $n$ is the same as that of the progenitor gas.

The third term on the right-hand side of Eq. \eqref{eq:onezone:gov} is the total
return mass due to feedback, namely SNe II/Ia, AGBs, and NSMs. 
The return mass at step $n$ is 
\begin{equation}
M_{\rm ej}^n = \sum_{k=1}^{k<n} M_{{\rm ret},n}^k,
\end{equation}
where $M_{{\rm ret},n}^k$ expresses the total return mass which is released in
step $n$ from the stellar component formed in step $k$.  We adopt the reference
models of feedback described in \S \ref{sec:modeling}.  The event time and the
amount of the released mass (and composition of the released mass) are evaluated
through the CELib APIs (see appendix \ref{sec:APIs}).  The important parameters
regarding feedback we fixed are $N_{\rm SNIa,c} = 3$ and $\Delta t_{\rm AGB} =
130~{\rm Myr}$.

From here we show several results obtained by our one-zone model. As mentioned
above, the aim of this section is to demonstrate the capability of this library and
thus not to compare the details with observational results. We compare results
obtained by different yields tables for SNe II and study the contribution of
SNe Ia, AGBs, and NSMs by turning on/off the feedback.

\begin{figure}
\centering
\epsscale{1.0}
\plotone{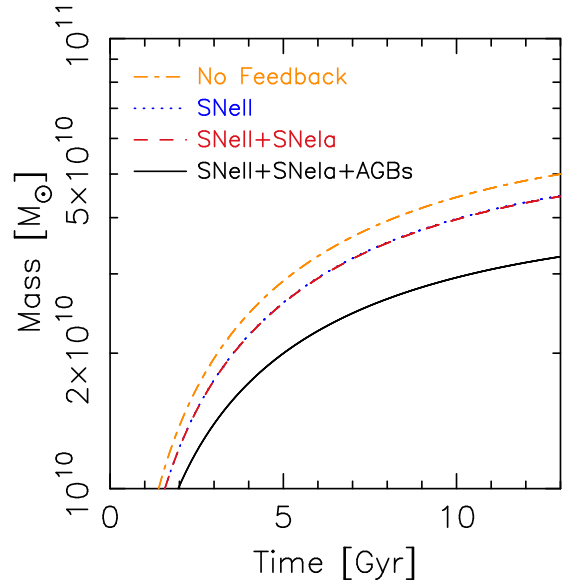}
\caption{Stellar mass evolution for four different models. The yields table of
\cite{Nomoto+2013} is used for SNe II. The N100 model in the yields table of
\cite{Seitenzahl+2013} is used for SNe Ia.  Both the normal and super AGB
yields are adopted.
}
\label{fig:OZ:MassEvolution:N}
\end{figure}

% age-Mass fraction
First, we compare the evolution of the stellar mass in four different models in
figure \ref{fig:OZ:MassEvolution:N}. We adopted the yields of \cite{Nomoto+2013}
for SNe II, those of the model N100 in \cite{Seitenzahl+2013} for SNe Ia, and
those obtained by the combination of yields of \cite{Karakas2010},
\cite{Doherty+2014}, \cite{CampbellLattanzio2008}, and \cite{Gil-Pons+2013}.
Note that HNe are not taken into account in this figure (i.e., $f_{\rm HN} =
0$).  From this figure, we can see that the SNe II and AGBs have significant
effects on the long-term evolution of stellar mass of the galaxy, whereas the
SNe Ia have less of an effect on galactic mass evolution.  This can be
understood from the tables of return mass fractions for SNe II/Ia and AGBs.
This is crucial to understanding the baryon recycling process in galaxies.

\begin{figure}
\centering
\epsscale{1.0}
\plotone{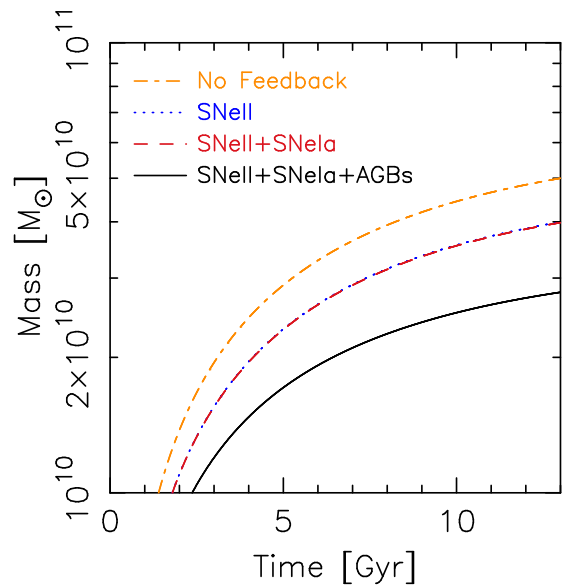}
\caption{Same as figure \ref{fig:OZ:MassEvolution:N}, but with the yields table
of \cite{Portinari+1998} used for SNe II.
}
\label{fig:OZ:MassEvolution:P}
\end{figure}

Figure \ref{fig:OZ:MassEvolution:P} shows the stellar mass evolution using the
yields table of \cite{Portinari+1998}.  The effect of SNe II is more prominent
since the return mass of the \cite{Portinari+1998}'s yields table is larger than
that of the \cite{Nomoto+2013}'s yields table (recall tables
\ref{tab:SNII:ReturnMass:N} and \ref{tab:SNII:ReturnMass:P}).  This increase
changes the total metallicity of the system.

\begin{figure}
\centering
\epsscale{1.0}
\plotone{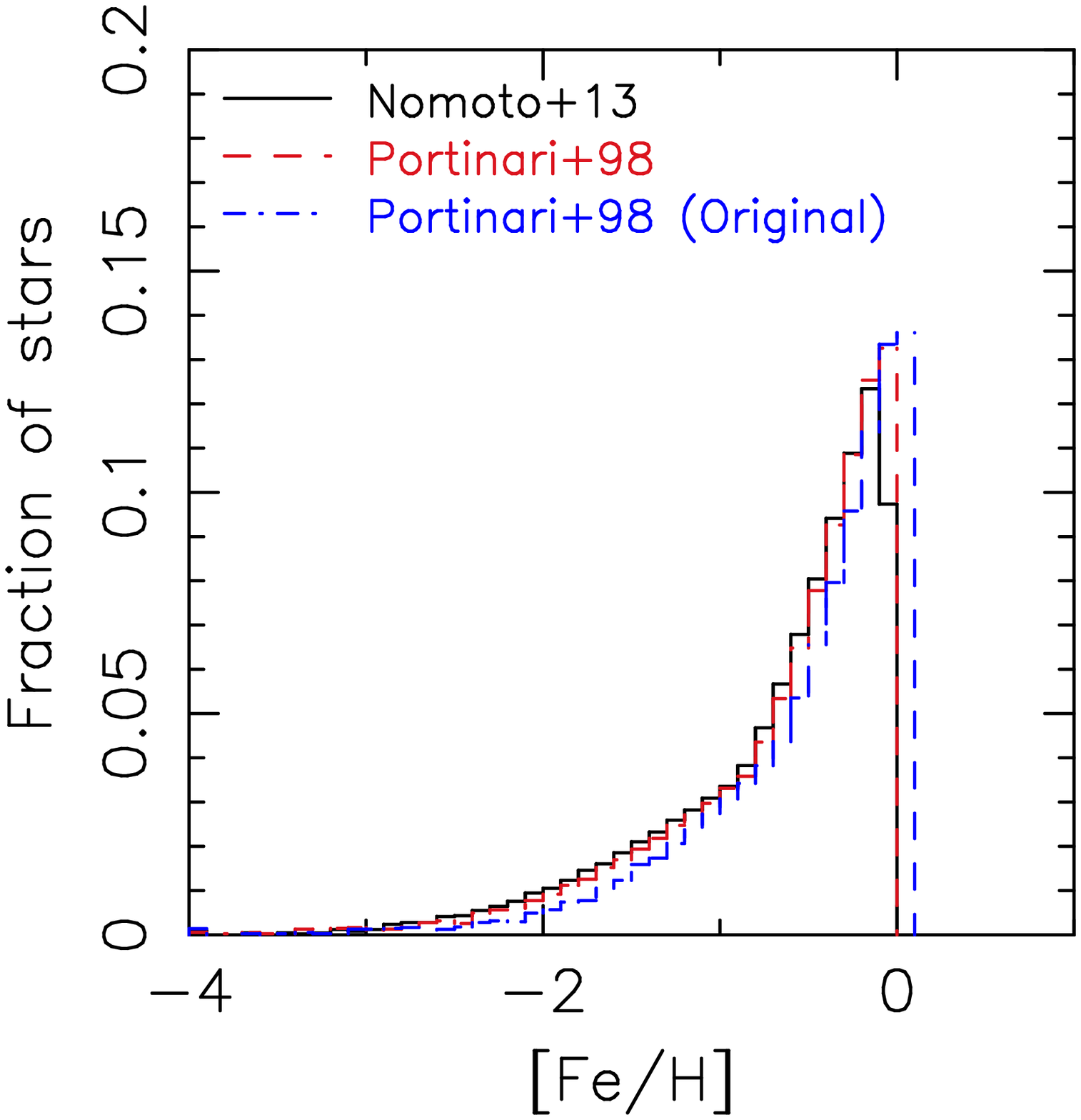}
\caption{Metallicity distribution functions. The histogram with black solid
lines shows the metallicity distribution function with \cite{Nomoto+2013} for
SNe II, whereas that with red dashed lines indicates that with
\cite{Portinari+1998}. The model with the \cite{Portinari+1998} yields table
without the modifications is also shown with the blue dotted histogram.  Type Ia
SNe and the mass loss due to AGB are also included and model parameters are
unchanged in both runs.
}
\label{fig:OZ:massdist}
\end{figure}

Figure \ref{fig:OZ:massdist} compares the metallicity distributions of the
closed system with two different yields for SNe II at $t=13~{\rm Gyr}$.  The two
results with the \cite{Nomoto+2013}'s yields table and the
\cite{Portinari+1998}'s yields table with the yields modifications is almost
comparable, although the latter is slightly enriched because of the larger return
mass fraction.  The result that adopted the yield table of \cite{Portinari+1998}
without the modifications is more enriched, reflecting the fact that Fe is
overproduced in the yield of \cite{Portinari+1998}.  Nonetheless, the overall
features, i.e., shapes, are not so different from each other because they are
dependent on the adopted star formation history in this case.

\begin{figure*}
\centering
\epsscale{1.0}
\plotone{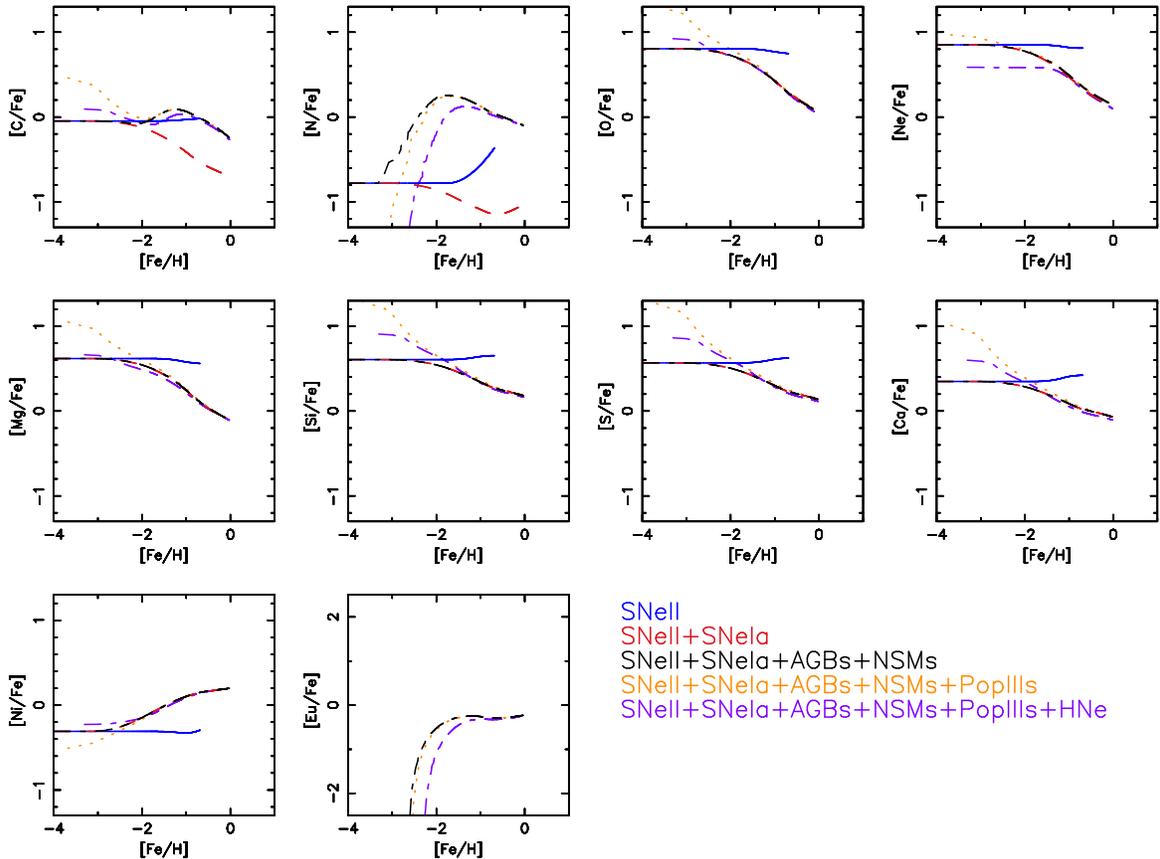}
\caption{[X/Fe]-[Fe/H] relations. Red sequences are results with only SNe II,
whereas blue sequences indicate those with both SNe II and Ia.  Black sequences
express results which include SNe II and Ia and the effect of the AGB mass loss.
The contribution of NSM is also involved.  Orange sequences indicate the results
taking into account the effect of Pop IIIs, while purple ones do both the
results of effect of Pop IIIs and HNe ($f_{\rm HN} = 0.5$) to the model
described with black sequences.  \cite{Nomoto+2013}'s yields table is used for
SNe II, whereas \cite{Seitenzahl+2013}'s yields table is adopted for SNe Ia
(N100).  The solar abundance pattern of \cite{Asplund+2009} is assumed.
}
\label{fig:OZ:AFeFeH:N}
\end{figure*}

\begin{figure*}
\centering
\epsscale{1.0}
\plotone{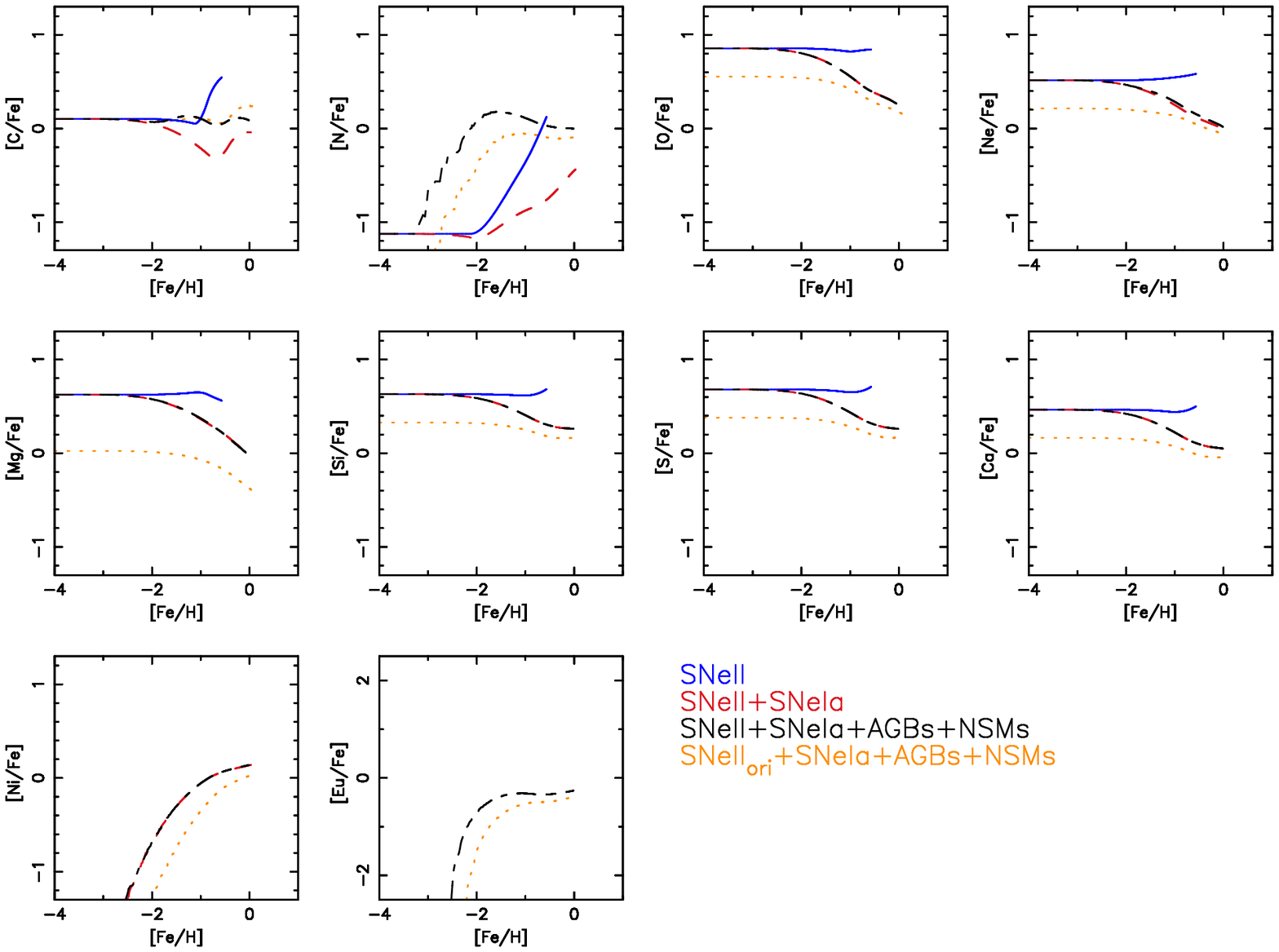}
\caption{Same as figure \ref{fig:OZ:AFeFeH:N}, but \cite{Portinari+1998}'s
yields table is used for SNe II.  The orange sequence is in the case that ad hoc
factors for C, Mg, and Fe are not multiplied.
}
\label{fig:OZ:AFeFeH:P}
\end{figure*}

We then show [X/Fe]-[Fe/H] relations, where X denotes an arbitrary element.
Figures \ref{fig:OZ:AFeFeH:N} and \ref{fig:OZ:AFeFeH:P} display the [X/Fe]-[Fe/H]
relations for ten primary elements.  The blue, red, black, orange and purple
sequences in figure \ref{fig:OZ:AFeFeH:N} are the distributions of stars in
[X/Fe]-[Fe/H] relations with different combinations of feedback.  By comparing
the blue sequence (the model with only SNe II) with the red sequence (that with
SNe II and Ia), we see that (1) the amount of Fe is increased by SNe Ia and thus
(2) the relative fraction of the other elements to Fe, i.e., [X/Fe], is
decreased.  In this configuration, the offset between the two models starts at
${\rm [Fe/H]} \sim -2.5$. This is owing to the adopted star formation history
and the starting time of SNe Ia (see below).

The contributions of the AGB stars are limited in the relatively lighter
elements such as C and N.  This is because the AGB yields are smaller than the
SNe II yields, even though the return mass fraction of AGBs is larger than that
of SNe II (recall figures \ref{fig:OZ:MassEvolution:N} and
\ref{fig:OZ:MassEvolution:P}, and tables \ref{tab:SNII:ReturnMass:N},
\ref{tab:SNII:ReturnMass:P}, and \ref{tab:AGB:ReturnMass}).

When the feedback from Pop III stars is taken into account, the evolution track
of the low metallicity regime changes.  Slight increases are observed in C, O,
Si, S and Ca while slight decreases are found in N in the low metallicity
regime.  Changes in others are almost negligible.  Their contributions are
insignificant in the high metallicity regime.  These changes are reduced when we
introduce HNe, because of HNe's high Fe yields.

It is worth noting that the overall features are similar to the figure 10 in
\cite{Nomoto+2013}, even though we adopted a different solar abundance pattern
[we adopted that of \cite{Asplund+2009} whereas they used that of
\cite{AndersGrevesse1989}] and a different one-zone model.  We note that they
did not show the results with the Pop III IMF case.

We see clear differences between results with \cite{Nomoto+2013} and \cite{Portinari+1998}.
In particular, the positions of plateaus of [X/Fe] where the contribution of SNe
II dominates are different, even with the modifications of \citep{Wiersma+2009}.
On the other hand, the transitions of the [X/Fe]-[Fe/H] relations with SNe Ia in
these two models are almost comparable. Typical differences are $\sim 0.2$ dex.
When we turn off the modifications of \cite{Wiersma+2009}, the transitions of
the [X/Fe]-[Fe/H] relations due to SNe Ia are shifted about 1 dex and the values
of [X/Fe] in low metallicity regimes with flat distributions ([Fe/H] $< -1.5$)
are shifted lower due to the overproduction of Fe in the yields of
\cite{Portinari+1998}.

\begin{figure}
\centering
\epsscale{1.0}
\plotone{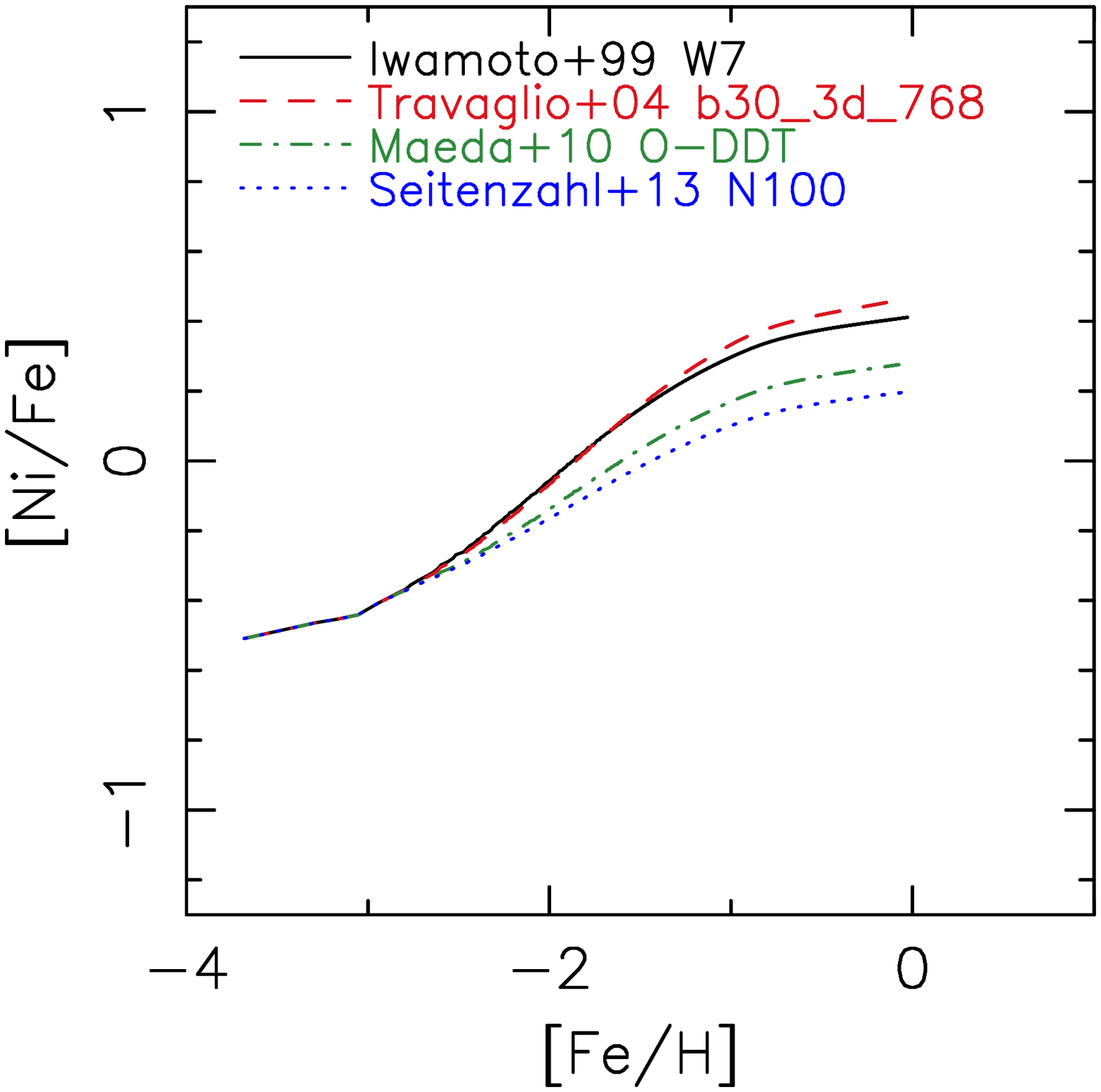}
\caption{[Ni/Fe]-[Fe/H] for different SN Ia yields.  Blue, green, red and
black sequences are results with yields of \cite{Iwamoto+1999} (W7),
\cite{Maeda+2010} (W7), \cite{Maeda+2010} (O-DDT), and \cite{Seitenzahl+2013}
(N100), respectively.}
\label{fig:OZ:SNIa:Ni:Dim}
\end{figure}

Here, we study the effects of SN Ia yields using the [Ni/Fe]-[Fe/H] diagram.
Figure \ref{fig:OZ:SNIa:Ni:Dim} shows results with four different models. The
four models are W7 in \cite{Iwamoto+1999}, b\_30\_3d\_768 in
\cite{Travaglio+2004}, O-DDT in \cite{Maeda+2010}, and N100 in
\cite{Seitenzahl+2013}. The variation of the final [Ni/Fe] is $\sim 0.3~{\rm
dex}$.  Models of O-DDT in \cite{Maeda+2010} and N100 in \cite{Seitenzahl+2013}
are characterized by the off-center ignition points, resulting in the
relatively low efficiency of Ni production.

\begin{figure}
\centering
\epsscale{1.0}
\plotone{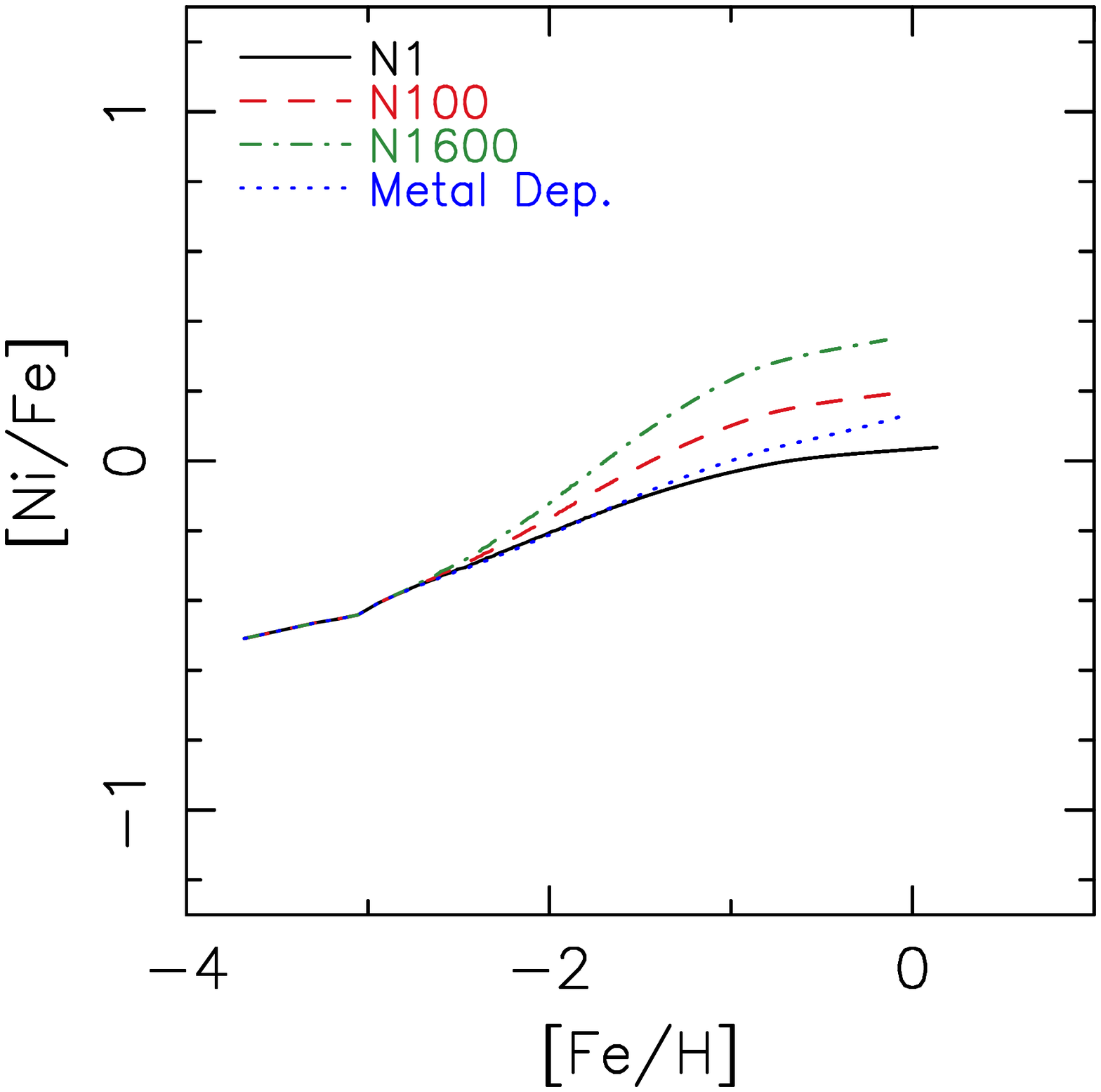}
\caption{Same as figure \ref{fig:OZ:SNIa:Ni:Dim}, but for different ignition
models of \cite{Seitenzahl+2013}.  Blue, green and red sequences are results
adopting N1, N100, and N1600, respectively. Black sequence is the result using
our metallicity dependent yields with models N100, N100\_Z0.5, N100\_Z0.1,
and N100\_Z0.01.}
\label{fig:OZ:SNIa:Ni:Seitenzahl}
\end{figure}

In figure \ref{fig:OZ:SNIa:Ni:Seitenzahl}, we show the [Ni/Fe]-[Fe/H] relations
of four different models based on \cite{Seitenzahl+2013}.  As is expected from
their yields tables, the Ni production rate increases with an increasing number
of ignition points.  When the metal-dependent model is used, the rise becomes
more moderate compared to those found in the model with N100. This tendency
would be preferable if we would like to reproduce the local [Ni/Fe]-[Fe/H]
relation \citep[e.g.,][]{Gratton+2003}.  This implies it is important to take
into account the metallicity dependence of SNe Ia yields.

\begin{figure}
\centering
\epsscale{1.0}
\plotone{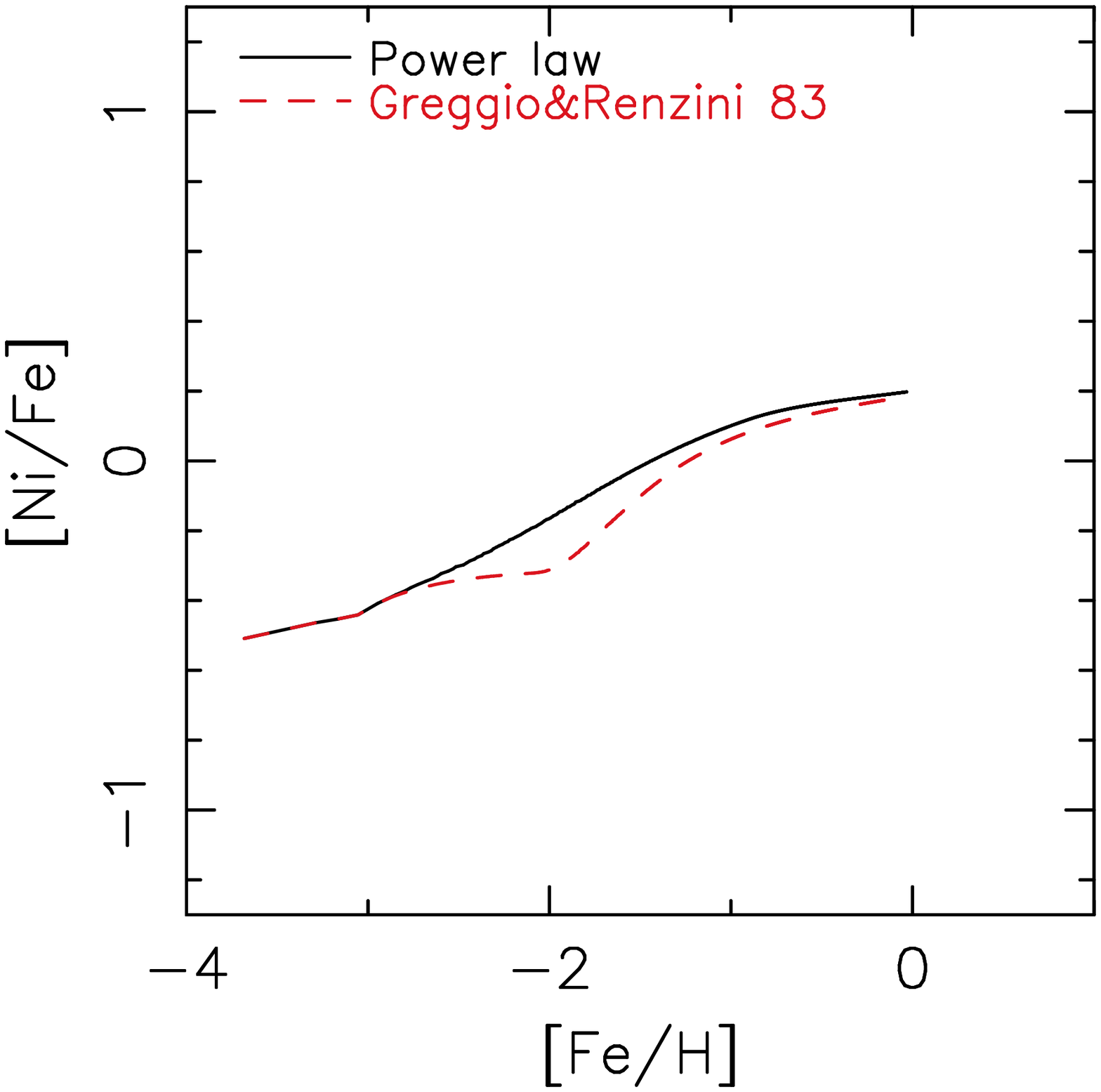}
\caption{Same as figure \ref{fig:OZ:SNIa:Ni:Dim}, but for different DTD models.
The yields of N100 in \cite{Seitenzahl+2013} are adopted for SNe Ia.  Red
and black sequences are results with the binary synthesis model and the
empirical power-law model, respectively. }
\label{fig:OZ:SNIa:Ni:DTD}
\end{figure}

The differences induced by different SNe Ia rates found in the [Ni/Fe]-[Fe/H]
relation are shown in figure \ref{fig:OZ:SNIa:Ni:DTD}. The most crucial
difference between the two models, the theoretical model based on the binary
synthesis and the observation based empirical model,
is the starting time of the enrichment by SNe Ia (see
figure \ref{fig:SNIa:Rate}).  The SNe Ia feedback which ignites relatively
earlier leads to an earlier increase of [Ni/Fe]. This effect is observed in this
figure.  Note that the time at which the effect of SNe Ia becomes prominent is
adjustable by changing the offset in DTD (Eq. \ref{eq:SNIa:DTD}).

\begin{figure}
\centering
\epsscale{1.0}
\plotone{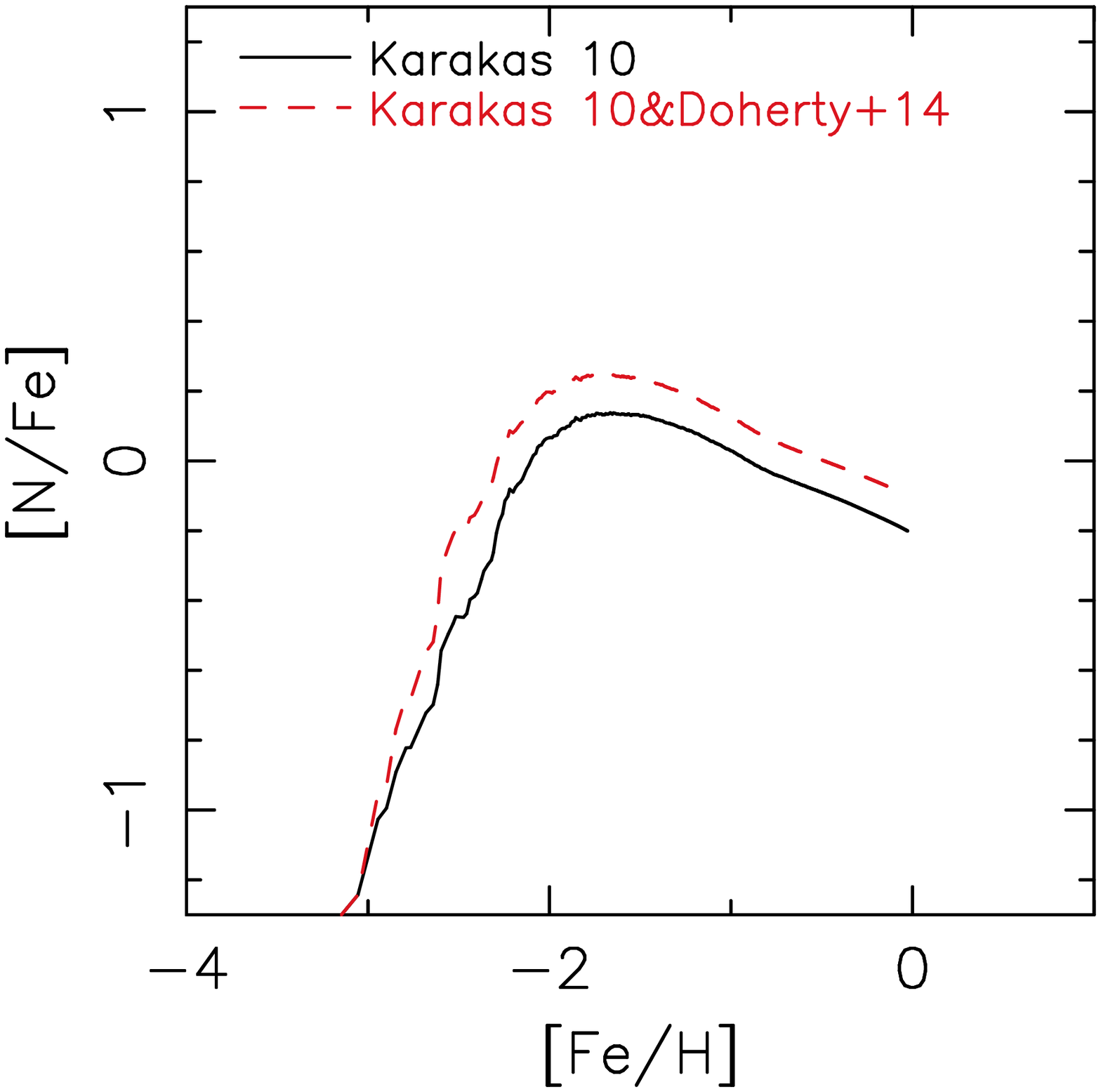}
\caption{[N/Fe] as a function of [Fe/H]. 
 }
\label{fig:OZ:AGB:N}
\end{figure}

% Super AGB model.
The introduction of the super-AGB yields is rather limited.  Since AGBs mainly
affect the light elements, such as C and N, the effects of the super AGBs are
found in these elements. Here, in figure \ref{fig:OZ:AGB:N}, we show the
[N/Fe]-[Fe/H] relation which shows the difference most clearly and we see
that the increase is only 0.1 dex in [N/Fe].

\begin{figure}
\centering
\epsscale{1.0}
\plotone{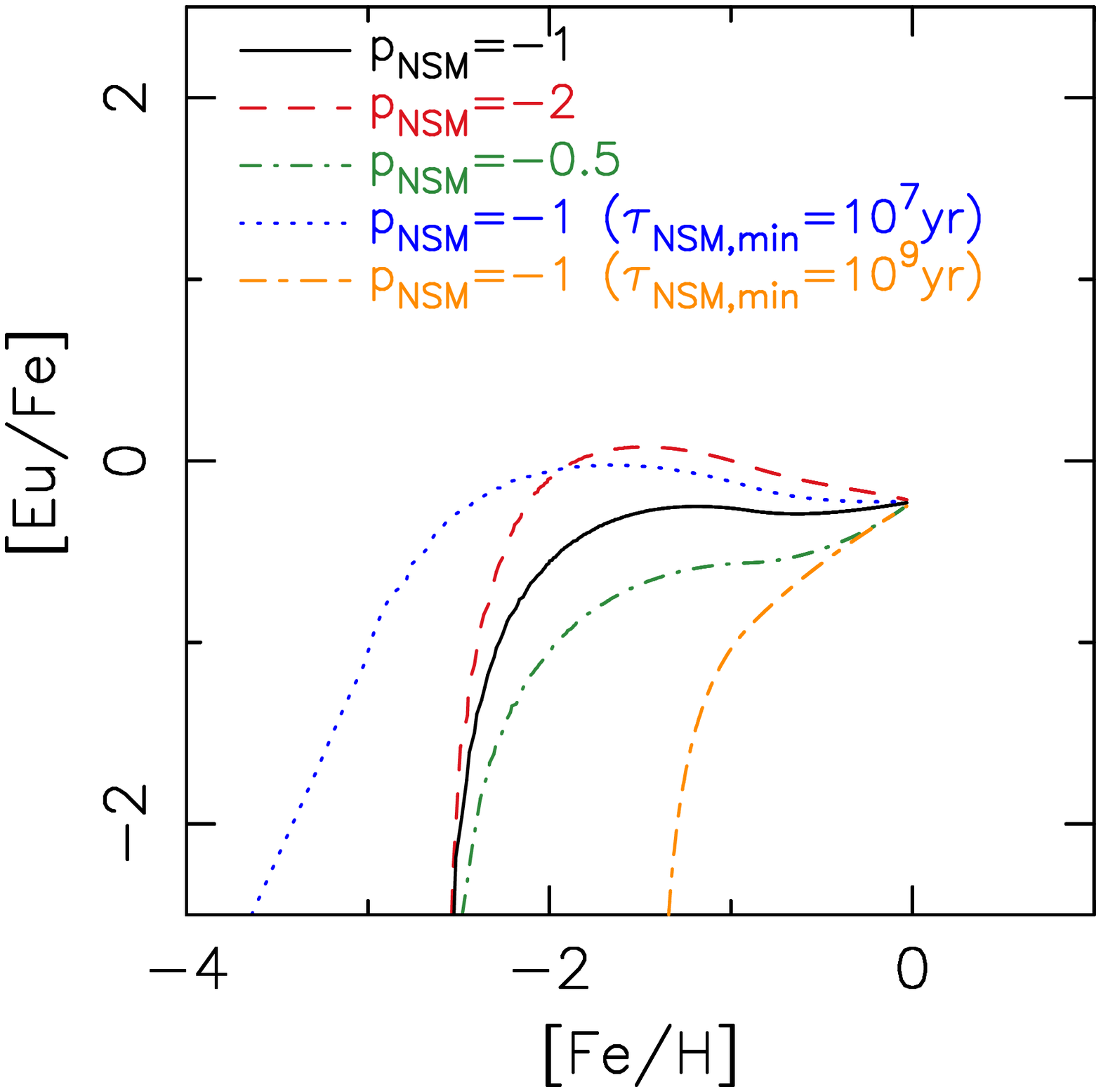}
\caption{[Eu/Fe] as a function of [Fe/H]. Three different power-law indexes,
$p_{\rm NSM}$, with $\tau_{\rm NSM,min} = 100~{\rm Myr}$ are depicted as well as
the models with the fiducial power law index $p_{\rm NSM} = -1$ and different $\tau_{\rm
NSM,min}$ ($10~{\rm Myr}$ and $1~{\rm Gyr}$).} 
\label{fig:OZ:NSMs:Eu}
\end{figure}

We close this section discussing the evolution of $r$-process element, Eu.  Figure
\ref{fig:OZ:NSMs:Eu} shows the [Eu/Fe]-[Fe/H] relations for three different
power-law indexes, $p_{\rm NSM} = -1$, $-2$, and $-0.5$.
When the index becomes shallower, the evolution track moves lower. This is
because a shallower index leads to a slower Eu release to the ISM and, as a
result, the enrichment of Fe due to SNe Ia progresses at the time of Eu
release.  Since our model is too simple, it might be difficult to make a
concrete conclusion.

There is a slight offset that [Eu/Fe] $\sim -0.3$ at [Fe/H] = 0 in our models,
when we refer to observations \cite[][references therein]{Suda+2011}.  The
offset at [Fe/H] = 0 comes from our model parameters of the Eu yield and the
fraction parameter of the event number of NSMs over that of core-collapse SNe.
These parameters are not observationally fixed yet and thus, it is adjustable.
As long as the normalization at [Fe/H] = 0, it can be done by multiplying a
factor so that it fits observations since here we adopt the time and metallicity
independent Eu yields and the contribution of Eu to dynamics is negligible.

When we change $\tau_{\rm NSM,min}$, the evolution track also changes.  For the
model with $\tau_{\rm NSM,min} = 10~{\rm Myr}$ the extremely low metal Eu
polluted stellar component appears and the evolution track follows that of
the fiducial model at the late epoch.  On the other hand, for the model with
$\tau_{\rm NSM,min} = 1~{\rm Gyr}$, the low metal Eu polluted component
disappears and the evolution track is far away from the observations.  It seems
that such a long duration model is unfavorable. However, it is again
difficult to mention the details of metal distribution because this model
assumed an instantaneous mixing model and did not take into account the mass
inflow and outflow.

The tendency obtained by changing $\tau_{\rm NSM,min}$ is comparable to that
reported by \cite{Matteucci+2014}; shorter delay-time models give a Eu enhanced
stellar component with lower metallicity. The clear difference between ours and
theirs is the raising point of [Fe/H].  We believe the difference comes from the
modeling (i.e., accretion, star formation, inflow, outflow, etc) and would not
be serious. Self-consistent modeling is important to understand the whole
distribution of Eu. For instance, \cite{Hirai+2015} can successfully reproduce
both the scatter of the Eu enhanced stars distribution in the low [Fe/H] regime
and the raising point of [Fe/H] in their three-dimensional $N$-body/SPH
simulations with a delay-time of $\sim 100~{\rm Myr}$.

\subsection{Chemical evolution in a NFW halo} \label{sec:Chemodyn}

\subsubsection{Initial setup} \label{sec:Chemodyn:InitialCondition}

For this test, we adopt a Milky Way size galaxy using a public code {\tt
dice}{\footnote {{\tt dice} is a public code which can generate particle
realizations of galaxy models and galaxy-galaxy merger configurations, as well
as these for mesh codes.  The distribution site is
{\tt https://bitbucket.org/vperret/dice/} .}} \citep{Perret+2014}. 
Here we use a system consisting of gas and dark matter. The virial velocity of
the system is set to $200~{\rm km~s^{-1}}$. Both gas and dark matter components
initially follow the NFW profile in which the concentration parameter is 13 and
spin parameter is 0.04.  We adopt a truncation radius of $150~{\rm kpc}$.  There
are no particles outside of the truncation radius.  The masses of gas and dark
matter particles are $8.5\times10^4~\Msun$ and $1.6\times10^6~\Msun$,
respectively.  The numbers of particles for gas and dark matter components are
$10^6$ and $3\times10^6$. The softening lengths of gas and DM particles are set
to $25~{\rm pc}$ and $50~{\rm pc}$, respectively.  We follow the system up to
$5~{\rm Gyr}$.

\subsubsection{Numerical techniques} \label{sec:Chemodyn:Technique}

% used physics schemes
We use {\tt ASURA} \citep{Saitoh+2008, Saitoh+2009} for simulations.  We take
into account the gravitational and hydrodynamical interactions, as well as the
radiative cooling, star formation, and stellar feedback.  The stellar feedback
is dealt with by CELib. We use the reference models described in section
\ref{sec:modeling}.  

The gravitational interactions are solved by using the tree method
\citep{BarnesHut1986}. The parallelization strategy of the tree method is the
same as \cite{Makino2004}. In order to deal with different gravitational
softening for different particle species, the symmetrized Plummer potential and
its multipole expansion are used \citep{SaitohMakino2012}. The opening angles
for the ordinary three dimensional spaces and the gravitational softening
lengths are 0.5 and 0.5, respectively.

The smoothed particle hydrodynamics (SPH) method \citep{Lucy1977,
GingoldMonaghan1977} is used to solve the evolution of the gas component.  
There are many variants.  Here we use the density independent formulation of the
SPH (DISPH) \citep{SaitohMakino2013} which adopts pressure instead of density
for the fundamental smoothed quantity for the formulation.  This formulation of
SPH drastically improves the treatment of the contact discontinuities. In order
to handle shocks, we use an artificial viscosity term with a functional form
which is the same as that proposed in \cite{Monaghan1997}. The number of neighbor
particles, $N_{\rm nb}$, for each SPH particle is kept in $N_{\rm nb} = 128\pm8$
and the Wendland C4 kernel \citep{Wendland1995} is used.

Time integration is carried out by the second-order scheme \citep[see section A1
in][]{SaitohMakino2016}.  The individual, block time step method is used
\citep{McMillan1986, Makino1991IndividualTimeStep}. In order to accelerate the
hydro simulation involving strong shocks and SNe, the FAST method
\citep{SaitohMakino2010} is adopted for the time integration of SPH particles
which allows SPH particles to have different time steps for the hydrodynamical
and gravitational interactions.  We also use the time-step limiter which keeps
the time-step difference among neighboring particles small enough to follow
strong shocks \citep{SaitohMakino2009}.

\begin{figure}
\centering
\epsscale{1.0}
\plotone{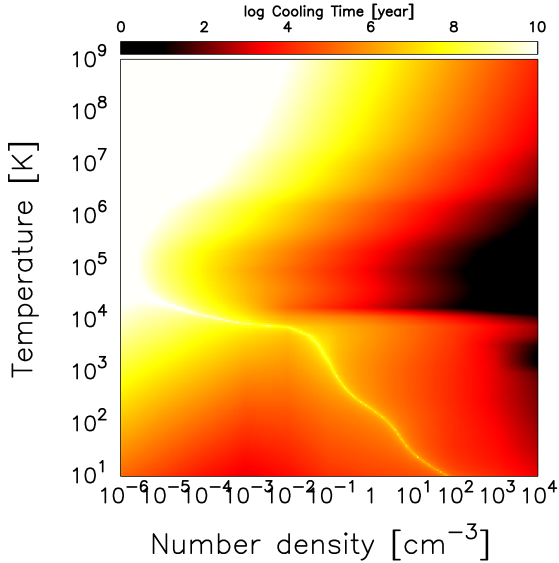}
\caption{Cooling/Heating time in the density-temperature plane.  The definition
is $E_{\rm th}(T)/\dot{E}_{\rm CH}(\rho,T,Z)$, where $E_{\rm th}(T)$ is the
thermal energy at a given temperature and $E_{\rm CH}(\rho,T,Z)$ is the absolute
value of the net energy change due to the radiative cooling and heating.  The
solar metallicity of \cite{Asplund+2009} is assumed. The redshift is set to
zero for this figure.  
}\label{fig:Chemodyn:CoolingHeatingTime}
\end{figure}

The radiative cooling is dealt with using a cooling function generated by {\tt
Cloudy} \citep{Ferland+1998, Ferland+2013}.  The UV background heating of
\cite{HaardtMadau2012} is taken into account. The self-shielding model of
\cite{Rahmati+2013}, which reduces the intrusion of the UV background flux to
the ISM, is also adopted. For radiative cooling, heating, and the
self-shielding, we use the models at the redshift zero.  Figure
\ref{fig:Chemodyn:CoolingHeatingTime} shows the cooling/heating time obtained
using our cooling/heating function. The red narrow region in $\rho > 10^{-3}~
{n_{\rm H}}~{\rm cm}^{-3}$ and $T\leq10^4~{\rm K}$ corresponds to the
equilibrium part of our cooling/heating function. Above this region, the
radiative cooling dominates over the heating, whereas below this, the heating
overcomes the cooling.

We set the high density ($> n_{\rm th}$) and low temperature ($< T_{\rm th}$)
regions as the star forming regions.  According to \cite{Saitoh+2008,
Saitoh+2009}, here we employ the following three conditions: (1) $\nabla \cdot v
< 0$, (1) $n_{\rm th} = 100~{\rm cc}$, and (1) $T_{\rm th} = 100~{\rm K}$. When
a gas particle satisfies all of the above conditions, the gas particle spawns a
collisionless star particle with a mass of 1/3 of the initial gas particle mass.
When the gas particle mass is less than 1/3 of the initial gas mass, the gas
particle is converted into a star particle.  The SSP approximation is applied to
the star particle.

Here, we adopt the Chabrier IMF with a mass range of $0.1~\Msun< M < 100~\Msun$.
Feedback energy from SNe II is calculated using adopted yields tables.  We
adopted the power law type DTD for SNe Ia. The power-law index is $-1$ and the
normalization of \cite{MaozMannucci2012} is used. We used the cluster mode for
the SNe Ia and regarded ten SNe Ia exploding at the same time. For the SNe Ia
yields, we always use the N100 model in \cite{Seitenzahl+2013}.  Two simulations
do not use the AGB feedback while others take it into account. The time interval
of this event, $\Delta t_{\rm AGB}$, is set to $10^8~{\rm yr}$.  NSMs are
considered. Here we use $p_{\rm NSM} = -1$ and $t_{\rm NSM,min} = 10^8~{\rm
yr}$.

% Feedback model.
As long as we use the SSP approximation and we put the released energy via
thermal energy, the feedback from SNe II is inefficient and this is a long
standing issue of galaxy formation \citep{DallaVecchiaSchaye2012}. Therefore, we
here implemented a stochastic model. This model is not the same as that proposed
by \cite{DallaVecchiaSchaye2012}, while the philosophy is the same.

According to the argument in section 2 of \cite{SaitohMakino2010}, if we use all
released energy from an SSP particle to the surrounding ISM, the typical
temperature of a heated region is impossible to exceed that of the thermal
instability region ($10^4~{\rm k} < T < 10^7~{\rm K}$) at the density of the
typical star forming regions (see figure \ref{fig:Chemodyn:CoolingHeatingTime}).
The averaged increase of the internal energy surrounding exploded SNe II,
$U_{\rm SN}$, is  
\begin{align}
U_{\rm SN} &= \frac{\epsilon_{\rm SN} m_{*} E_{\rm SN} }{N_{\rm NB} m_{\rm SPH}} \label{eq:SN:U1} \\
&\simeq 5\times 10^{48} \times \frac{m_{*}}{N_{\rm NB} m_{\rm SPH}}
~[{\rm ergs~M_{\odot}^{-1}}] \label{eq:SN:U2}\\
&\simeq \frac{2.5 \times 10^{15}}{N_{\rm NB}} [{\rm ergs~g^{-1}}],
\label{eq:SN:U3}
\end{align}
where $\epsilon_{\rm SN}$, $m_{*}$ and $m_{\rm SPH}$ are the SNe II fraction per
$1~\Msun$ SSP particle, masses of star and gas particles, respectively. $N_{\rm
NB}$ is the number of neighboring particles and the feedback energy is injected
to the particles. From Eq \eqref{eq:SN:U1} to Eq.  \eqref{eq:SN:U2}, we assume
that $\epsilon_{\rm SN} = 0.005$ as a typical value {\footnote{In
\cite{SaitohMakino2010}, $\epsilon_{\rm SN} = 0.0074$ is used.  The values
obtained here are slightly different from those in \cite{SaitohMakino2010}.}}
with the standard IMF and the mass range of SNe II (see \S \ref{sec:Misc}).
In addition, we assume that the typical mass of a single star particle is
identical to that of a single gas particle.  When we assume an ideal gas with
the specific heat ratio of $5/3$ and the mean molecular weight is $\sim 0.6$,
Eq. \eqref{eq:SN:U3} finally becomes
\begin{equation}
T_{\rm SN} \sim 3.78 \times 10^{5}~{(N_{\rm NB}/32)}~{\rm [K]}. 
\end{equation}
This clearly tells us that the thermal feedback is inefficient and this comes
from the limitation of the numerical resolution that $m_{\rm gas} \sim m_{*}$.
For mesh codes, this problem does not occur intrinsically. However, in the mesh
cases, there is no limitation for the gas temperature and thus, it is possible
to reach unreasonably high temperatures by accident.

Here, for the feedback of SNe II, we adopt a probabilistic injection model so
that the temperature of a heated region can reach a threshold temperature
$T_{\rm SN,th}$.  First of all, the total energy release from SSP particles is 
\begin{equation}
E_{\rm SN,tot} = \sum_i \epsilon_{\rm SN} E_{\rm SN} m_{*,i}, \label{eq:Et}
\end{equation}
where the index $i$ runs all star particles, and $m_{*,i}$ is the mass of the
$i$-th SSP particle. We should construct a stochastic model which recovers this
total energy even in a probabilistic manner. The temperature increase induced
by the energy released from a SSP particle is written as
\begin{equation}
T_i = \frac{2 \mu m_{\rm P}}{3 k_{\rm B}} 
\frac{\epsilon_{\rm SN} E_{\rm SN} m_{*,i}}{{M_{{\rm gas},i}}}, \label{eq:Ti}
\end{equation}
where $\mu$, $m_{\rm P}$, and $k_{\rm B}$ are the mean molecular weight, the
proton mass, and the Boltzmann constant, respectively, and 
$M_{{\rm gas},i}$ is the typical gas mass surrounding the SSP particles and it
is typically $N_{\rm NB}\times m_{\rm gas}$, where $m_{\rm gas}$ indicates the
mass of the gas particle.  Here we introduce a new quantity $\mathcal P_{{\rm
SN},i}$ and $\mathcal P_{{\rm SN},i} \equiv T_i/T_{\rm SN,th}$.  If we assume a
sufficiently high $T_{\rm {SN,th}}$, $\mathcal P_{{\rm SN},i} < 1$ and it can be
regraded as a probability, as we show below.  We use the acceptance-rejection
method to evaluate the probability $\mathcal P_{{\rm SN},i}$:
\begin{equation}
E_{{\rm SN},i} = 
\begin{cases}
E_{{\rm SN,th},i} & (A_{\mathcal R} \le \mathcal P_{{\rm SN},i}),\\
0 & (A_{\mathcal R} > \mathcal P_{{\rm SN},i}),
\end{cases}
\end{equation}
where $A_{\mathcal R}$ is a random real number in $[0,1)$ and 
$E_{\rm SN,th}$ is the energy where the averaged temperature reaches $T_{\rm
SN,th}$ and its functional form is 
\begin{align}
E_{{\rm SN,th},i} &=  U_{{\rm SN,th},i} M_{{\rm gas},i}, \\
&= \frac{3 k_{\rm B} T_{\rm SN,th}}{2\mu m_{\rm p}} M_{{\rm gas},i}.
\end{align}
The total injection energy using this model is 
\begin{align}
E_{\rm SN,tot,new} &= \sum_i E_{{\rm SN,th},i} \mathcal P_{{\rm SN},i} \\
&= \sum_i \frac{3 k_{\rm B} T_{\rm SN,th}}{2 \mu m_{\rm P}} 
    M_{{\rm gas},i} T_i/T_{\rm SN,th} \\
&= \sum_i \epsilon_{\rm SN} E_{\rm SN} m_{*,i}
\end{align}
This is the same as Eq. \eqref{eq:Et}.  In the following, we assume $T_{\rm
SN,th} = 5\times 10^7~{\rm K}$. Even though a failed case, $E_{{\rm SN},i}$, the
metal redistribution takes place in our model (see below). Only released energy
is probabilistically redistributed.

When a star reaches the feedback time, the metallicity and energy released from
the star are distributed to the surrounding ISM. Here we follow the
implementation of \cite{Mosconi+2001}. We distribute the metals to the
neighboring particles of the star particles of $N_{\rm nb} = 128$ with the
weight of the SPH kernel.

Metal diffusion is taken into account. Our implementation is based on that of
\cite{Shen+2010}, in which the sub-grid turbulence model is used.  The
diffusion equation for $k$-th element we use is 
\begin{equation}
\frac{dZ_{k,i}}{dt} =  - \sum_j \frac{m_i}{(\rho_i+\rho_j)/2} \frac{4 D_i D_j}{D_i+D_j} 
\frac{(Z_{k,i}-Z_{k,j})}{|r_{ij}|^2} \boldsymbol {r}_{ij} \cdot \nabla W_{ij}
\label{eq:metaldiffusion}
\end{equation}
where 
\begin{equation}
\hat{S}_{ab,i} = \frac{1}{q_i} \sum_j U_j (v_{b,i}-v_{a,j}) \nabla_{a} W_{ij},
\end{equation}
\begin{equation}
S_{ab,i} = \frac{1}{2} (\hat{S}_{ab,i}+\hat{S}_{ba,i}) - \delta_{ab} \frac{1}{3}
\rm{Trace}~\hat{S}_{ab,i},
\end{equation}
\begin{equation}
D_i = C_{\rm diff} |S_{ab,i}| h_i^2.
\end{equation}
Here, $i$ and $j$ are particle indexes, $a$ and $b$ denote $x$, $y$, and $z$
directions, $h$ is the kernel size, and $\delta_{ab}$ is the Kronecker's delta.
$W_{ij}$ is the kernel function.  The diffusion coefficient, $C_{\rm diff}$,
depends on the structure of flow (typically $\sim 0.1$).  It is set to $0.1$ in
this study.  The mass exchange of $k$-th element using this equation has an
antisymmetric form for particles $i$ and $j$;
\begin{equation}
\frac{dM_{k,i}}{dt} = m_i \frac{dZ_{k,i}}{dt}
= -m_j \frac{dZ_{k,i}}{dt} = -\frac{dM_{k,i}}{dt},
\end{equation}
and thus this formulation can conserve the gas mass.

We carry out eleven runs with the different combinations of yields and models
(with and without the Pop III IMF and metal mixing).  The models are summarized
in table \ref{tab:Chemodyn:Runs}.

\begin{table*}[htb]
\begin{center}
\caption{Models for chemodynamical simulations.}\label{tab:Chemodyn:Runs}
\scriptsize
\begin{tabular}{lcccccc}
\hline
\hline
&  SNII yields & SNIa yields & AGB yields & NSM yields  & Pop III & Mixing\\
\hline
Run A & \cite{Nomoto+2013} & \cite{Seitenzahl+2013} & $\times$ & $\times$  & $\times$ & $\times$\\
      & ($f_{\rm HN}=0.0$) & (N100) & & & &\\
Run B & \cite{Portinari+1998}$^a$ & \cite{Seitenzahl+2013}& $\times$ & $\times$  & $\times$ & $\times$\\
      &  & (N100) & & & &\\
Run C & \cite{Nomoto+2013} & \cite{Seitenzahl+2013}& \cite{Karakas2010}+\cite{Doherty+2014}  & $\times$  & $\times$ & $\times$\\
      & ($f_{\rm HN}=0.0$) & (N100) & & & &\\
Run D & \cite{Portinari+1998}$^a$ & \cite{Seitenzahl+2013}& \cite{Karakas2010} & $\times$  & $\times$ & $\times$\\
      &   & (N100) & +\cite{Doherty+2014} & & &\\
Run E & \cite{Nomoto+2013} & \cite{Seitenzahl+2013}& \cite{Karakas2010}+\cite{Doherty+2014}  & $\times$  & $\checkmark$ & $\times$\\
      & ($f_{\rm HN}=0$) & (N100) & +\cite{CampbellLattanzio2008} & & &\\
      & & & +\cite{Gil-Pons+2013} & & &\\
Run F & \cite{Nomoto+2013} & \cite{Seitenzahl+2013}& \cite{Karakas2010}+\cite{Doherty+2014}  & $\times$  & $\times$ & $\times$\\
      & ($f_{\rm HN}=0.05$) & (N100) & +\cite{CampbellLattanzio2008}  & & &\\
      & & & +\cite{Gil-Pons+2013} & & &\\
Run G & \cite{Nomoto+2013} & \cite{Seitenzahl+2013}& \cite{Karakas2010}+\cite{Doherty+2014}  & $\times$  & $\times$ & $\times$\\
      & ($f_{\rm HN}=0.5$) & (N100) & +\cite{CampbellLattanzio2008}  & & &\\
      & & & +\cite{Gil-Pons+2013} & & &\\
Run H & \cite{Nomoto+2013} & \cite{Seitenzahl+2013}& \cite{Karakas2010}+\cite{Doherty+2014}  & \cite{Wanajo+2014}  & $\checkmark$ & $\times$\\
      & ($f_{\rm HN}=0.05$) & (N100) & +\cite{CampbellLattanzio2008}   & & &\\
      & & & +\cite{Gil-Pons+2013} & & &\\
Run I & \cite{Nomoto+2013} & \cite{Seitenzahl+2013}& \cite{Karakas2010}+\cite{Doherty+2014}  & \cite{Wanajo+2014}  & $\checkmark$ & $\times$\\
      & ($f_{\rm HN}=0.5$) & (N100) & +\cite{CampbellLattanzio2008}  & & &\\
      & & & +\cite{Gil-Pons+2013} & & &\\
Run J & \cite{Nomoto+2013} & \cite{Seitenzahl+2013}& \cite{Karakas2010}+\cite{Doherty+2014}  & \cite{Wanajo+2014}  & $\checkmark$ & $\checkmark$\\
      & ($f_{\rm HN}=0.05$) & (N100) & +\cite{CampbellLattanzio2008}  & & &\\
      & & & +\cite{Gil-Pons+2013} & & &\\
Run K & \cite{Nomoto+2013} & \cite{Seitenzahl+2013}& \cite{Karakas2010}+\cite{Doherty+2014}  & \cite{Wanajo+2014}  & $\checkmark$ & $\checkmark$\\
      & ($f_{\rm HN}=0.5$) & (N100) & +\cite{CampbellLattanzio2008}  & & &\\
      & & & +\cite{Gil-Pons+2013} & & &\\
\hline
\end{tabular}\\
\end{center} 
$^a$ Note that the modifications for the
yields of C, Mg, and Fe are applied.  See also section \ref{sec:SNII}.
\end{table*}

\subsubsection{General features of galaxies}

\begin{figure*}
\centering
\epsscale{1.0}
\plotone{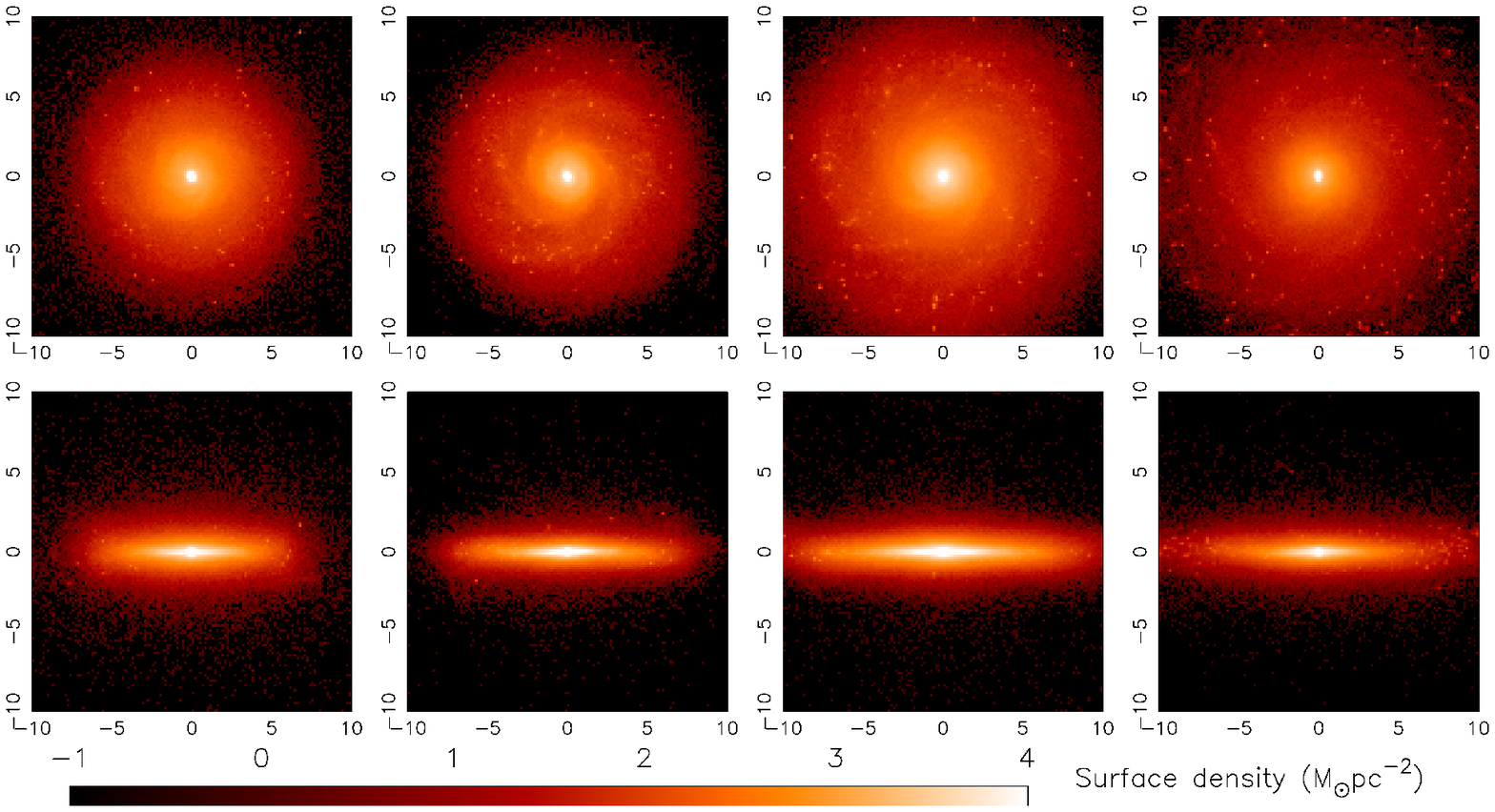}
\caption{Face-on and edge-on maps of the stellar disks at $5~{\rm Gyr}$.  From
left to right columns, we show the maps of runs A, B, C, and D. Each panel shows
the $20~{\rm kpc} \times 20~{\rm kpc}$ region.
} 
\label{fig:Chemodyn:Disks}
\end{figure*}

Figure \ref{fig:Chemodyn:Disks} displays the face-on and edge-on maps of the
surface stellar density of runs A, B, C and D at $t=5~{\rm Gyr}$.  When we
compare the results with the yields tables of \cite{Nomoto+2013} and
\cite{Portinari+1998} and without AGBs, i.e., runs A and B, we see that the
stellar disk of run A is more compact and thicker than that of run B.  The reason
for this difference is that there is a relatively smaller amount of energy
release from SNe II in run A compared to run B, as is expected from table
\ref{tab:SNIIEnergy}.  With this less efficient feedback, a dissipation-less
nature is emphasized further. 

To clarify the contribution of AGBs, we turn on the AGB feedback, which is
mainly the mass recycling processes involving newly synthesized metals (runs C
and D). We see from figure \ref{fig:Chemodyn:Disks} that the stellar disk
becomes larger in the radial direction.  The return mass via AGBs enhances the
late epoch star formation, resulting in the extended disk. Reflecting the fact
that run C forms more stars, the contribution of AGBs is also more prominent,
resulting in a larger disk compared to that in run D.

\begin{figure*}
\centering
\epsscale{0.75}
\plotone{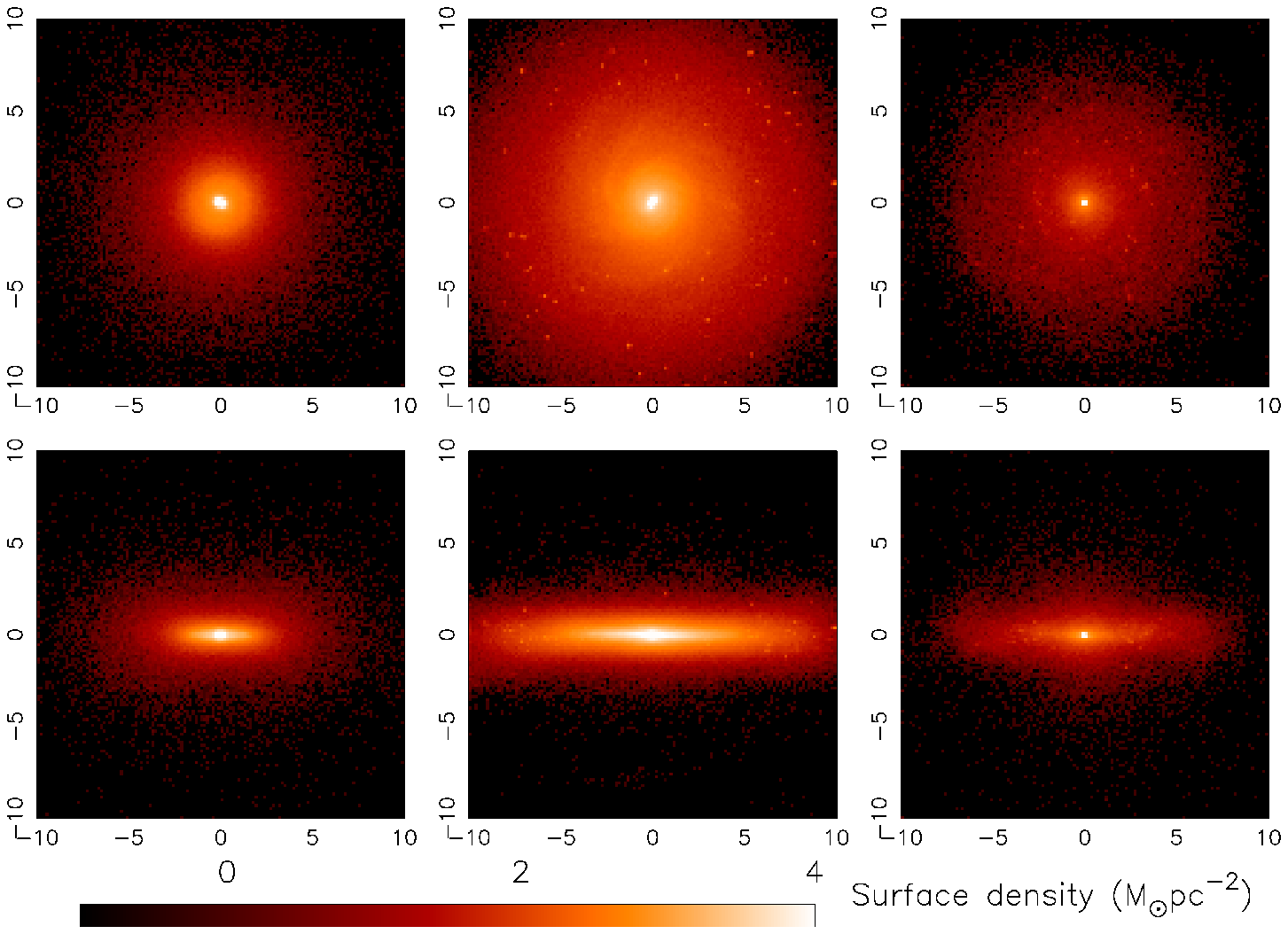}
\caption{Same as figure \ref{fig:Chemodyn:Disks}, but for runs E, F, G.
} 
\label{fig:Chemodyn:Disks:EFG}
\end{figure*}

The impacts of the Pop III IMF (run E) and HNe (runs F and G) on galactic
structures are found in figure \ref{fig:Chemodyn:Disks:EFG}. With the Pop III IMF
and \cite{Nomoto+2013}'s yields table, the released energy due to SNe II
increases $\sim 9$ times larger than that without the Pop III IMF (see table
\ref{tab:SNIIEnergy}).  Thus there are striking effects on the galactic
structure.  When we use the Pop III IMF, the initial burst of star formation is
suppressed significantly (we argue it below).  In this case, the stellar disk
becomes smaller because the initial burst induced by the Pop III feedback
removes the gas component from the galaxy.

As is expected, the model with HNe ($f_{\rm HN} = 0.05$) is almost no difference
from the run without HNe (run C).  The model with HNe ($f_{\rm HN} = 0.5$), on
the other hand, has large impacts on the galactic structure since the amount of
the released energy with $f_{\rm HN} = 0.5$ is about four times larger than that
with $f_{\rm HN} = 0$ (see table \ref{tab:SNIIEnergy}). This larger feedback
energy results in the strong suppression of star formation and the removal of
the log-angular momentum gas \citep[e.g.,][]{Brook+2012AM}.
The disk becomes much smaller compared to runs C and F.

\begin{figure*}
\centering
\epsscale{1.0}
\plotone{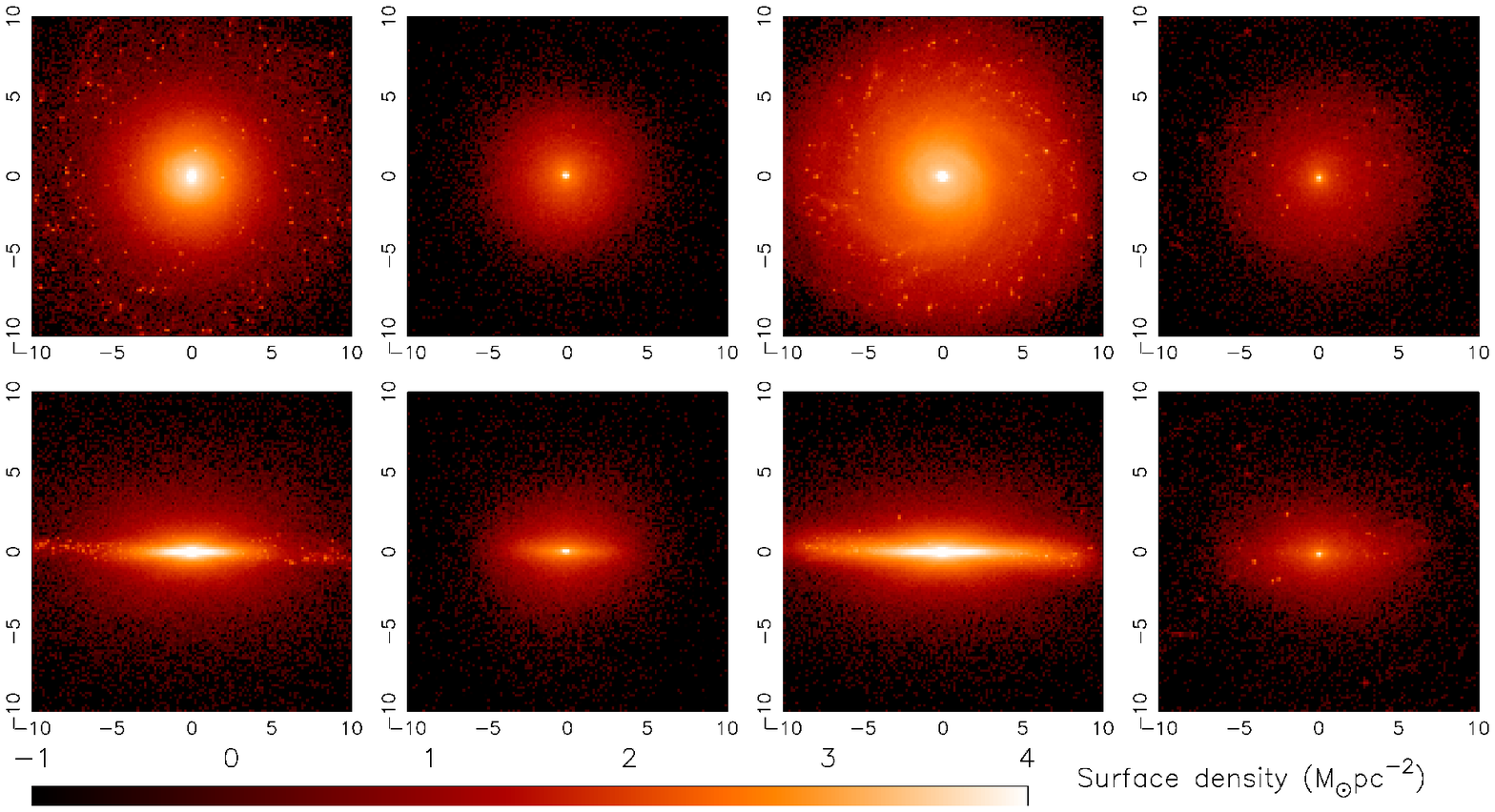}
\caption{Same as figure \ref{fig:Chemodyn:Disks}, but for runs H, I, J, and K.
} 
\label{fig:Chemodyn:Disks:HIJK}
\end{figure*}

Figure \ref{fig:Chemodyn:Disks:HIJK} shows the runs involving both HNe ($f_{\rm
HN} = 0.05$ and $0.5$) and the Pop III IMF, without and with metal mixing (runs
H, I, J, and K).   In the model with the metal mixing (run J), metals spread to
the surrounding ISM and the radiative cooling becomes more efficient compared to
that without mixing (run H). Thus, the star formation is significantly enhanced
and the disk size becomes larger when the metal mixing is involved.  In the
models with $f_{\rm HN} = 0.5$ (runs I and K), the difference is hard to see
because of their strong feedback.  Note that these models are not models of a
realistic galaxy formation through hierarchical mergers and hence, the
contribution of metal mixing might be more complicated in the realistic model.
We will study this in the future.

\subsubsection{Star formation histories}

\begin{figure*}
\centering
\epsscale{0.8}
\plotone{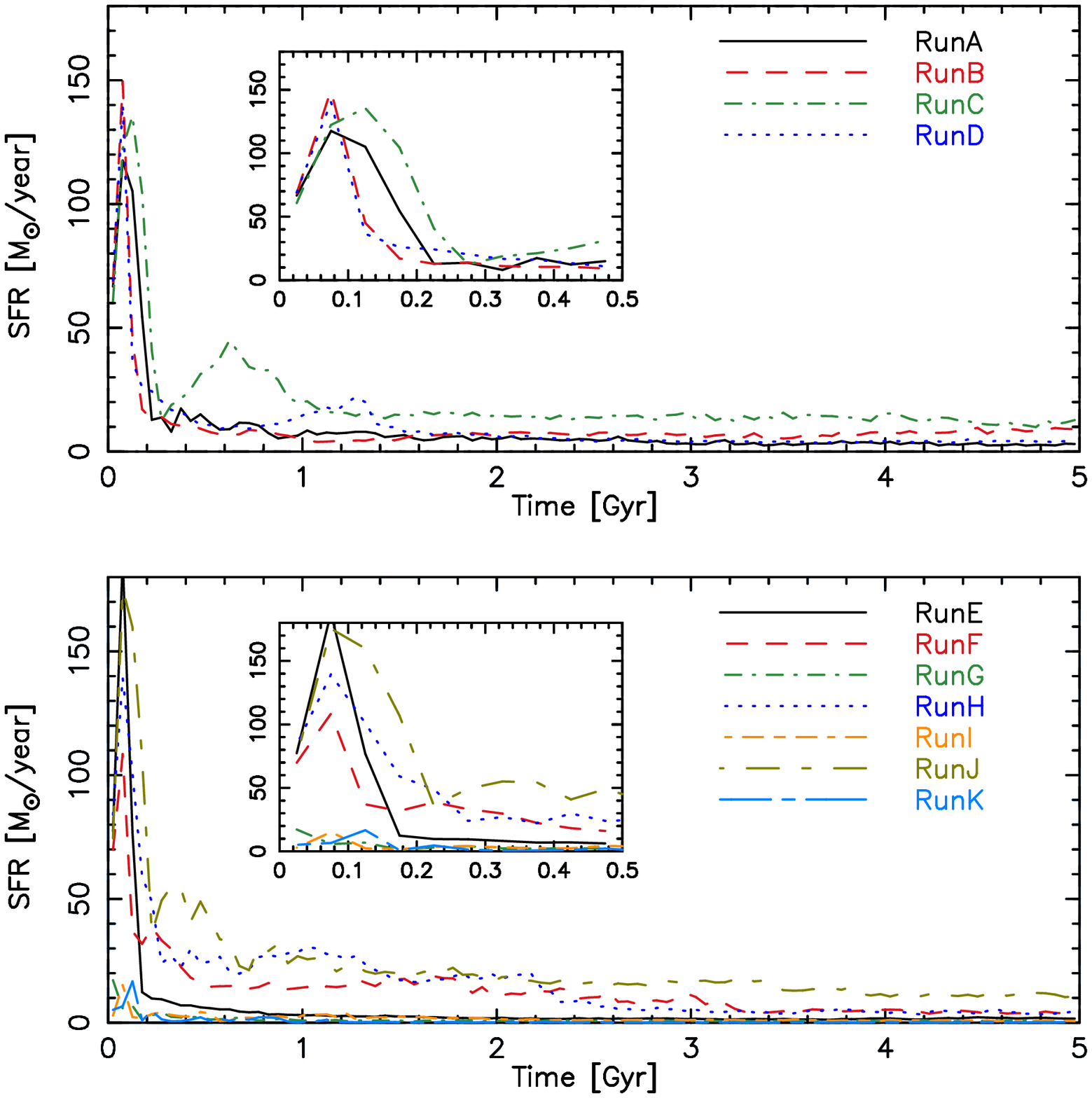}
\caption{Star formation rate as a function of time.  The close up of SFRs until
$t = 0.5~{\rm Gyr}$ is shown in the inset.
} 
\label{fig:Chemodyn:SFR}
\end{figure*}

Star formation rates (SFRs) as a function of time for all runs are shown in
figure \ref{fig:Chemodyn:SFR}. All of the models shown the upper panel have
strong initial bursts of star formation and then they rapidly quench. The
duration of the initial starburst depends on the adopted yields tables and model
parameters.  Models whose released energy are large have shorter duration times:
the duration time of the initial starburst of run B (D) is shorter than that of
run A (C).  We find the contribution of AGBs is easily seen in run C while not
in run D. The reason why the run D does not show the enhancement of star
formation at the late stage would be non-linear effects.

The bottom panel of figure \ref{fig:Chemodyn:SFR} summarizes star formation
histories with Pop III/HNe/metal mixing.  We see that the initial burst of the
star formation in the model with Pop III (run E) is strongly enhanced: this
enhancement of star formation is due to a large amount of return mass and
metals. When the released energy reaches the typical potential energy of the central region, 
the gas component leaks from the galaxy and thus the star formation quenches.
This effect is also found in the models involving both Pop III and
HNe (for instance, runs H and J).  When we turn on the HN mode (runs F and G),
the peaks of SFRs decrease. In the case with $f_{\rm HN} = 0.5$, the star
formation is strongly suppressed and the peak of the SFR becomes less than $20~{\rm
\Msun~yr^{-1}}$, because of an extremely large release energy.

The metal mixing model enhances star formation in the whole time. The peak SFR
increases from $140~{\rm \Msun~yr^{-1}}$ (run H, without mixing) to $170~{\rm
Msun~yr^{-1}}$ (run J, with mixing). The star formation in the late stage is kept
a slightly higher SFR of $\sim 10~{\rm \Msun~yr^{-1}}$ if the metal mixing model
is used. For the models with $f_{\rm HN} = 0.5$, the contribution of metal
mixing is hard to see.

\subsubsection{Distributions of Elements}

\begin{figure*}
\centering
\epsscale{1.0}
\plotone{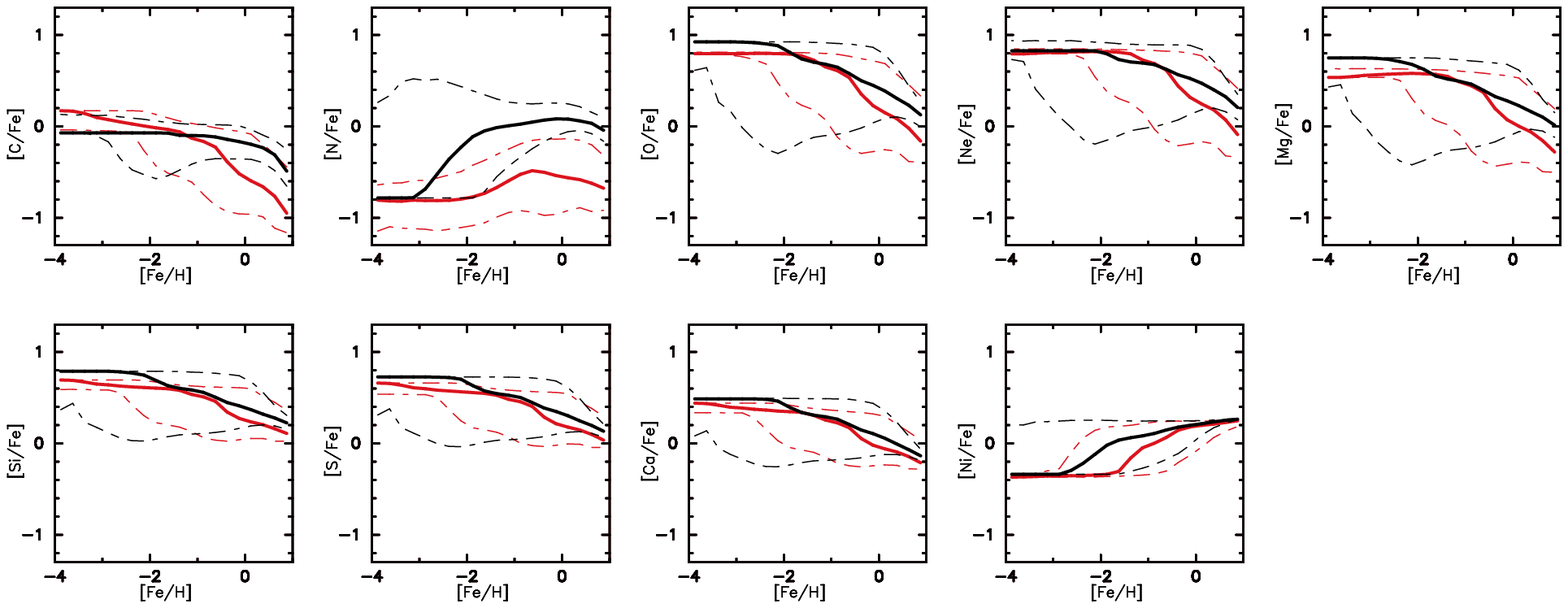}
\caption{[X/Fe]-[Fe/H] relations of nine elements with \cite{Nomoto+2013}'s
yields table (for runs A and C). The data at $5~{\rm Gyr}$ is used. Thick curves
indicate the median value whereas thin dashed curves show the 10\% and 90\%
values of each [Fe/H] bin. Red and black curves are for runs A and C,
respectively.  
} 
\label{fig:Chemodyn:XFeFeH:N13}
\end{figure*}

\begin{figure*}
\centering
\epsscale{1.0}
\plotone{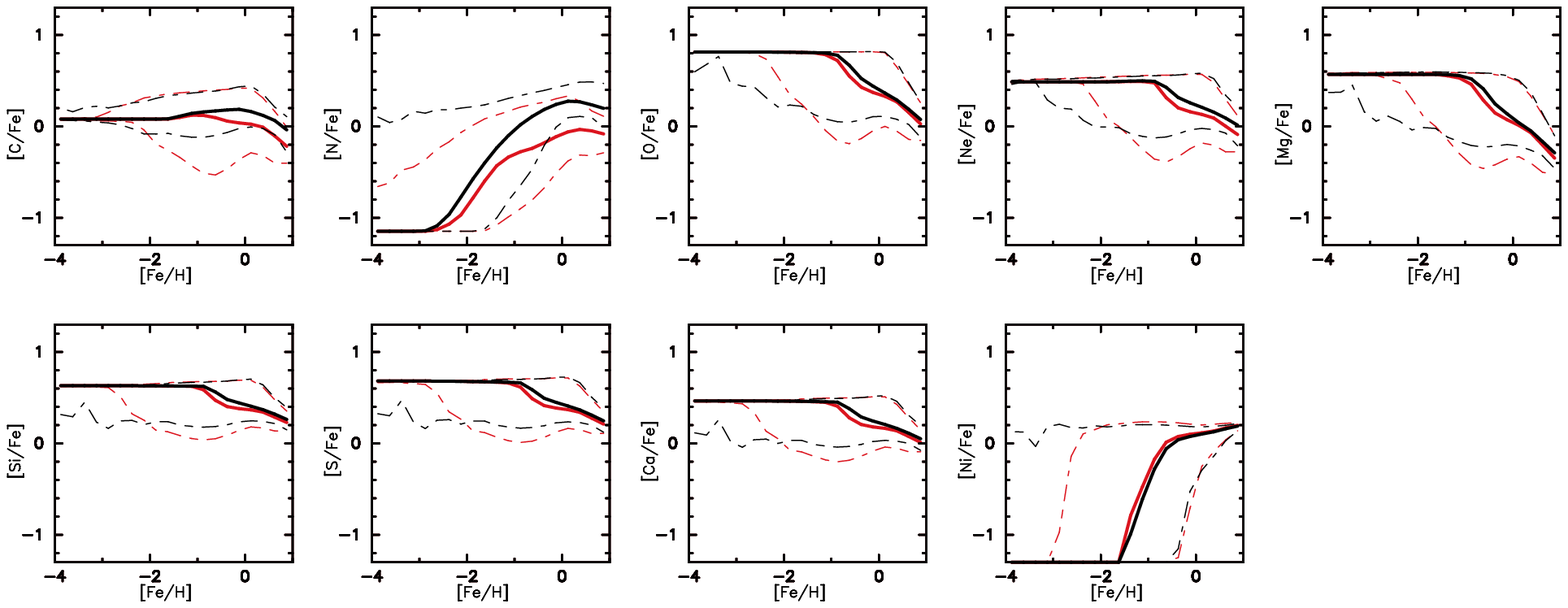}
\caption{Same as figure \ref{fig:Chemodyn:XFeFeH:N13}, but with
\cite{Portinari+1998}'s yields table.  Red and black curves are for runs B and
D, respectively.
} 
\label{fig:Chemodyn:XFeFeH:P98}
\end{figure*}

Figures \ref{fig:Chemodyn:XFeFeH:N13} and \ref{fig:Chemodyn:XFeFeH:P98} show
[X/Fe]-[Fe/H] relations of nine elements for runs A, B, C, and D.  We draw the
median and 10\% and 90\% values as a function of [Fe/H] in these figures.  The
evolution of [X/Fe] in these figures is basically similar to those obtained by
our one-zone models (see figures \ref{fig:OZ:AFeFeH:N} and
\ref{fig:OZ:AFeFeH:P}), although the metal redistribution scale is completely
different.  Initially, there are plateaus consisting of the SNe II yields. Then
the values of [X/Fe] become mixtures of SNe II, SNe Ia and AGBs yields.  
In the cases with AGBs in 3-dimensional simulations, the contribution of AGBs
are more prominent compared to the one-zone models and they are not only limited
in light elements. This effect is more prominent in the run with the
\cite{Nomoto+2013}'s yields table. This is because of the localization effect.  The
breaking points are shifted to +0.5 dex from those of one-zone models due to the
rapid evolution of SFR (See figure \ref{fig:Chemodyn:SFR}). 
When the AGBs are taken into account, not only the median but also the scatter
has changed. In particular in the low metal stars, the scatter increases because
of the early AGBs (recall figure \ref{fig:AGBs:CumulativeReturnMass}).

\begin{figure*}
\centering
\epsscale{1.0}
\plotone{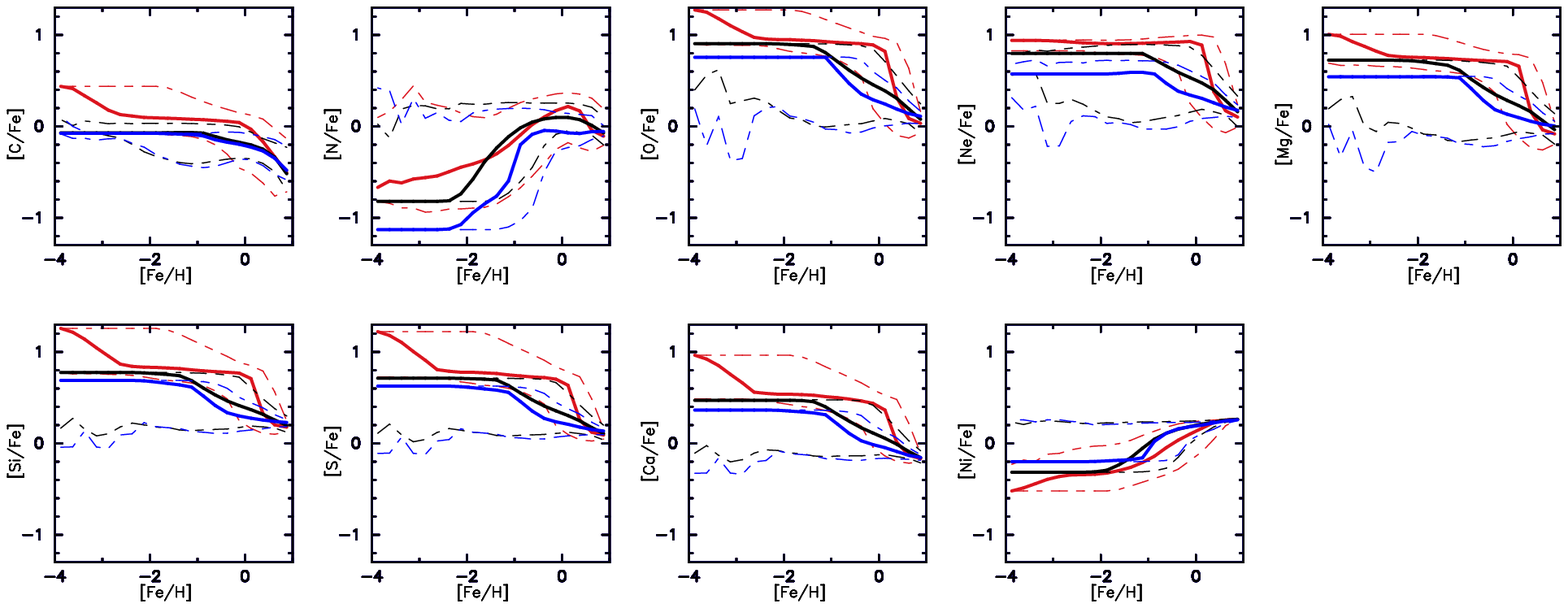} \caption{Same as figure
\ref{fig:Chemodyn:XFeFeH:N13}, but with the Pop III IMF (run E), HNe ($f_{\rm
HN} = 0.05$: run F), and HNe ($f_{\rm HN} = 0.5$: run G).  For the Pop III IMF
case, the corresponding yields table of \cite{Nomoto+2013} is used.  Red, black,
and blue curves are for runs E, F, and G, respectively.
} 
\label{fig:Chemodyn:XFeFeH:PopIIIHNe}
\end{figure*}

Figure \ref{fig:Chemodyn:XFeFeH:PopIIIHNe} displays the [X/Fe]-[Fe/H] relations
with the Pop III stars (run E) and with HNe (runs F and G). In these runs, the
\cite{Nomoto+2013}'s yields table is used.  The impact of the Pop III stars 
on [X/Fe]-[Fe/H] relations are significant.  The effect of the Pop III stars is
observed up to [Fe/H] = $-2.5$ (Note that $Z_{\rm popIII} = 10^{-5}$).  With
the Pop III stars, the amounts of [$\alpha$/Fe] increase (see figure
\ref{fig:SNII:Yields:N13}) and thus, [$\alpha$/Fe]s at [Fe/H] $<-2.5$ shift
upward.  This behavior is comparable to that found in the one-zone model in
figure \ref{fig:OZ:AFeFeH:N}.  However, these rapid changes in [X/Fe]-[Fe/H]
relations are not observationally confirmed.
We will see in the next figures, figures \ref{fig:Chemodyn:XFeFeH:NSMs:005} and
\ref{fig:Chemodyn:XFeFeH:NSMs}, that these gaps diminish/disappear when we adopt
both the Pop III IMF and HNe. 

The contributions of HNe to [X/Fe]-[Fe/H] relations can be seen as the timescale
of the Fe pollution.  Comparing [X/Fe]-[Fe/H] relations of runs F and G to those
of run C, we find that the breaking points of [X/Fe]-[Fe/H] relations change
from $\sim -2$ (run C) to $\sim -1$ (runs F and G). This is because the larger
Fe yield of HNe (recall figure \ref{fig:SNII:Yields:N13HN}).  In the case with
$f_{\rm HN} = 0.5$, the values of [X/Fe] at SNe II plateaus decrease $\sim 0.2$
dex in [O/Fe], [Ne/Fe], [Mg/Fe], [Si/Fe], [S/Fe] and [Ca/Fe], while that
increases $\sim 0.2$ dex in [Ni/Fe].  We can see similar tendencies in the case
with $f_{\rm HN} = 0.05$, although the changes are much smaller than the case
with $f_{\rm HN} = 0.5$.

From figure \ref{fig:Chemodyn:XFeFeH:NSMs:005}, we can see the impact of metal
mixing on the [X/Fe]-[Fe/H] relations. As is expected, the metal mixing greatly
reduces scatters, which is identical to the result obtained in
\citet{Shen+2015}. The transition points from the Pop III SNe to the normal SNe
II and the breaking points of SNe II/Ia are slightly moved toward high [Fe/H]
regions when the mixing is adopted. This is because that the self-enrichment is
more efficient in the run with the mixing model. With a small amount of HNe, the
values of [X/Fe] of the most metal-rich stars are almost identical to those
without HNe (runs A and C).

Figure \ref{fig:Chemodyn:XFeFeH:NSMs} shows the [X/Fe]-{Fe/H} relations with
$f_{\rm HN} = 0.5$ and with/without the metal mixing.  Although the star
formation histories are completely different between cases with $f_{\rm HN} =
0.5$ and $f_{\rm HN} = 0.05$, the positions of breaking points with $f_{\rm HN}
= 0.5$ are comparable to those found in the case with $f_{\rm HN} = 0.05$.  The
scatters in the run I are much wider than those in the run H. This would be that
stars in the run I distribute farther from the galactic center and AGBs affect
more in the outer fresh gas, due to the more energetic feedback.  When the metal
mixing turns on, the scatters become narrow (run K).  The black curves (run K)
have sudden rises at $Z>0.5$.  Their values of [X/Fe]s are identical to those
expected by yields of SNe II.  Hence, the contribution of SNe II dominates in
stars at the metallicity range.  This might be an accidental case.

The [Eu/Fe]-[Fe/H] relations are shown in figures
\ref{fig:Chemodyn:XFeFeH:NSMs:005} and \ref{fig:Chemodyn:XFeFeH:NSMs}.  The
median values of [Eu/Fe] rise at [Fe/H ]$\sim -1$ and they saturate [Eu/Fe]
$\sim -0.2$ at [Fe/H] $\sim +1$.  These evolution tracks are expected by
one-zone simulations.  While we adopted $p_{\rm NSM} = -1$ and $\tau_{\rm
NSM,min} = 10^8~{\rm yr}$, the evolution tracks on the plane in figures
\ref{fig:Chemodyn:XFeFeH:NSMs:005} and \ref{fig:Chemodyn:XFeFeH:NSMs} are
similar to the case with $p_{\rm NSM} = -1$ and $\tau_{\rm NSM,min} = 10^9~{\rm
yr}$ in figure \ref{fig:OZ:NSMs:Eu}. This is because star-formation time scales
in our three-dimensional simulations are shorter than those used in one-zone
simulations. As a result, stars with low-Z and high [Eu/Fe] are hard to see in
our model, whereas stars with [Eu/Fe] $> 0$ and [Fe/H] $< -3$ are found in
observations \citep[see][]{Suda+2011}.

The gap between our results and observations originate from the models we used
in this paper. Our models are too simple to express detailed distributions of
chemical composition generated by rare events.  We expect that the scatter will
revert in a cosmological simulation because of its hierarchical nature.  We will
investigate the evolution of r-process elements in the cosmological context
elsewhere in the near future.

\begin{figure*}
\centering
\epsscale{1.0}
\plotone{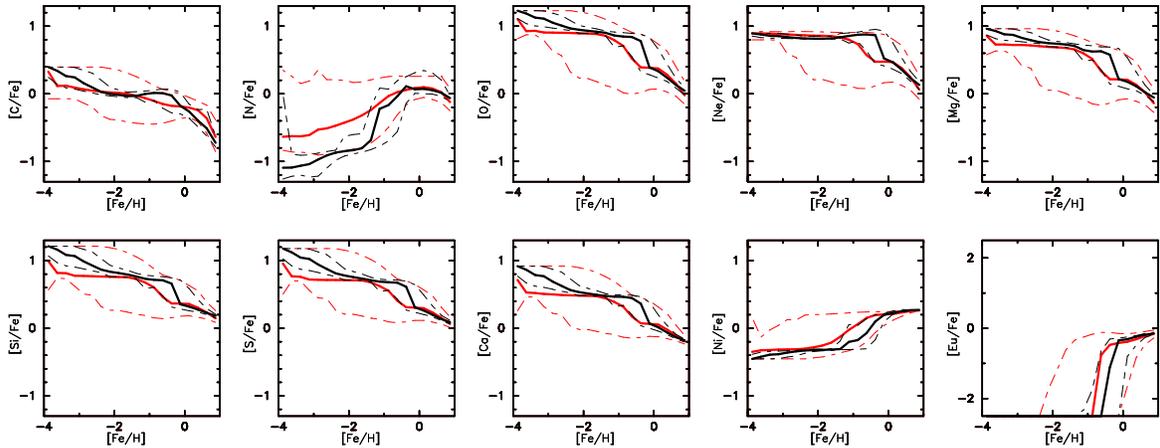}
\caption{Same as figure \ref{fig:Chemodyn:XFeFeH:N13}, but with HNe ($f_{\rm HN}
= 0.05$), the Pop III IMF and NSMs.  Red and black curves are for runs H and J,
respectively.  Run J adopts the metal diffusion model of Eq
\eqref{eq:metaldiffusion}.
} 
\label{fig:Chemodyn:XFeFeH:NSMs:005}
\end{figure*}

\begin{figure*}
\centering
\epsscale{1.0}
\plotone{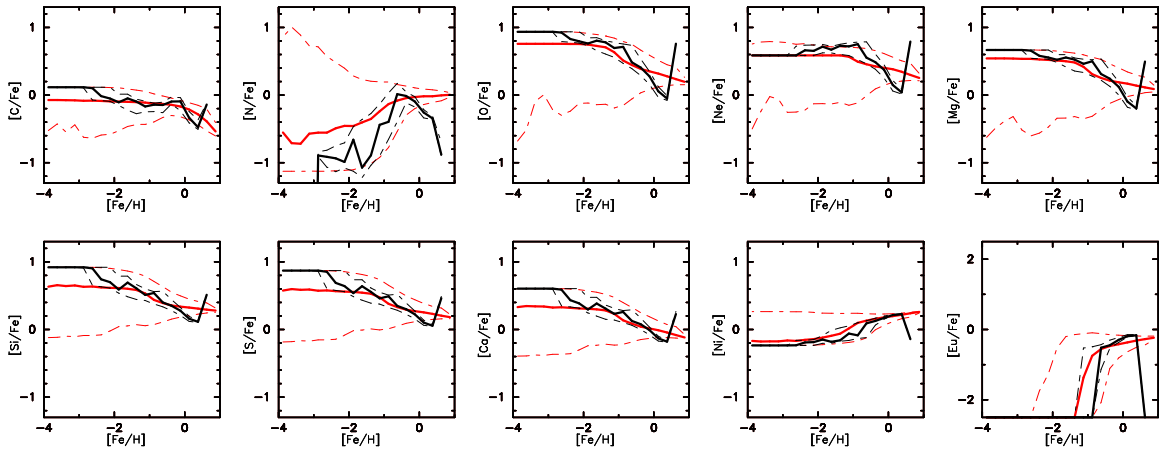}
\caption{Same as figure \ref{fig:Chemodyn:XFeFeH:N13}, but with HNe ($f_{\rm HN}
= 0.5$), the Pop III IMF and NSMs.  Red and black curves are for runs I and K,
respectively.  Run K adopts the metal diffusion model of Eq
\eqref{eq:metaldiffusion}.
\vspace{0.1cm}
} 
\label{fig:Chemodyn:XFeFeH:NSMs}
\end{figure*}

\section{Summary and future prospects} \label{sec:Summary}

We developed a software library for the chemical evolution simulation of galaxy
formation, named ``CELib''.  This library adopts the SSP approximation and,
under this approximation, it provides the return mass of each element and
released energy from an SSP particle depending on feedback type. How these
quantities are redistributed is left to the user's simulation code.

All of the necessary data, such as IMFs, stellar lifetime data, and yields are
implemented in CELib, as well as reference feedback models.  The data is
obtained from the literature.  Most functions are selectable at the runtime,
and hence it is easy to compare the contribution of each model. The use of
CELib is easy if users follow the standard way because it consists of a limited
number of APIs.  Since CELib is a simple software library, it is easy to carry
out simulations of chemical evolution even for a beginner of this field.  Using
internal functions, it is also possible to construct one's favorite model of
chemical evolution.

As demonstrations of CELib, we showed the results of our simple one-zone models
and three-dimensional chemodynamical simulations in a dark matter halo.  The
yields for SNe II have a large impact on chemical evolution, as is pointed out
in previous studies.  When we compare the results with the yields table of
\cite{Nomoto+2013} and that of \cite{Portinari+1998} with slight modifications,
these two results are almost comparable except for the light elements, C and N.
On the other hand, if we use the original yields of \cite{Portinari+1998}, the
[X/Fe]-[Fe/H] relations become inconsistent with other results.

SNe Ia affect the position of the breaking point of the plateaus.  A different
yields table gives different final amounts of metals released by SNe Ia.  This
implies that we need to be careful with models and yields of SNe Ia when we
compare results with observations.

AGBs affect the evolutions of relatively light elements, such as C and N.  These
effects can be seen in both one-zone and three-dimensional simulations.  The
feedback from AGBs is sometimes not taken into account even in current
simulations of galaxy formation. However, they cannot be ignored since they are
very common in the ISM.

We provide the community with this library ({\tt
https://bitbucket.org/tsaitoh/celib}) and it is also archived on Zenodo
(10.5281/zenodo.190830).  We believe that this library will accelerate the
understanding of galaxy formation from the perspective of chemical evolution.

Following are the future prospects of this library. 

\begin{itemize}
\item Further flexibility would be necessary for IMF shapes.  It is more
convenient if each stellar particle can have its own IMF and mass range. The
importance of the top-heavy IMF in galaxy formation is well understood
\citep[e.g.,][]{Baugh+2005,Nagashima+2005}.

\item Further extension of yields tables is desirable.  For instance, recently,
the NuGrid collaboration published their yields tables \citep{Pignatari+2016}
which are based on the {\tt MESA} and {\tt GENEC} codes \citep{Paxton+2011,
Eggenberger+2008}.  The published yields are $Z=0.01$ and $0.02$. We will take
their yields when all data is published.  In the current version of this
library, all yields tables are obtained from models without stellar rotations.
It is pointed out that stellar rotation changes yields
\citep[e.g.,][]{Heger+2000, MeynetMaeder2002}.  The effect of the rotating zero
metal stars was studied in \cite{Kobayashi+2011} and they showed these stars
have significant effects on [C/Fe]-[Fe/H] and [N/Fe]-[Fe/H] relations and a
moderate effect on the [O/Fe]-[Fe/H] relation.

\item Further sophisticated treatment of the chemical enrichment might be
important. Strictly speaking, the yield of each element depends on the abundance
of other elements.  However, the current formulation implemented on CELib
ignores this effect.  As such, the $Q_{ij}$ formalism has been proposed
\citep{TalbotArnett1973, Ferrini+1992, Portinari+1998}.  With and without this
formulation, there might be certain differences in chemical enrichment
\citep{Martinez-Serrano+2008}. Note, however, that they assumed the solar
proportions, which is not ideal for understanding the $Q_{ij}$ formulation.

\item Further extension of available yields and isotopes are important to
connect studies of galaxy formation and planet formation. In the current version
of CELib, the distributions of isotopes are not considered. It is also important
to distinguish the contributions from different types of feedback. Generally,
only the data of long lived isotopes are provided.  {\footnote {The yields
tables of SNe Ia usually provides not only the amounts of stable elements but
also those of short-lived elements.}}.  This data is sufficient for galactic
chemical evolution studies.  However, in the case that the distributions of the
short lived isotopes are important, the current treatment of the isotopes is
insufficient. For example, it is pointed out that the radio isotope of
aluminum, $^{26}$Al, is abundant during the early age of the solar system
\citep[e.g.,][]{Lee+1976, Russell+1996, Jacobsen+2008, BouvierWadhwa2010,
Larsen+2011} and the decay heat of $^{26}$Al, whose half-time period is
$0.72~{\rm Myr}$, is considered to be a primary source of the Earth's early
evolution \citep[e.g.,][]{Urey1955, Castillo-Rogez+2009, Elkins-Tanton+2011}, as
well as these of $^{235}$U, $^{238}$U, $^{232}$Th, and $^{40}$K. To deal with
the decays of all of radioactive isotopes is unrealistic.  At least it is
necessary to deal with the decays of some of the important radioactivate
isotopes to understand the formation history of stars and planets from the
perspective of galactic chemodynamical simulations.
\end{itemize}

It is certain that star-by-star simulations are the next breakthrough of galaxy
formation simulation since there are an enormous amount of evidence that massive
stars have crucial impacts on the galaxy formation and evolution.  In some first
star and first galaxy simulations, Pop III stars are dealt with discrete stars
sampled from a Pop III IMF instead of using the SSP approximation
\citep[e.g.,][]{WiseAbel2008, Greif+2010, Wise+2012, Wise+2014, Ritter+2015,
O'Shea+2015, Smith+2015} or its formation is directly followed
\citep[e.g.,][]{Hirano+2014, Hosokawa+2016}.  While such treatment of stars is
beyond the original scope of CELib, the all necessary data used for the chemical
evolution of star-by-star simulations have been implemented. Hence, CELib
provides APIs which can be used by star-by-star simulation (see appendix
\ref{sec:APIs:starbystar}).

\bigskip

\acknowledgements

The author thanks the anonymous referee who gave constructive and helpful
comments that improved this study.  The author also thanks Yutaka Hirai, Takashi
Okamoto, Junichi Baba, Daisuke Kawata, Ko Nakamura, Takuma Suda and Yutaka
Katsuta who gave important input for this study.  A part of numerical
simulations was carried out on the Cray XC30 system in the Center for
Computational Astrophysics at the National Astronomical Observatory of Japan.
This work is supported by a Grant-in-Aid for Scientific Research (26707007) of
Japan Society for the Promotion of Science and Strategic Programs for Innovative
Research of the Ministry of Education, Culture, Sports, Science and Technology
(SPIRE).

\appendix 

\section{Application Interfaces} \label{sec:APIs}

Here we describe the major functions of CELib.  All available functions are
defined in {\tt CELib.h}.

\subsection{Initialize CELib}

This library is initialized by just calling this function;
\begin{verbatim}
  void CELibInit(void);
\end{verbatim}
When this function is called, the IMF and its mass range are fixed.  Then under
these conditions, the IMF weighted yields of the adopted yield tables are
computed. The smoothed lifetime functions and necessary data for reference
feedback models are also computed.

This function should be called at the beginning of a simulation using this
library. Every model's parameters are fixed at this time.  If a user wants to use
different model parameters, the user needs to call this function again after
resetting new parameters.

\subsection{Get event time}

With reference feedback models, the user can easily obtain event times.  In
order to obtain an event time of an SSP particle, the user has to use the
following function;
\begin{verbatim}
double CELibGetNextEventTime(struct CELibStructNextEventTimeInput Input, const int Type);
\end{verbatim}
This function returns the event time of a target event in units of year.

This function requires two arguments. The first argument is a structure defined
in \verb+CELib.h+ and the definition of it is
\begin{verbatim}
struct CELibStructNextEventTimeInput{
    double R;                   // A random real number in [0,1)
    double InitialMass_in_Msun; // An initial mass of the target SSP particle
                                //   in units of the solar mass
    double Metallicity;         // A metallicity of the target SSP particle
    int Count;                  // A counter for a target event 
                                // This is used in SNe Ia/AGBs/NSMs
                                // Count should start zero.
}; 
\end{verbatim}

The second argument of this reference API is used to specify the feedback type.
Feedback types are defined as \verb+enum+ and the user should select one out of
four;
\begin{verbatim}
enum {
    CELibFeedbackType_SNII,
    CELibFeedbackType_SNIa,
    CELibFeedbackType_AGB,
    CELibFeedbackType_NSM,
    CELibFeedbackType_Number,
};
\end{verbatim}

For example, we consider a case in which the user wants to obtain the explosion
time of SNe II using the reference model. In this case, the user has to call the
reference API like this; 
\begin{verbatim}
struct CELibStructNextEventTimeInput Input = {
        .R = A_r,                      // A random real number in [0,1)
        .InitialMass_in_Msun = M_ssp,  // The initial mass of the target SSP particle
        .Metallicity = Z_ssp,          // The metallicity of the target SSP particle
        }; 
double t_snII = CELibGetNextEventTime(Input,CELibFeedbackType_SNII);
\end{verbatim}
\verb+Count+ is not used to obtain the feedback time of SNe II and thus it
is ignored in this case. The return value \verb+t_snII+ is the explosion time
which follows figures \ref{fig:SNII:Rate:N13} or \ref{fig:SNII:Rate:P98}. If the
simulation time is $t_{\rm sim}$, the feedback event takes place at $t_{\rm sim}
+ t_{\rm snII}$.

\subsection{Get released masses of metals and energy}

The released masses of metals and energy are also easily obtained by using a
reference API. When the time in a simulation reaches the event time, we need to
call the following function; 
\begin{verbatim}
struct CELibStructFeedbackOutput 
                CELibGetFeedback(struct CELibStructFeedbackInput Input, const int Type);
\end{verbatim}

The first argument is the structure defined in \verb+CELib.h+ and it holds 
all the necessary data to evaluate the feedback event. The structure is  
\begin{verbatim}
struct CELibStructFeedbackInput{
    double Mass;                 // The mass of the target SSP particle in simulation unit
    double Metallicity;          // The metallicity of the target SSP particle
    double MassConversionFactor; // A factor to convert Elements[] from 
                                 //   the simulation mass unit to the solar mass
    double *Elements;            // The pointer to the array of elements for 
                                 //   the target SSP particle in the simulation mass unit
    int Count;                   // A counter of the target event
};
\end{verbatim}

The second argument is the type of feedback and it is the same as that used to
obtain the feedback time. 

The results are loaded to a structure whose type is \verb+struct CELibStructFeedbackOutput+.
The definition of it is as follows:
\begin{verbatim}
struct CELibStructFeedbackOutput{
    double Energy;                      // The released energy in units of erg
    double EjectaMass;                  // The ejecta mass in units of the solar mass
    double RemnantMass;                 // The remnant mass in units of the solar mass
    double Elements[CELibYield_Number]; // The mass of released metals in units of the solar mass
};
\end{verbatim}
Note that \verb+CELibYield_Number+ is defined in {\tt CELib.h} and is thirteen
in the current version.

For example, we show the way to obtain the results of SNe II feedback.
First the user needs to put all necessary data to the structure 
\verb+struct CELibStructFeedbackInput+ and then call \verb+CELibGetFeedback+
with \verb+CELibFeedbackType_SNII+.
\begin{verbatim}
struct CELibStructFeedbackInput Input = {
        .Mass = M_ssp,
        .Metallicity = Z_ssp,
        .MassConversionFactor = Mass_solar_mass/Mass_sim_unit,
        .Elements = Elements_ssp,
    }
struct CELibStructFeedbackOutput SNII = 
                CELibGetFeedback(Input,CELibFeedbackType_SNII);
\end{verbatim}
Units of energy and mass in \verb+SNII+ are erg for energy and $\Msun$.  If the
simulation units are different from these units, the user needs to convert them
into the simulation units.

\subsection{Select models and set model parameters}

All of the control parameters are stored in the 
structure \verb+struct CELibStructRunParameters+. CELib prepares a structure
\verb+CELibRunParameters+ and it is used to manage this library. The fiducial
model is the same as that used in \S \ref{sec:Chemodyn} as run F (see table
\ref{tab:Chemodyn:Runs}).  When the values in \verb+CELibRunParameters+ are
changed and the initializer is called, CELib recomputes all data and thus a new
simulation is ready to start.

\section{Application interfaces for star by a star simulation} \label{sec:APIs:starbystar}

CELib provides APIs for star-by-star simulations.  So far, CELib supports only
feedback from massive stars.

\subsection{Get event time for star by star simulations}

The function shown below is used in order to get the event time of a star:
\begin{verbatim}
double CELibGetNextEventTimeStarbyStar
          (struct CELibStructNextEventTimeStarbyStarInput Input, const int Type);
\end{verbatim}
This function returns the lifetime of a star in units of year, referring the
lifetime tables built in \S \ref{sec:LifeTime}.
The structure of the first argument, which is also defined in \verb+CELib.h+, is as follows:
\begin{verbatim}
struct CELibStructNextEventTimeStarbyStarInput{
    double InitialMass_in_Msun;  // Mass of the star
    double Metallicity;          // Metallicity of the star, Z
};
\end{verbatim} 
The second argument is the type of feedback.  Although only feedback from
massive stars is supported in the current version of CELib, we prepare this
argument for the future extension.  In the current version, \verb+Type+ should
be \verb+CELibFeedbackType_SNII+.

\subsection{Get released masses of metals and energy for star by star simulations}

Mass, metals, and energy released from a single event can be obtained by using
the following function:
\begin{verbatim}
struct CELibStructFeedbackStarbyStarOutput 
        CELibGetFeedbackStarbyStar(struct CELibStructFeedbackStarbyStarInput Input, const int Type);
\end{verbatim}
Again, the first argument is the structure defined in \verb+CELib.h+.  This
structure holds all the necessary data to evaluate the feedback event. Here is 
the members of the structure:
\begin{verbatim}
struct CELibStructFeedbackStarbyStarInput{
    double Mass;                 // Mass of the star in simulation unit.
    double Metallicity;          // Metallicity of the star Z
    double MassConversionFactor; // A factor to convert Elements[] from
                                 //   the simulation mass unit to Msun.
    double *Elements;            // Star particle's elements composition in simulation unit.
};
\end{verbatim}
The second argument used in \verb+CELibStructFeedbackStarbyStarOutput+ is the
feedback type and is also reserved for the future extension.

The results are stored to a structure, \verb+struct CELibStructFeedbackStarbyStarOutput+, 
of which definition is 
\begin{verbatim}
struct CELibStructFeedbackStarbyStarOutput{
    double Energy;                      // The released energy in units of erg
    double EjectaMass;                  // The ejecta mass in units of the solar mass
    double RemnantMass;                 // The remnant mass in units of the solar mass
    double Elements[CELibYield_Number]; // The mass of released metals in units of the solar mass
};
\end{verbatim}

% \bibliographystyle{apj}
% \bibliography{ms}

\end{document}